\documentclass[a4paper,11pt]{article}
\pdfoutput=1  

\usepackage{jheppub}  

\usepackage[T1]{fontenc} 

\usepackage{lipsum}  

\usepackage{multirow}
\usepackage{amsmath,mathrsfs}
\usepackage{amsfonts,amssymb}
\usepackage{accents}

\usepackage[x11names]{xcolor}

\usepackage{comment}

\usepackage{enumitem}

\usepackage{mwe,tikz}
\usepackage[percent]{overpic}
\usepackage{tikz-cd}

\usepackage{array,ragged2e}

\usepackage{bm}

\usepackage[normalem]{ulem}
\usepackage{caption}
\usepackage{subcaption}

\newcommand{\be}{\begin{equation}}
\newcommand{\ee}{\end{equation}}

\newcommand{\ba}{\begin{aligned}}
\newcommand{\ea}{\end{aligned}}

\newcommand{\nn}{\nonumber}

\newcommand{\cA}{\mathcal{A}}
\newcommand{\cB}{\mathcal{B}}
\newcommand{\cC}{\mathcal{C}}
\newcommand{\cD}{\mathcal{D}}
\newcommand{\cE}{\mathcal{E}}
\newcommand{\cF}{\mathcal{F}}

\newcommand{\cI}{\mathcal{I}}

\newcommand{\cK}{\mathcal{K}}
\newcommand{\cL}{\mathcal{L}}

\newcommand{\cO}{\mathcal{O}}

\newcommand{\cT}{\mathcal{T}}
\newcommand{\cU}{\mathcal{U}}
\newcommand{\cV}{\mathcal{V}}
\newcommand{\cW}{\mathcal{W}}

\title{\boldmath SymTFT for Continuous Symmetries: Non-linear Realizations and Spontaneous Breaking}

\author[a]{Federico Bonetti,}
\author[b,c,d]{Michele Del Zotto,}
\author[e]{Ruben Minasian}

\affiliation[a]{Departamento de Electromagnetismo y Electr\'onica, Universidad de Murcia, Campus de Espinardo, 30100 Murcia, Spain}
\affiliation[b]{Department of Mathematics, Uppsala University, Box 480, SE-75106, Uppsala, Sweden}
\affiliation[c]{Department of Physics and Astronomy, Uppsala University, Box 516, SE-75120, Uppsala, Sweden}
\affiliation[d]{Center for Geometry and Physics, Uppsala University, Box 480, SE-75106, Uppsala, Sweden}
\affiliation[e]{Institut de Physique Th\'{e}orique, Universit\'{e} Paris Saclay, CNRS, CEA, F-91191, Gif-sur-Yvette, France} 

\emailAdd{f.bonetti@um.es, michele.delzotto@math.uu.se, ruben.minasian@ipht.fr}

\abstract{It is well known that continuous symmetries of quantum fields can be realized non-linearly, e.g.~in the context of sigma models, and can also be spontaneously broken on non-compact spacetimes. In this note we study how these effects are realized in the context of the topological symmetry theory for continuous symmetries. In particular, we explain coset realizations and their higher $p$-form symmetry versions from this perspective, as well as uplifts to higher groups and non-invertible symmetries. Moreover, using a setup with boundaries and corners, we explore spontaneous symmetry breaking scenarios for higher $p$-form symmetries as well as non-Abelian $0$-form symmetries.}

\begin{document} 
\maketitle
\flushbottom

\newpage

\section{Introduction}

Symmetry plays a foundational role in modern theoretical physics, governing the structure of fundamental interactions and the behavior of phases in quantum and classical systems. In particular, continuous symmetries and their spontaneous breaking underpin a wide range of physical phenomena -- from chiral symmetry breaking in particle physics to Goldstone modes in condensed matter systems. Traditionally, the non-linear realization of spontaneously broken symmetries has been described using coset constructions developed by Coleman, Wess, and Zumino \cite{Coleman:1969sm}, and Callan et al.~\cite{Callan:1969sn}, which explicitly parameterize the space of Goldstone fields associated with the symmetry breaking pattern $G \to H$.

\smallskip

More recently, a shift in perspective has emerged, wherein symmetries of quantum field theory are generalized by topological defects and operators of various codimensions
\cite{Gaiotto:2014kfa}. Within this framework, symmetries generalize widely beyond their mere actions on local fields, leading to an effort to understand symmetries via their global and categorical structure 
-- see e.g.~\cite{Gomes:2023ahz,Schafer-Nameki:2023jdn,Brennan:2023mmt,Bhardwaj:2023kri,Shao:2023gho,Carqueville:2023jhb,Luo:2023ive,Costa:2024wks,Iqbal:2024pee} for some recent reviews and further references on this rapidly developing subject. A powerful strategy to capture topological defects and operators is to exploit an isomorphism of the theory of interest, $\cT_d$, with a bulk-boundary system \cite{Kapustin:2014gua,Ji:2019jhk,Gaiotto:2020iye,Apruzzi:2021nmk,Freed:2022qnc,Kaidi:2022cpf}, consisting of a bulk topological field theory in one higher dimension, $\mathcal S_{d+1}$, the \textit{topological symmetry theory} (or SymTFT for short), placed along a finite interval with two boundaries 
\be\label{eq:symTFT}
\begin{tikzpicture}
\draw [LightBlue2, fill=LightBlue2] (0,0) -- (0,1.5) -- (3,1.5) -- (3,0) -- (0,0) ;
\draw [very thick] (0,0) -- (0,1.5) ; 
\draw [very thick] (3,0) -- (3,1.5) ; 
\node at (1.5,0.75) {$\mathcal S_{d+1}$} ;
\node at (3.75,0.75) {$\cong$} ;
\draw [very thick] (4.5,0) -- (4.5,1.5) ; 
\node[below] at (0,0) {$B^d_{\rm sym}$}; 
\node[below] at (3,0) {$B^d_{\rm phys}$};
\node[below] at (4.5,0) {$\cT_d$}; 
\end{tikzpicture}
\ee
where $B^d_{\rm sym}$ is a topological boundary condition encoding the generalized symmetry of the theory and $B^d_{\rm phys}$ is a boundary condition that typically supports a relative field theory that couples to the topological bulk, responsible for realizing the dynamics of $\cT_d$. Originally developed in the context of finite symmetries, recently some proposals to include the continuous symmetries in this formalism have appeared in the literature \cite{Brennan:2024fgj,Antinucci:2024zjp,Bonetti:2024cjk,Antinucci:2024bcm} -- see also \cite{Argurio:2024ewp,Paznokas:2025auw,Arbalestrier:2025poq,Jia:2025jmn}. The SymTFT constructions so far have dealt with linearly realized continuous symmetries (with some notable exceptions, e.g.~\cite{Argurio:2024ewp,Paznokas:2025auw,Apruzzi:2025mdl}), so it is natural to ask how to generalize the construction in the context of non-linear realization of the symmetries, which happens 
e.g.~when $\mathcal T_d$ is a non-linear $\sigma$-model (which are also known to have interesting higher symmetry structures \cite{Chen:2022cyw,Hsin:2022heo,Chen:2023czk,Pace:2023kyi,Pace:2023mdo,Sheckler:2025rlk}). In this work we address this question and recover the Callan-Coleman-Wess-Zumino universal effective actions from a SymTFT perspective, leading to the non-linearly realized symmetries for all coset models of type~$G/H$.

\smallskip

Another interesting aspect of the structure of symmetries in quantum field theory is given by their spontaneous breaking, which can happen when the field theory is placed on a non-compact spacetime. In this context, the fields acquire boundary conditions at infinity which transform non-trivially with respect to the symmetry action, leading to a vacuum degeneracy that can be detected by suitable order parameters picking non-trivial vevs. In the theory of generalized symmetries the spontaneous breaking is among the key applications, leading to hierarchies associated with higher structures \cite{Cordova:2018cvg,Cordova:2022rer,Cordova:2022ieu} as well as interpreting various massless fields with higher spins in terms of generalized Nambu-Goldstone bosons \cite{Gaiotto:2014kfa} -- see also \cite{Lake:2018dqm,Hofman:2018lfz}. However, the isomorphisms leading to the SymTFT for continuous symmetries discussed in the literature so far typically involve only compact spacetimes, thus defying the symmetry breaking scenarios. In the case of finite symmetries, SymTFT with boundaries and corners have been introduced which served as a strong inspiration for our work here \cite{
Copetti:2024onh,
Cordova:2024iti,
Cvetic:2024dzu,
Copetti:2024dcz,
GarciaEtxebarria:2024jfv,
Choi:2024tri,
Das:2024qdx,
Bhardwaj:2024igy,
Heymann:2024vvf,
Choi:2024wfm}. In this paper we begin an exploration of spontaneous symmetry breaking (SSB) of continuous symmetries from the perspective of the SymTFT precisely by studying it in a setup with boundaries and corners. We stress this is just the tip of an iceberg and we expect many further applications from the methods we begin developing here. In particular even if we limit ourselves to considering the simplest cases of higher form symmetries and non-Abelian zero form symmetries, our techniques apply uniformly to models with more complicated symmetry categories. We expect to be able to learn more about the SSB of non-invertible symmetries using this approach in future work, focusing on the implications of SSB on the higher structures building on \cite{Copetti:2023mcq,DelZotto:2024ngj,DelZotto:2024arv,Cordova:2025eim,Gagliano:2025gwr}.

\smallskip

This work is organized as follows. In Section \ref{sec_SymTFT_review} we revisit the construction of the SymTFT for continuous zero-form symmetries. These models are non-Abelian BF theories \cite{Horowitz:1989ng}, which in particular have topological Wilson line operators and codimension two defects, whose worldvolume theory is a non-trivial BF theory of its own \cite{Cattaneo:1996pz,Cattaneo:2000mc,Cattaneo:2002tk}. We present alternative derivations for several of the properties of these topological defects, including their linking, which we bring to a test in some examples of interest -- we explain how in presence of Chern-Simons terms, higher linking invariants can be detected by the Horowitz theory. We also discuss a proposal to describe non-flat backgrounds in the context of the non-Abelian BF theory. Along the way we also present some preliminary remarks about the fusion ring for the codimension two defects that these theories have. In Section \ref{sec:nonlinearly} we present our first core result, describing boundary conditions for the SymTFTs with continuous symmetries giving rise to non-linearly realized symmetries. We present several examples where our formalism can be used to derive well-known effective actions \textit{\`a la} Callan-Coleman-Wess-Zumino. In particular, we recover EM dualities (and T-dualities) from the action of a bulk EM duality defect in the SymTFT. Finally in Section \ref{sec:SSB} we present some comments about SSB. To this end we consider a setup where the continuous SymTFT is placed on a space with boundaries and corners. This gives rise to further possible boundary conditions at infinity. We find boundary conditions corresponding to unbroken symmetries that are left invariant by the symmetry action, as well as boundary conditions associated to spontaneous breaking that indeed come in families organized by the action of the symmetry (and leading to the correct geometry expected for Nambu-Goldstone degrees of freedom). In particular, this setup allows to recover the well-known spontaneous symmetry breaking Ward identities, forcing one-point functions of order parameters to zero when the symmetry is unbroken by the choice of boundary conditions. 

\smallskip 

\noindent
\textit{Note added}: While this work was being finalized, the paper \cite{Apruzzi:2025hvs} appeared, which has some minor overlap with our results in Section \ref{sec:nonlinearly}.

\section{Revisiting the continuous SymTFT}
\label{sec_SymTFT_review}

In this section we revisit the construction of the SymTFT with continuous symmetries. The purpose is to streamline properties of the bulk non-Abelian BF theory and its topological operators, and present some alternative derivations. Among the results which follow from this analysis we recover the interpretation suggested in \cite{Bonetti:2024cjk} that codimension two defects of BF theory   act as Gukov-Witten operators on Wilson lines. Along the way we also discuss some slight generalizations. For example we discuss the topological defects that arise when the bulk BF theory couple to a non-Abelian Chern-Simons term, and give a proposal how to couple the non-Abelian BF theory to non-flat backgrounds.

\subsection{Bulk topological operators}
\label{sec_bulk_top_operators}

\subsubsection{BF action in the bulk}
\label{sec_define_BF}

Let $G$ be a compact Lie group. Its Lie algebra $\mathfrak g$ admits an Ad-invariant positive definite inner product, which we denote Tr
and
which gives an isomorphism
$\mathfrak g^* \cong \mathfrak g$ between the Lie algebra and its dual.  
The action for the BF theory based on the gauge group $G$ reads \cite{Horowitz:1989ng}
\be \label{eq_BF}
S = \frac{1}{2\pi} \int_{X^{d+1}}
{\rm Tr} (B \wedge F_A) \ .
\ee 
The field $A$ is a $G$-connection with field strength
\be 
F_A = dA + \tfrac 12 [A,A] \ . 
\ee 
The field $B$ is a $\mathfrak g$-valued $(d-1)$-form. The action 
\eqref{eq_BF} is invariant under the following gauge transformations of $A$ and $B$,
\be \label{eq_BF_gauge_trans}
\ba
A  &\mapsto g(d+A) g^{-1} \ , &
&\qquad \qquad \text{hence} & 
F_A  & \mapsto g F_A g^{-1} \ , \\ 
B & \mapsto g(B - d_A \tau ) g^{-1} \ .
\ea 
\ee 
In the above expressions, $g$ is a $G$-valued 0-form, $\tau$ is a $\mathfrak g$-valued
$(d-2)$-form, and 
\be 
d_A \tau = d\tau + [A,\tau] \ . 
\ee 
Invariance of \eqref{eq_BF} under    transformations
\eqref{eq_BF_gauge_trans}
with $\tau \neq 0$
hinges on the Bianchi identity 
\be 
0 = d_A F_A = dF_A + [A,F_A] \ . 
\ee 
The bulk equations of motion derived from varying
\eqref{eq_BF} read
\be 
F_A = 0 \ , \qquad 
0=d_A B = dB + [A,B]  \ . 
\ee 
The field $B$ 
can be rescaled by any nonzero constant.
We have chosen a convenient normalization.

\subsubsection{Wilson line operators}
 
The theory \eqref{eq_BF} admits topological Wilson loop operators 
\be \label{eq_Wilson_loop}
{\mathbf W}_{\mathbf R} (\gamma)= {\rm Tr}_{\mathbf R}{\rm Pexp} \int_{\gamma} (-A) \ ,
\ee 
where $\gamma$ is a loop in $X^{d+1}$
and $\mathbf R$ is an irreducible representation of $G$. The operator
\eqref{eq_Wilson_loop} is topological because the connection $A$ is flat on-shell.

For future reference, we recall that
the path-ordered exponential
in \eqref{eq_Wilson_loop} is defined in terms of
the parallel transport operator.
For  $\gamma: [0,1] \rightarrow X^{d+1}$
a piecewise smooth path, we introduce the $G$-valued quantity
\be 
\Gamma(\gamma)_0^s \; : \;\;
\text{parallel transport 
with connection $A$ from $\gamma(0)$ to $\gamma(s)$.}
\ee 
The quantity $\Gamma(\gamma)_0^s$ satisfies the 
$\mathfrak g$-valued ODE
\be \label{eq_transport_ODE}
\frac{d\Gamma(\gamma)_0^s}{ds}   \, 
\Big( \Gamma(\gamma)_0^s \Big)^{-1}
+ A_\mu (x(s)) \frac{dx^\mu(s)}{ds} = 0 
 \ , 
\ee 
where $x^\mu(s)$ are the coordinates of
the point
$\gamma(s)$,
together with the initial condition
\be 
\Gamma(\gamma)_0^0 = 1 \ , 
\ee 
where $1$ denotes the identity element in $G$.
Under a gauge transformation
\eqref{eq_BF_gauge_trans},
\be 
\Gamma(\gamma)^s_0 \mapsto  g(\gamma(s)) \;
\Gamma(\gamma)^s_0 \;
g(\gamma(0))^{-1} \ . 
\ee 
For an infinitesimal displacement,
\be 
\Gamma(\gamma)_0^s = 1 - s A_\mu(x(0)) \frac{dx^\mu(0)}{ds} + \cO(s^2) \ . 
\ee 
If we specialize to the case of a loop
with basepoint
$\gamma(0) = \gamma(1)= P$,
we have
\be 
{\rm Pexp} \int_{\gamma} (-A)
= \Gamma(\gamma)^1_0 \ , \qquad 
{\rm Pexp} \int_{\gamma} (-A) \mapsto 
g(P) \bigg[ {\rm Pexp} \int_{\gamma} (-A) 
\bigg] g(P)^{-1} 
\ . 
\ee 

\paragraph{Non-Abelian Stokes' formula.}
Below we will need a formula
for the variation of the parallel transport operator under a deformation of the path $\gamma$, which is sometimes referred to as
the non-Abelian Stokes' formula \cite{Polyakov:1980ca,Bralic:1980ra,Fishbane:1980eq}.
To make our exposition self-contained,
we briefly review its derivation.

Let us
fix two points $P$, $Q$ on $X^{d+1}$
(not necessarily distinct)
and let us
consider 
a homotopy
$\gamma_t(s)$ between two paths $\gamma_0(s)$,
$\gamma_1(s)$ connecting $P$ and $Q$,
\be 
t\in [0,1] \ , \qquad 
s\in [0,1] \ , \qquad 
\gamma_t(s) \in X^{d+1} \ , \qquad 
\gamma_t(0) = P \ , \qquad 
\gamma_t(1) = Q \ . 
\ee 
We   define 
\be 
\Gamma(\gamma_t)_0^s  \; = \;  
\begin{array}{l}
\text{parallel transport with connection $A$ }
\\
\text{along the path $s \mapsto \gamma_t(s)$
from $\gamma_t(0) $
to $\gamma_t(s)$.}
\end{array}
\ee 
From \eqref{eq_transport_ODE} we know that, for each fixed $t$,
\be 
\frac{\partial \Gamma(\gamma_t)_0^s}{\partial s}   \, 
\Big( \Gamma(\gamma_t)_0^s \Big)^{-1}
+ A_\mu (x(s,t)) \frac{\partial x^\mu(s,t)}{\partial s} = 0 
 \ , 
\ee 
where $x^\mu(s,t)$ are the coordinates
of the point $\gamma_t(s)$.
Taking a derivative with respect to $t$, and rearranging some terms by collecting a total $s$ derivative, we get 
\be 
\ba 
0 & = 
\frac{\partial}{\partial s} \bigg[
\Big( \Gamma(\gamma_t)_0^s \Big)^{-1} \frac{\partial  \Gamma(\gamma_t)_0^s}{\partial t}
+ \frac{\partial x^\mu(s,t)}{\partial t}
\Big( \Gamma(\gamma_t)_0^s \Big)^{-1}  A_\mu \Gamma(\gamma_t)_0^s
\bigg]
\\
& + 
\frac{\partial x^\mu(s,t)}{ \partial t} 
\frac{\partial x^\nu(s,t)}{ \partial s} 
\Big( \Gamma(\gamma_t)_0^s \Big)^{-1}
 F_{\mu\nu}(x(s,t))
 \Gamma(\gamma_t)_0^s   \ ,
 \ea
\ee 
where $F_{\mu\nu} = \partial_\mu A_\nu - \partial_\nu A_\mu +[A_\mu,A_\nu]$
are the components of the field strength of $A$.
Next, we integrate in $s$ from $0$ to $1$,
\be \label{eq_nonAb_Stokes}
\ba 
0 & = 
\Big( \Gamma(\gamma_t)_0^1 \Big)^{-1} \frac{\partial  \Gamma(\gamma_t)_0^1}{\partial t}
 + \int_0^1 ds
\frac{\partial x^\mu(s,t)}{ \partial t} 
\frac{\partial x^\nu(s,t)}{ \partial s} 
\Big( \Gamma(\gamma_t)_0^s \Big)^{-1}
 F_{\mu\nu}(x(s,t))
 \Gamma(\gamma_t)_0^s   \ .
 \ea
\ee 
We have recalled that
$\Gamma(\gamma_t)_0^0 = 1$ for any $t$,
hence $\partial_t \Gamma(\gamma_t)_0^0 = 0$,
and similarly that
$\gamma_t(0) = P$, $\gamma_t(1) = Q$ for any $t$, hence
$\partial_t x^\mu(0,t)=0$,
$\partial_t x^\mu(1,t)=0$.
Equation
\eqref{eq_nonAb_Stokes} is the sought-for
non-Abelian Stokes' formula.

As a trivial application,
let us specialize to a flat connection.
From \eqref{eq_nonAb_Stokes}
we recover the well-known fact that
the parallel transport 
$\Gamma(\gamma_t)_0^1$ is independent of $t$, namely, that the result of parallel transport depends on the initial and final points $P$, $Q$,
but does not change if we perform a small deformation of the path connecting them.
A slightly less trivial application of \eqref{eq_nonAb_Stokes}
is discussed in Section \ref{sec_Bop_as_GW}.

\subsubsection{$B$-operators}

The theory \eqref{eq_BF} also admits 
 a class of codimension-2 topological operators, which we refer to as $B$-operators. They are
supported on submanifolds without boundary, 
are labeled by an element
$X_0 \in \mathfrak g$, and we denote them
$\mathbf Q_{X_0}(\Sigma^{d-1})$.
They can be described by the following exponentiated action \cite{Cattaneo:2002tk,Jia:2025jmn},
\be \label{eq_def_B_op}
\mathbf Q_{X_0} (\Sigma^{d-1}) \;: \;\; 
\int [\cD U] [\cD \beta] \exp \bigg \{
i \int_{\Sigma^{d-1}} {\rm Tr}
\bigg[ 
(B + d_A \beta) U X_0 U^{-1}
\bigg]
\bigg \} \ . 
\ee 
The bulk fields $A$, $B$ are pulled back from
$X^{d+1}$ to $\Sigma^{d-1}$, but in our notation the pullback is implicit.
The fields $U$, $\beta$ are localized on the support $\Sigma^{d-1}$ of the $B$-operator.
In particular, $U$ is a $G$-valued 0-form
and $\beta$ is a $\mathfrak g$-valued $(d-2)$-form.
The action in \eqref{eq_def_B_op} is invariant under bulk gauge transformations 
\eqref{eq_BF_gauge_trans} accompanied by the following transformations of the localized fields $U$, $\beta$,
\be \label{eq_U_and_beta_gauge}
\ba
U  & \mapsto g U \ ,  & 
\beta & \mapsto  g( \beta + \tau ) g^{-1} \ , &
&\qquad  \text{hence} & 
d_A\beta  & \mapsto g (d_A\beta + d_A \tau) g^{-1}  \ . 
\ea 
\ee 
We assume that the path integral measure $[\cD U]$ is left-invariant,
and that $[\cD \beta]$ is translation and Ad-invariant.\footnote{\, Heuristically, we can describe the measure $[\cD U]$ as 
$\prod_{x\in \Sigma^{d-1}} d\mu(U(x))$,
where $\mu$ denotes the Haar measure on $G$.
Incidentally, we focus on the case of compact $G$, in which the Haar measure is both left and right invariant.
The field $\beta$ takes values in $\mathfrak g$,
which is a vector space equipped with an Ad invariant positive definite inner product, for compact $G$.
The measure $[\cD \beta]$ is heuristically the
product over $x\in \Sigma^{d-1}$
of the standard measure on 
a Euclidean vector space 
with a positive definite inner product.
}
The action \eqref{eq_def_B_op} enjoys another local redundancy.
To describe it, we define 
the subgroup of $G$
that stabilizes $X_0$ under the adjoint action
$X_0 \mapsto   h X_0 h^{-1}$, 
\be \label{eq_Stab_defined}
{\rm Stab}(X_0) := \{ h\in G| h X_0 h^{-1}= X_0 \} \ .  
\ee 
With this notation, 
\eqref{eq_def_B_op} is invariant under 
\be \label{eq_local_U_redundancy}
U \mapsto Uh \ , 
\ee 
where $h$ is any ${\rm Stab}(X_0)$-valued 
0-form on $\Sigma^{d-1}$.

\paragraph{$B$-operators are labeled by adjoint orbits.}
Let us now discuss in greater detail the  constant parameter $X_0$ labeling the operator
$\mathbf Q_{X_0}(\Sigma^{d-1})$.
We can always perform a field redefinition of $U$
of the form $U =   \widetilde U g_0$, where $g_0$ is a constant element of $G$.
After this redefinition, the action
in terms of $\widetilde U$ has the same form
as
\eqref{eq_def_B_op} with $X_0$
replaced by  $g_0 X_0 g_0^{-1}$.
This shows that the operator
$\mathbf Q_{X_0} (\Sigma^{d-1})$
depends on $X_0$
only via its adjoint orbit
\be 
\cO_{X_0} := \{ g_0 X_0 g_0^{-1}, g_0\in G \} \subset \mathfrak g \ . 
\ee 
The adjoint orbit $\cO_{X_0}$
is isomorphic to the coset  
$G/{\rm Stab}(X_0)$,
with ${\rm Stab}(X_0)$ as in 
\eqref{eq_Stab_defined}.

We are mainly interested in the case in which $G$ is a compact connected  Lie group.
In this setting, it is known (see e.g.~Chapter 5, Lemma 3 in \cite{kirillov2004lectures})
that every adjoint orbit of $G$
intersects each Weyl chamber in the
Cartan subalgebra $\mathfrak t$ of $\mathfrak g$
in exactly one point. 
Thus, without loss of generality, we can 
choose the fundamental Weyl chamber and take 
\be 
X_0 \in \text{fundamental Weyl chamber}\subset \mathfrak t  \ .  
\ee

\paragraph{Worldvolume theory on a $B$-operator.}
Due to the local redundancy
\eqref{eq_local_U_redundancy},
the $G$-valued scalar field $U$ in the 
$(d-1)$-dimensional system with action 
\eqref{eq_def_B_op} 
describes
a sigma model with  
 target space $G/{\rm Stab}(X_0)$.
In particular, we have the map
\be \label{eq_orbit_map}
G/{\rm Stab}(X_0) \xrightarrow{\sim} \cO_{X_0} \ , \qquad 
\text{(coset of $U$)}  \mapsto  U X_0 U^{-1} \ , 
\ee 
which describes the embedding of the abstract coset 
$G/{\rm Stab}(X_0)$ into the 
adjoint orbit $\cO_{X_0}$
as a submanifold of $\mathfrak g$.
For example, for $G=SU(2)$
the Lie algebra $\mathfrak g$ is 
$\mathbb R^3$ as a vector space 
and the adjoint orbits are
2-spheres centered at the origin of
$\mathbb R^3$.

We can also describe this system in 
the language that is often used for
sigma models onto a coset space. 
To this end, we choose a decomposition
of  $\mathfrak g$   of the form
\be 
\mathfrak g = \mathfrak {stab}(X_0) \oplus 
\mathfrak f \ , 
\ee 
where $\mathfrak {stab}(X_0)$ is
the Lie algebra of the subgroup
${\rm Stab}(X_0)$ of $G$,
thus a subalgebra of $\mathfrak g$.
The summand $\mathfrak f$ is 
the orthogonal complement of $\mathfrak {stab}(X_0)$ in $\mathfrak g$
with respect to the Cartan-Killing metric. (Here we assume that 
$G$ is a connected, simple, compact Lie group.)
We decompose the $\mathfrak g$-valued
1-form $U^{-1}(d+A)U$ as
\be \label{eq_stabH_decomp}
U^{-1}(d+A)U = Q +P \ , \qquad 
Q \in \mathfrak {stab}(X_0) \ , \qquad 
P \in \mathfrak f \ . 
\ee 
Under 
a combined gauge transformation
\eqref{eq_U_and_beta_gauge} and local redundancy
\eqref{eq_local_U_redundancy},
$Q$ transforms as
a ${\rm Stab}(X_0)$-connection
while
$P$  transforms homogeneously,
\be \label{eq_P_and_Q_transform}
U \mapsto gUh \ , \qquad
P \mapsto  h^{-1} P h \ , \qquad 
Q \mapsto h^{-1} (d+Q) h \ . 
\ee 
Notice how $P$ and $Q$ are inert under $g$ transformations.

We now discuss the equations of motions for $U$ and $\beta$.
Let us first consider the variation 
of \eqref{eq_def_B_op}
with respect to $\beta$. Upon integration by parts, we get
\be  \label{eq_beta_EOM}
d_A (U X_0 U^{-1}) = 0 \ . 
\ee 
This condition shows that,
on-shell, the profile of the sigma model scalars 
corresponds to a 
 (covariantly) constant position $U X_0 U^{-1}$ on the orbit $\cO_{X_0}$,
 see \eqref{eq_orbit_map}.
Now, we compute 
\be 
d_A(U X_0 U^{-1}) = U[ U^{-1}(d+A)U , X_0] U^{-1} \ , 
\ee 
where we have used the fact that $X_0$ is a constant parameter.
We learn that $U^{-1}(d+A)U$ commutes with $X_0$, hence
$U^{-1}(d+A)U \in \mathfrak{stab}(X_0)$. As a result, in the decomposition \eqref{eq_stabH_decomp}, on-shell we have  
\be 
P=0 \ . 
\ee 
In other words, \eqref{eq_def_B_op} describes a topological sigma model, without the familiar
${\rm Tr}(P \wedge *P)$ kinetic term for the scalars parametrizing the coset. Correspondingly, the scalars 
satisfy the  ``flatness''
condition $P=0$ on-shell.

Finally, let us record  the equation of motion
deriving from the variation
of
\eqref{eq_def_B_op} with respect to $U$.
It can be written as
\be 
[ B+ d_A \beta  , U X_0 U^{-1}   ] = 0 \ . 
\ee

\paragraph{Alternative description of  $B$-operators.}

We have seen above that the $\beta$ equation of motion imposes
$P=0$. 
For simplicity, let us consider the case in which $A$ restricted to the support of the $B$-defect is zero.
Then, the condition $P=0$ 
states that $U^{-1} dU$ has no component along $\mathfrak f$.
This
implies that $U$ can be written as
$U = U_0 h$, where $U_0 \in G$ is
constant and $h\in {\rm Stab}(X_0)$. 
(While we do not have a proof of the last statement, we have checked several examples.)
By means of a local ${\rm Stab}(X_0)$ transformation, we can then set $U = U_0$.
The path integral over $U$ then reduces to an ordinary integral over the group $G$ with the Haar measure $d\mu$.
We arrive at the following presentation of the $B$-operator
(see also \cite{Cordova:2022rer,Jia:2025jmn}),
\be \label{eq_average_presentation}
\mathbf Q_{X_0}(\Sigma^{d-1}) \; : \;\; 
\int_G d\mu(U) \exp  \bigg\{ i \int_{\Sigma^{d-1}}
{\rm Tr} (B \, UX_0 U^{-1})
\bigg\}
 \ .
\ee 
We refrain from fixing the precise normalization of the operator $\mathbf Q_{X_0}(\Sigma^{d-1})$.
Notice that we could alternatively integrate over constant $U$ valued in $G/{\rm Stab}(X_0)$, since the integrand is invariant under $U \rightarrow U h_0$ with $h_0\in {\rm Stab}(X_0)$ constant.

In the case of general $A$,
we  expect 
that
$P=0$ implies that $U$ can be written in the form $U = U_0 h$ where $h\in {\rm Stab}(X_0)$
and $U_0 \in G$ is covariantly constant, in the sense that it satisfies
$U_0^{-1}(d+A)U_0 =0$.
Then, the path integral over $U$ reduces to integrating over covariantly constant $U_0$
configurations. 
The resulting operator is still of the form \eqref{eq_average_presentation}, but now 
$\int_G d\mu(U)$ is understood
as the integral   over the space of solutions to the classical differential equation $U_0^{-1}(d+A) U _0= 0$, modulo local ${\rm Stab}(X_0)$-invariance.

\subsubsection{$B$-operators as Gukov-Witten operators} \label{sec_Bop_as_GW}
Let us insert a $B$-operator
$\mathbf Q_{X_0}(\Sigma^{d-1})$ and consider the total action of the bulk plus operator system,
\be 
S = \int_{X^{d+1}} \bigg\{ 
\frac{1}{2\pi} 
{\rm Tr}(B \wedge F_A)
+ {\rm Tr}
\bigg[ 
(B + d_A \beta) U X_0 U^{-1} \, \delta_2(\Sigma^{d-1} \subset X^{d+1})
\bigg]
\bigg\} \ . 
\ee 
We have made use of the Poincar\'e dual 
$\delta_2(\Sigma^{d-1} \subset X^{d+1})$
to $\Sigma^{d-1}$ in $X^{d+1}$.
Varying the total action with respect to $B$ gives
\be \label{eq_insert_B_op}
F_A + 2\pi U X_0 U^{-1} \, \delta_2(\Sigma^{d-1} \subset X^{d+1}) = 0  \ . 
\ee 
We see that the $B$-operator sources a delta-function localized profile for the field strength, equal to $-2\pi X_0$ up to
adjoint action of $G$.

\begin{figure} \centering
\includegraphics[width=8.5cm]{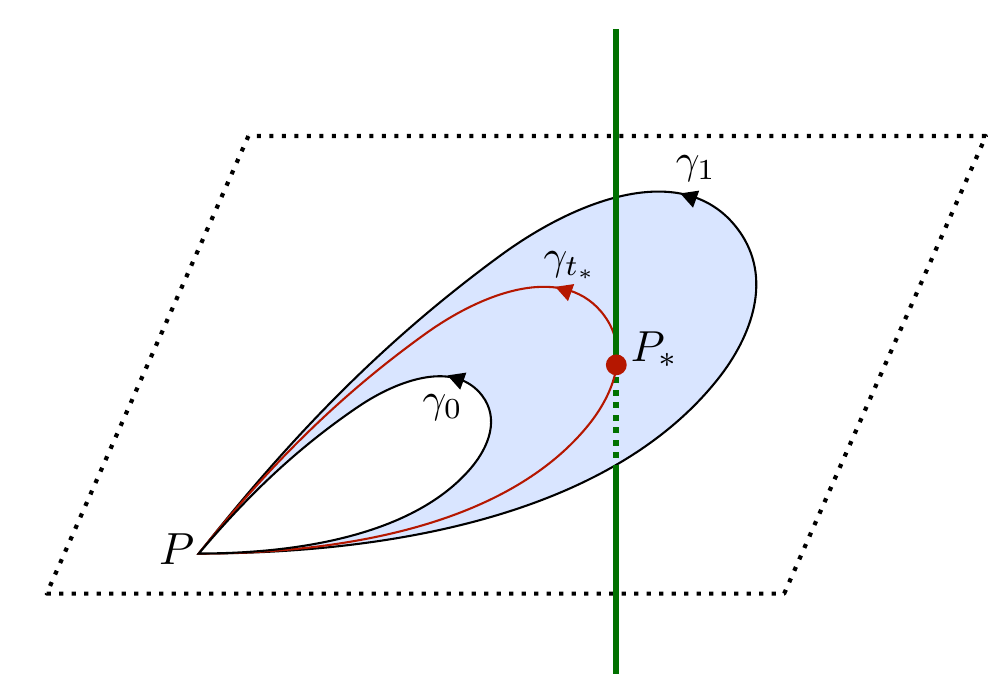}
\caption{A one-parameter family $\gamma_t$ of loops based at $P$ interpolates between $\gamma_0$ and $\gamma_1$ in the 2-plane transverse to a $B$-operator,
represented by the thick vertical line.
We depict in red the path $\gamma_{t_*}$
corresponding to the critical value
$t_*$ of $t$ for which the loop hits the $B$-operator.
}
\label{fig_Wilson_link}
\end{figure}

We now argue that \eqref{eq_insert_B_op}
allows us to characterize
the operator 
 $\mathbf Q_{X_0}(\Sigma^{d-1})$
as a disorder operator of Gukov-Witten type, namely, a codimension-2 operator defined
by requiring a prescribed holonomy of the gauge field $A$ in a small loop encircling it in the transverse space.
To this end we fix a point $P$ outside the support $\Sigma^{d-1}$ of the operator
and we consider a 1-parameter family 
$\gamma_t(s)$
of loops based at $P$,
\be 
t\in [0,1] \ , \quad 
s\in [0,1] \ , \quad 
\gamma_t(s) \in X^{d+1} \ , \quad 
\gamma_t(0) = P =\gamma_t(1)  \ . 
\ee 
We choose the family $\gamma_t$
as depicted in Figure \ref{fig_Wilson_link}.
In particular, $\gamma_0$ is homotopically 
trivial, while $\gamma_1$ links $\Sigma^{d-1}$ once.

As before, we use the notation
$\Gamma(\gamma_t)_0^s$
for the parallel transport along $\gamma_t$ from the point $s=0$ to the point with parameter $s$.
We recall that the holonomy of a flat connection along a homotopically trivial loop is trivial, hence
$\Gamma(\gamma_0)_0^1=1$.
Next, we apply the non-Abelian Stokes' formula \eqref{eq_nonAb_Stokes}. 
As we vary  $(s,t)$
parametrizing the  surface 
swept by the 1-parameter family of loops
$\gamma_t(s)$,
$F_{\mu \nu}(x(s,t))$ is everywhere zero,
except for a single point
$(s_*,t_*)$.
Indeed, there is a unique value $t_*\in (0,1)$ of $t$ such that the loop
$\gamma_{t_*}$ touches the support $\Sigma^{d-1}$ of the $B$-operator;
we use $s_*$ to denote the value of the
$s$ parameter along the loop $\gamma_{t_*}(s)$
such that $\gamma_{t_*}(s_*)$ is the point
of contact with $\Sigma^{d-1}$.
In other words,
$F_{\mu\nu}(s,t)$ in 
\eqref{eq_nonAb_Stokes}
is proportional to $\delta(s-s_*)\delta(t - t_*)$.
The factor $\delta(s-s_*)$ allows us to do immediately the integral in the variable $s$.
We are left with the following result,
\be  \label{eq_derivation_GW}
\ba  
\Big( \Gamma(\gamma_t)_0^1 \Big)^{-1} \frac{\partial  \Gamma(\gamma_t)_0^1}{\partial t}
 = 2\pi \delta(t-t_*)
\Big( \Gamma(\gamma_{t_*})_0^{s_*} \Big)^{-1}
U(x(s_*,t_*)) X_0 U(x(s_*,t_*))^{-1}
 \Gamma(\gamma_{t_*})_0^{s_*}   \ .
 \ea
\ee 
We notice that the RHS is $\delta(t-t_*)$
times a $t$-independent element in $\mathfrak g$.
This suggests the parametrization 
\be 
\Gamma(\gamma_t)_0^1 =\exp(f(t) Y) \ , 
\ee 
for some function $f(t)$ and some constant $Y\in \mathfrak g$.
Then,
$(\Gamma(\gamma_t)_0^1)^{-1} \partial_t \Gamma(\gamma_t)_0^1 = \partial_t f Y$.
We can satisfy
\eqref{eq_derivation_GW} 
together with the boundary condition
$\Gamma(\gamma_0)_0^1 = 1$
if we select
\be 
f(t) = \theta(t-t_*) \ , \qquad 
Y = 2\pi \Big( \Gamma(\gamma_{t_*})_0^{s_*} \Big)^{-1}
U(x(s_*,t_*)) X_0 U(x(s_*,t_*))^{-1}
 \Gamma(\gamma_{t_*})_0^{s_*}    \ .
\ee 
where $\theta$ is the Heaviside theta function.
With these identifications we conclude that
\be \label{eq_parallel_transport_two_cases}
\ba 
\Gamma(\gamma_t)_0^1 & = 1 \ , & 
&t < t_* \ , \\ 
\Gamma(\gamma_t)_0^1 & =  \Big( \Gamma(\gamma_{t_*})_0^{s_*} \Big)^{-1}
U(x(s_*,t_*)) e^{2\pi X_0} U(x(s_*,t_*))^{-1}
 \Gamma(\gamma_{t_*})_0^{s_*}
 \ , &
&t > t_*  \ . 
\ea 
\ee 
We have recalled $e^{gXg^{-1}}=ge^{X} g^{-1}$
for any $g \in G$, $X\in \mathfrak g$.
The second line encodes the desired result,
but we can formulate it in a more transparent way by introducing a more convenient notation. First, let us write
\be 
\Gamma(\gamma_t)_0^1 = {\rm hol}_{\gamma_t} \ , 
\ee 
in order to emphasize that the 
parallel transport operator
$\Gamma(\gamma_t)_0^1$ along the loop 
$\gamma_t$ is
the same as the holonomy of the $A$ connection along $\gamma_t$.
Moreover, let us denote as $P_*$ the contact point
$\gamma_{t_*}(s_*)$, and let us write 
\be 
\Gamma_P^{P_*} :=  \Gamma(\gamma_{t_*})_0^{s_*} \ . 
\ee 
Here we are using the fact that the connection is flat away from the support of the $B$-operator and hence 
$\Gamma(\gamma_{t_*})_0^{s_*}$ 
depends on $P$, $P_*$, but
does not depend on the specific shape of the 
portion of path $\gamma_{t_*}$ connecting
$P$ ($s=0$) to $P_*$ ($s=s_*$).
Finally, let us write
\be 
U(P_*) = U(x(s_*,t_*)) \ , 
\ee 
and consider $t=1$.
In conclusion, we get the equation
\be \label{eq_hol_firstway}
{\rm hol}_{\gamma_1} =  (\Gamma_P^{P_*} )^{-1}
U(P_*) e^{2\pi X_0} U(P_*)^{-1} \Gamma_P^{P_*} \ . 
\ee 
This confirms that 
 inserting the $B$-operator implies that the connection $A$ has a non-trivial holonomy along a loop linking with it,
\be 
{\rm hol}_{\gamma_1} \sim e^{2\pi X_0} \ ,
\ee 
where $\sim$ is equality up to $G$-conjugation.
This is the defining property of a disorder operator of Gukov-Witten type
labeled by the conjugacy class $[e^{2\pi X_0}]$.

We recall that for $G$ a connected and compact Lie group,
the exponential map $\exp :\mathfrak g \rightarrow G$ is surjective.
In particular, this implies that any conjugacy class
$[g_0]$ in $G$ can be written in the form
\be 
[g_0] = [e^{2\pi X_0}] \ , 
\ee 
for some $X_0\in \mathfrak g$.
In fact, $X_0$ is only defined up to
the adjoint action of $G$.
As a result, as already noted above,
without loss of generality we can take $X_0$
to lie in the fundamental Weyl chamber inside
the Cartan subalgebra of $\mathfrak g$.

\paragraph{Remark.}
In the above derivation
we set out to compute the holonomy around the loop 
$\gamma_1$ based at $P$,
but the final formula \eqref{eq_hol_firstway}
apparently depends on more data, namely
the choice of interpolating
loops via the point $P_*$
where the critical loop
$\gamma_{t_*}$ meets the $B$-operator.
In fact, we can argue that
the holonomy 
of $\gamma$ based at $P$
is independent of   this extra data
(not merely up to conjugation, which is manifest, but as a $G$-valued object).
To see this, 
we imagine repeating the same computation that led to 
\eqref{eq_hol_firstway}
choosing a different one-parameter family
of interpolating loops.
 In general, we will get an expression of the form
\be \label{eq_second_way}
{\rm hol}_{\gamma_1} =  (\Gamma_P^{P_*'} )^{-1}
U(P_*') e^{2\pi X_0} U(P_*')^{-1} \Gamma_P^{P_*'} \ , 
\ee 
for some point $P_*'$ on the $B$-operator.
Now, we have argued above that,
on the worldvolume of the $B$-operator, 
the $\beta$ equation of motion and  local ${\rm Stab}(X_0)$ invariance 
allow us to set
$U^{-1}(d+A)U=0$. In other words, $U$ is covariantly constant. 
This implies that, if we connect
$P_*$ and $P_*'$ using a path inside the worldvolume of the $B$-operator,
we have the simple relation
\be \label{eq_Q_parallel_transport}
U(P_*') = (\Gamma_{P_*'}^{P_*})^{-1} U(P_*) \ .
\ee 
Now, we plug
\eqref{eq_Q_parallel_transport} inside 
\eqref{eq_second_way}  to arrive at the expression
\be 
{\rm hol}_{\gamma_1}  = (\Gamma_P^{P_*'} )^{-1}
(\Gamma_{P_*'}^{P_*})^{-1} U(P_*) e^{2\pi X_0} U(P_*) \Gamma_{P_*'}^{P_*} \Gamma_P^{P_*'}
= (\Gamma_P^{P_*} )^{-1}
 U(P_*) e^{2\pi X_0} U(P_*)  \Gamma_P^{P_*} \ , 
\ee 
which is the same $G$ element as in 
\eqref{eq_hol_firstway}.

\subsubsection{Linking factor between $B$-operators and Wilson loops}

We have seen above that inserting
a $B$-operator with parameter $X_0$
induces a non-trivial holonomy 
\eqref{eq_hol_firstway} for the gauge field $A$ around the $B$-operator. With this result we can immediately write down the linking factor between the $B$-operator and 
a Wilson loop labeled by an irreducible
representation $\mathbf R$ of $G$. It reads
\cite{Cattaneo:2002tk, Heidenreich:2021xpr}  
\be \label{eq_linking_factor}
\langle \mathbf Q_{X_0}(\Sigma^{d-1})
\mathbf W_{\mathbf R}(\gamma)
\rangle
=
\frac{\chi_{\mathbf R}(e^{ 2\pi X_0})}{
\chi_{\mathbf R}(1)
}
\langle \mathbf Q_{X_0}(\Sigma^{d-1})
\rangle
\langle 
\mathbf W_{\mathbf R}(\gamma)
\rangle \ , 
\ee 
where the character $\chi_{\mathbf R}$ is defined as
\be 
\chi_{\mathbf R}(g) := {\rm Tr}_{\mathbf R}(g)  \;\;\; \text{for }g\in G \ . 
\ee 
The relation 
\eqref{eq_linking_factor} encodes the fact that, if we go from a configuration with $B$-operator and Wilson loop linking, and we transition to a configuration in which they are not linked,
we pick up a factor $\chi_{\mathbf R}(e^{2\pi X_0})/\chi_{\mathbf R}(1)$.
In this language, trivial linking 
corresponds to the case 
$\chi_{\mathbf R}(e^{ 2\pi X_0})/\chi_{\mathbf R}(1)=1$.

\subsubsection{Example: $G=SU(N)$}
\label{sec_example_SUN}

In this section we consider the example of $G = SU(N)$ and we describe explicit expressions for
the parameter $X_0$ entering $B$-operators and for 
the conjugacy class $[e^{2\pi X_0}]$.

\paragraph{Conventions.}

Our conventions for $G=SU(N)$ are as follows.
The Lie algebra $\mathfrak g = \mathfrak{su}(N)$ consists of
traceless antihermitian matrices.
The standard maximal torus $T\subset G$
is comprised of
 diagonal 
$SU(N)$ matrices. 
The corresponding Cartan subalgebra $\mathfrak t$
consists of traceless 
diagonal antihermitian matrices.
Any element of $\mathfrak t$ can be written as
\be \label{eq_SUN_a_presentation}
X_0 =  i\, {\rm diag} (a_1, \dots, a_N) \ , 
\qquad 
a_1 + \dots + a_N = 0 \ , \qquad a_i \in \mathbb R \ .  
\ee 
Alternatively, we can expand $X_0$ onto the basis of $\mathfrak t$ provided by the fundamental weights $\omega_i$, $i=1,\dots,N-1$,
\be  \label{eq_SUN_lambda_presentation}
X_0 = \sum_{i=1}^{N-1} \lambda_i \omega_i \ , \qquad 
\lambda_i \in \mathbb R \ . 
\ee 
The dictionary between \eqref{eq_SUN_a_presentation} and \eqref{eq_SUN_lambda_presentation} is encoded in the relations
\be 
\lambda_i = a_{i+1} - a_i  \ , \qquad 
i=1, \dots, N-1 \ . 
\ee 
This is equivalent to the following definition
of the fundamental weights,
\be 
\omega_j = \frac{1}{i N} {\rm diag}\Big( 
\underbrace{N-j , \dots, N-j}_{\text{$j$ times}}
, 
\underbrace{-j , \dots, -j}_{\text{$N-j$ times}}
\Big) \ , \qquad j = 1,\dots, N-1 \ . 
\ee 
We take the positive definite Ad-invariant 
inner product on 
$\mathfrak g$ to be 
\be 
g(X,Y) = -   {\rm tr}(X Y) \ , 
\ee 
where on the RHS tr is the trace in the defining representation (trace of $N\times N$ matrices).
With this normalization, one can verify
\be 
g(\omega_i, \omega_j) = (A^{-1})_{ij} \ ,
\qquad i,j = 1, \dots, N-1 \ , 
\ee 
where $A$ is the  Cartan matrix
of $\mathfrak{su}(N)$
(the square matrix of dimension $N-1$ with $2$'s on the diagonals and $-1$'s 
right below and above).

The simple roots are 
\be 
\alpha_i = \sum_{j=1}^{N-1} A_{ij} \omega_j \ ,\qquad i=1,\dots,N-1 \ , 
\qquad \text{with normalization $g(\alpha_i,\alpha_i)=2$} \ . 
\ee 
The Weyl group is generated by reflections
with respect to the hyperplanes orthogonal to the
simple roots, and is isomorphic
to the symmetric group $S_N$.
In terms of the parameters $a_i$,
$S_N$ acts by permuting $(a_1, \dots, a_N)$.
By acting with a suitable element of 
the Weyl group, any element $X_0 \in \mathfrak t$
can be mapped into the fundamental Weyl chamber.
The latter 
can be written 
in terms of the parametrization \eqref{eq_SUN_a_presentation}
as
\be 
\text{fund.~Weyl chamber:} 
\qquad 
a_1 \le a_2 \le  \dots \le a_N  \ , \qquad 
a_1 + \dots + a_N = 0  \ . 
\ee

\paragraph{Conjugacy classes and Weyl alcove.}

Any element in $G$ is conjugate to an element in
$T$. Hence,
any conjugacy class in $G$ can be presented as 
\be 
[g]= [e^{2 \pi X_0}] \ , \qquad 
\text{for some $X_0  \in \mathfrak t$} \ .
\ee 
Generically, there are infinitely many
$X_0  \in \mathfrak t$ that yield the same conjugacy class $[e^{2\pi X_0}]$.  
In fact, given 
\be 
X_0 =  i {\rm diag}(a_1 , \dots, a_N) \ , \quad 
\sum_{i=1}^N a_i = 0  \ , \quad 
X_0' =  i {\rm diag}(a_1' , \dots, a_N') \ , \quad 
\sum_{i=1}^N a_i' = 0 \ , 
\ee 
$X_0$ and $X_0'$ yield the same $[e^{2\pi X_0}]$ if and only if
\be \label{eq_same_conj_class}
a'_{i}= a_{\sigma(i)} + n_i \ , \qquad
i = 1,\dots, N \  , \qquad 
\text{for some $\sigma \in S_N$, $n_i \in \mathbb Z$} \ . 
\ee 
One can prove that elements
$X_0, X_0' \in \mathfrak t$
satisfy \eqref{eq_same_conj_class} if and only if they are related by an affine  reflection
$r_{\alpha,k}$,
namely, the reflection  with respect to 
a hyperplane of the form
\be \label{eq_affine_reflection}
H_{\alpha,k} = \{  \lambda \in \mathfrak t  | 
g(\lambda,\alpha)=k
\} \ , 
\ee 
where $\alpha$ is a root and $k\in \mathbb Z$.
(The case $k=0$ corresponds to a usual
Weyl reflection with respect to the hyperplane
orthogonal to the root $\alpha$).
The group of reflections
$r_{\alpha,k}$ is the affine Weyl group.
It is generated by $r_{\alpha_i,0}$,
where $\alpha_i$ runs over the simple roots,
together with $r_{\theta,1}$, where $\theta$ is the highest root.

\begin{figure}
    \centering
    \includegraphics[width=0.4\linewidth]{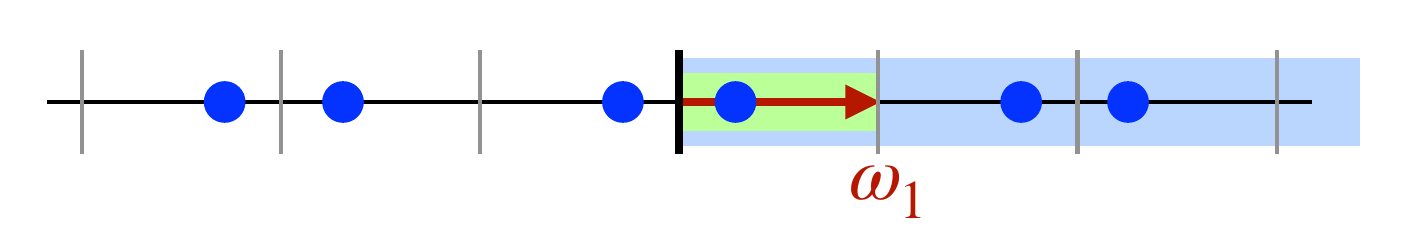}
    \caption{The Cartan algebra $\mathfrak t$ of $G=SU(2)$. The half-line shaded  in blue is the fundamental Weyl chamber. The interval shaded  in green is the fundamental Weyl alcove. 
    We include also the fundamental weight $\omega_1$. The blue points represent different elements  $X_0 \in \mathfrak t$ that yield the same conjugacy class $[e^{2\pi X_0}]$. They are all related by affine Weyl reflections, namely reflections with respect to the points
    $k \omega_1$, $k\in 
    \mathbb Z$.}
    \label{fig_SU2_alcove}
\end{figure}

\begin{figure}
    \centering
    \includegraphics[width=8. cm]{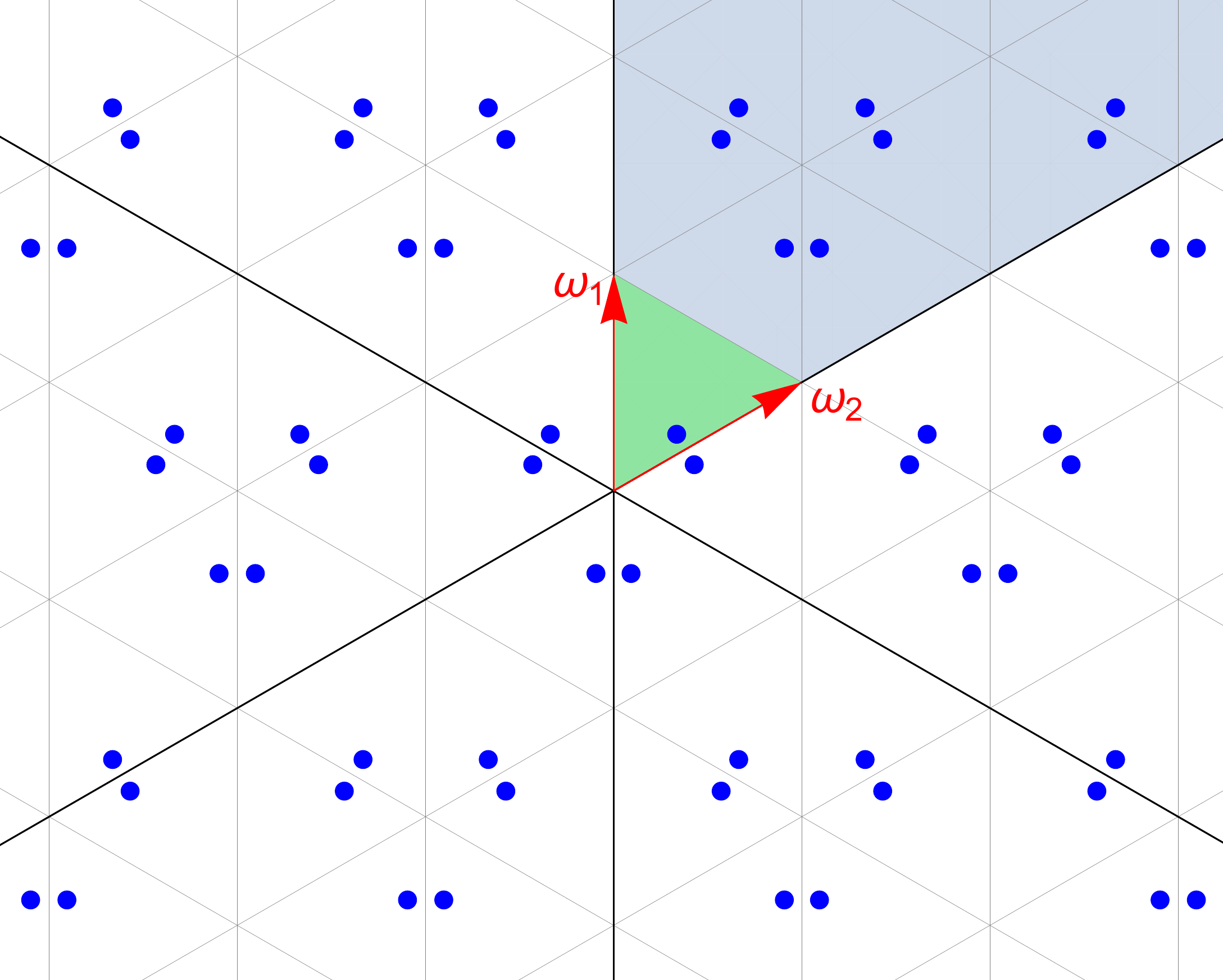}
    \caption{The Cartan algebra $\mathfrak t$ of $G=SU(3)$. The unbounded shaded region in blue is the fundamental Weyl chamber. The bounded shaded region in green is the fundamental Weyl alcove. 
    We include also the fundamental weights $\omega_1$, $\omega_2$. The blue points represent different elements  $X_0 \in \mathfrak t$ that yield the same conjugacy class $[e^{2\pi X_0}]$. They are all related by affine Weyl reflections, namely reflections with respect to the hyperplanes
    $H_{\alpha,k}$ as in \eqref{eq_affine_reflection},
    depicted here as gray and black lines.
    }
    \label{fig_SU3_alcove}
\end{figure}

A fundamental domain for the action of the affine Weyl group is the fundamental Weyl alcove
(of size 1).
In terms of the parametrization 
\eqref{eq_SUN_a_presentation}, it is described as
\be 
\text{fund.~Weyl alcove:} 
\qquad 
a_1 \le a_2 \le  \dots \le a_N \le a_1+1 \ , \qquad 
a_1 + \dots + a_N = 0  \ . 
\ee 
The fundamental Weyl alcove is a bounded 
subset of the fundamental Weyl chamber.

The notion of Weyl alcove is useful for our discussion because any conjugacy class in $SU(N)$ can be presented in a unique way as
\be 
[e^{2\pi X_0}] \qquad 
\text{with $X_0\in$ fundamental Weyl alcove $\subset \mathfrak t$} \ . 
\ee 
We depict the fundamental Weyl chambers and alcoves for $G=SU(2)$ and $G=SU(3)$ in Figures 
\ref{fig_SU2_alcove} and \ref{fig_SU3_alcove}.

\subsection{Chern-Simons terms and higher linking}
\label{sec_higher_linking}

Depending on the gauge group $G$ and the spacetime dimension, one may be able to add
extra topological couplings to the
BF action \eqref{eq_BF}.
In particular, one may add Chern-Simons couplings constructed with the connection 1-form.
These additional couplings modify some properties of the $B$-operators of the theory, for example they can induce non-trivial 
higher-linking correlators. 
Correspondingly, we shall see that the 
there can be more information in the label
$X_0$ of a $B$-operator, than merely the
conjugacy class $[e^{2\pi X_0}]$
that is detected by linking with Wilson loop operators.

In this section we demonstrate these phenomena
with the  case study $d=4$, $G = U(1) \times SU(2)$, adding a mixed five-dimensional Chern-Simons term to the BF action.
We refer the reader to \cite{Antinucci:2024zjp} for 
similar analyses in the Abelian case.

\subsubsection{Case study: $d=4$, $G=U(1) \times SU(2)$ with mixed Chern-Simons term}
The 5d topological BF action for 
$G=U(1) \times SU(2)$ can be written in terms of two pairs of fields,
$(b,a)$ for the $U(1)$ factor,
and $(B,A)$ for the $SU(2)$ factor.
More precisely, $b$ is an $\mathbb R$-valued
3-form, $a$ is a $U(1)$ connection
with field strength $f=da$ whose periods 
are in $2\pi \mathbb Z$.
The fields $A$, $B$ are as in
Section \ref{sec_define_BF} with $d=5$, $G=SU(2)$.
The action reads
\be \label{eq_example_mixed_CS}
S = \int_{X^{5}} \bigg[ 
\frac{1}{2\pi} b \wedge f
+ \frac{1}{2\pi} {\rm Tr}(B \wedge F_A)
+ k a \wedge  c_2(F_A)
\bigg]  \ , 
\ee 
where $f=da$, $k\in \mathbb Z$,
and the second Chern form constructed with the $SU(2)$ curvature
$F_A$ reads 
\be \label{eq_c2_def}
c_2(F_A) = \frac{1}{2(2\pi)^2} {\rm Tr}(F_A \wedge F_A) \ . 
\ee 
This theory admits topological $b$-operators
and $B$-operators,\footnote{\;These operators are topological by virtue of the perturbative equations of motion for $a$, $A$, $b$, $B$, which collectively imply $f=0$, $F_A=0$, $db=0$, $d_A B=0$.}
\be 
\ba 
\text{$b$-operator $\mathbf Q^{[b]}_{x_0}(\Sigma^3)$:}& & 
& \exp \bigg \{
i x_0 \int_{\Sigma^{3}} 
 b
\bigg \} \ , \\
\text{$B$-operator $\mathbf Q^{[B]}_{X_0}(\Sigma^3)$:}& & 
 \int [\cD U] [\cD \beta] &\exp \bigg \{
i \int_{\Sigma^{3}} {\rm Tr}
\bigg[ 
(B + d_A \beta) U X_0 U^{-1}
\bigg]
\bigg \} \ ,
\ea 
\ee 
where $x_0 \in \mathbb R$
and $X_0$ can be taken in the fundamental
Weyl chamber of the Cartan algebra of $\mathfrak{su}(2)$.
We also have Wilson loop operators for $a$ and $A$,
\be 
\mathbf W^{[a]}_n(\gamma) = \exp i n \int_{\gamma} a   \ , \qquad 
{\mathbf W}^{[A]}_{\mathbf R} (\gamma)= {\rm Tr}_{\mathbf R}{\rm Pexp} \int_{\gamma} (-A) \ , 
\ee 
with $n \in \mathbb Z$ and $\mathbf R$ an irreducible representation of $SU(2)$.

The linking factors of the pairs
$(\mathbf Q^{[b]}_{x_0}(\Sigma^3) , \mathbf W^{[a]}_n(\gamma))$
and
$(\mathbf Q^{[B]}_{X_0}(\Sigma^3),
{\mathbf W}^{[A]}_{\mathbf R} (\gamma)
)$
are 
\be 
\ba 
\langle \mathbf Q^{[b]}_{x_0}(\Sigma^{3})
\mathbf W^{[a]}_{\mathbf R}(\gamma)
\rangle 
& = 
e^{-2\pi i n x_0}
\langle \mathbf Q^{[b]}_{x_0}(\Sigma^{3})
\rangle 
\langle 
\mathbf W^{[a]}_{\mathbf R}(\gamma)
\rangle 
\ , \\
\langle \mathbf Q^{[B]}_{X_0}(\Sigma^{3})
\mathbf W^{[A]}_{\mathbf R}(\gamma)
\rangle
&=
\frac{\chi_{\mathbf R}(e^{ 2\pi X_0})}{
\chi_{\mathbf R}(1)
}
\langle \mathbf Q^{[B]}_{X_0}(\Sigma^{3})
\rangle
\langle 
\mathbf W^{[A]}_{\mathbf R}(\gamma)
\rangle \ . 
\ea 
\ee 
We have assumed that the supports $\Sigma^3$, $\gamma$ have linking number one in $X^5$.

We now turn to an example of higher linking
which detects the level $k$ of the Chern-Simons term in \eqref{eq_example_mixed_CS}.
To this end, we insert the operator 
$\mathbf Q^{[B]}_{X_0}(\Sigma^{3})$
and we consider the equations of motion of the bulk plus operator system.
In particular, we focus on the $B$ and $a$ equations of motion,
\be 
F_A + 2\pi UX_0 U^{-1} \delta_2(\Sigma^3) = 0 \ ,\qquad 
db + 2\pi k c_2(F_A) = 0 \ . 
\ee 
These relations together with 
\eqref{eq_c2_def} imply 
\be \label{eq_mixedCS_self_derivation1}
db = - 2\pi k \tfrac 12 {\rm Tr}(X_0^2) \delta_4(\cC^1)  \ , \qquad 
\delta_4(\cC^1) :=   \delta_2(\Sigma^3) \wedge \delta_2(\Sigma^3) \ . 
\ee 
Here we interpret $\delta_2(\Sigma^3) \wedge \delta_2(\Sigma^3)$ as describing the Poincar\'e dual to
the self-intersection $\cC^1 = \Sigma^3 \cap \Sigma^3$.
We observe that, for non-zero $k$,
the insertion of the operator
$\mathbf Q^{[B]}_{X_0}(\Sigma^{3})$ induces a source for the field $b$. We can probe it
by introducing an operator 
$\mathbf Q^{[b]}_{x_0}(\widetilde \Sigma^{3})$.
More precisely, we choose
$\widetilde \Sigma^{3} = \partial \cB^4$,
with $\cB^4$ a small 4-disk intersecting $\cC^1$ once transversely. 
In other words, $\widetilde \Sigma^3$
has linking number one with the self-intersection
$\cC^1$.
By an application of the usual Stokes' theorem,
the action for the operator 
$\mathbf Q^{[b]}_{x_0}(\widetilde \Sigma^{3})$
in the background sourced by
$\mathbf Q^{[B]}_{X_0}( \Sigma^{3})$
reads
\be 
\int_{\widetilde \Sigma^3} x_0 b
= \int_{\cB^4} x_0 db
= - 2\pi k \tfrac 12 x_0 {\rm Tr}(X_0^2) \int_{\cB^4} \delta_4(\cC^1)
=  - 2\pi k \tfrac 12 x_0 {\rm Tr}(X_0^2) \ .
\ee 
In conclusion, for non-zero $k$ we have a non-trivial higher linking involving the self-intersection of a $B$-operator and a $b$-operator,
with linking factor 
\be \label{eq_mixed_CS_linking}
\begin{array}{l}
\text{$\mathbf Q^{[b]}_{x_0}(\widetilde \Sigma^3)$, $\mathbf Q^{[B]}_{X_0}(\Sigma^3)$,} \\
\text{$\widetilde \Sigma^3$ links self-intersection of $\Sigma^3$:}
\end{array}
\qquad 
\exp \Big[ - 2\pi i k \tfrac 12 x_0 {\rm Tr}(X_0^2) \Big] \ . 
\ee 
As a small sanity check, we recall that
the label $X_0$ of the $B$-operator is only defined up to the adjoint action of $SU(2)$.
The trace 
in \eqref{eq_mixed_CS_linking} 
is indeed invariant under $X_0 \mapsto g X_0 g^{-1}$.

Further comments on the labels of $B$- and $b$-operators is in order.
If we  probe a $b$-operator 
by linking it with an $a$ Wilson
line of arbitrary charge $n$, the result depends on the label $x_0$ of the $b$-operator
only up to a shift $x_0 \sim x_0 +1$,
namely only via $e^{2\pi i x_0}$.
Similarly, if we  probe a $B$-operator 
by linking it with an $A$ Wilson
line in an arbitrary irrep $\mathbf R$,
the result depends on the label $X_0$ of the $B$-operator only via the conjugacy class
$[e^{2\pi X_0}]$. 
In contrast, for non-zero $k$
the expression 
\eqref{eq_mixed_CS_linking} for the higher linking
depends on $x_0$, $X_0$, and not only
on $e^{2\pi i x_0}$ and $[e^{2\pi X_0}]$.

One may wonder if we can also have a higher linking involving two distinct $B$-operators, and one $b$-operator.
Let us retrace the above steps in this case.
Suppose we insert the operators
$\mathbf Q^{[B]}_{X_0^{(1)}}( \Sigma^{3}_{(1)})$ and
$\mathbf Q^{[B]}_{X_0^{(2)}}( \Sigma^{3}_{(2)})$.
By reasoning as above, we see that for non-zero $k$ the quantity $db$ contains a source term of the form
\be \label{eq_mixed_CS_cross_term}
db \supset - 2\pi k {\rm Tr} 
\Big\{ (U_{(1)} X_0^{(1)} U_{(1)}^{-1}  
U_{(2)} X_0^{(2)} U_{(2)}^{-1}
\Big\} \delta_2(\Sigma^3_{(1)}) \wedge 
\delta_2(\Sigma^3_{(2)}) \ .
\ee 
Here $U_{(i)}$, $i=1,2$ are the two independent $SU(2)$-valued 0-forms on the two
$B$-operators. 
Following the same steps from 
\eqref{eq_mixedCS_self_derivation1} to \eqref{eq_mixed_CS_linking},
we infer a linking factor 
\be \label{eq_mixedCS_other_linking}
e^{i x_0 \int_{\cB^4} db} = 
\exp \Big[ - 2\pi i k   x_0 {\rm Tr} 
\Big\{ X_0^{(1)} \cU  X_0^{(2)}
\cU^{-1}
\Big\}  \Big] \ . 
\ee 
In the last step we have defined 
$\cU = U_{(1)}^{-1} U_{(2)}$.
The fields $U_{(i)}$ are integrated over.
Recall that in the worldvolume theory of a $B$-operator we can take $U_{(i)}$ to be covariantly constant. 
The path integral over $U_{(i)}$
reduces effectively to a finite-dimensional integral over $SU(2)$ with the Haar measure.
For this reason, the final expression for the sought-for 
triple linking factor is 
the average of \eqref{eq_mixedCS_other_linking} in $\cU$,
\be \label{eq_mixedCS_second_triple_linking}
\begin{array}{l}
\text{$\mathbf Q^{[b]}_{x_0}(\widetilde \Sigma^3)$, $\mathbf Q^{[B]}_{X_0^{(1)}}(\Sigma^3_{(1)})$, $\mathbf Q^{[B]}_{X_0^{(2)}}(\Sigma^3_{(2)})$,} \\
\text{$\widetilde \Sigma^3$ links intersection $\Sigma^3_{(1)} \cap \Sigma^3_{(2)}$:}
\end{array}
\qquad
\int_{SU(2)} d\mu(\cU) \exp \Big[ - 2\pi i k   x_0 {\rm Tr} 
\Big\{ X_0^{(1)} \cU  X_0^{(2)}
\cU^{-1}
\Big\}  \Big] \ . 
\ee 
Since the labels $X_0^{(i)}$
are only defined up to adjoint action,
the linking factor ought to be invariant under
$X_0^{(i)} \mapsto g_{(i)} X_0^{(i)} g_{(i)}^{-1}$ with constant $g_{(i)} \in SU(2)$.
This is indeed the case, thanks to the fact that we are averaging over $\cU$
with a left- and right-invariant measure.

We may say that the effects of the two $B$-operators combine \emph{coherently}
in the terminology
introduced in \cite{Alford:1992yx} for the discussion of non-Abelian vortices.

Finally, let us provide more explicit expressions for the results 
\eqref{eq_mixed_CS_linking}, \eqref{eq_mixedCS_second_triple_linking}.
We set 
\be 
X_0 = i {\rm diag}(a,-a) \ , \qquad 
X_0^{(1)} = i {\rm diag}(a^{(1)},-a^{(1)}) \ , \qquad
X_0^{(2)} = i {\rm diag}(a^{(2)},-a^{(2)}) \ .
\ee 
A generic element of $SU(2)$ is parametrized as
\be 
\cU = \exp(i \tfrac 12   \psi   \sigma_3)
\exp(i \tfrac 12   \theta   \sigma_1)
\exp(i \tfrac 12   \phi  \sigma_3) \ , 
\ee 
with $\theta \in [0,\pi]$,
$\phi \sim \phi + 2\pi$,
$\psi \sim \psi + 4 \pi$.
The Haar measure is the (normalized) measure induced by the bi-invariant metric on $SU(2)$,
namely the round metric on the 3-sphere,
\be 
ds^2 = - \frac 12 {\rm Tr}((\cU^{-1} d\cU)^2)
= \frac 14 (d\theta^2 + \sin^2 \theta d\phi^2)
+ \frac 14 D\psi^2 \ , \qquad 
D\psi = d\psi + \cos \theta d\phi \ . 
\ee 
With this notation, we compute 
\be 
\ba 
\exp \Big[ - 2\pi i k \tfrac 12 x_0 {\rm Tr}(X_0^2) \Big]
& = e^{2\pi i k x_0 a^2}
 \ , \\
\int_{SU(2)} d\mu(\cU) \exp \Big[ - 2\pi i k   x_0 {\rm Tr} 
\Big\{ X_0^{(1)} \cU  X_0^{(2)}
\cU^{-1}
\Big\}  \Big]
& = \frac{\sin(4\pi k x_0 a^{(1)} a^{(2)})}{4\pi k x_0 a^{(1)} a^{(2)}} \ . 
\ea 
\ee 
These expressions pass a couple of simple consistency checks. First, they are invariant under a sign flip of $a$, $a^{(1)}$, and/or
$a^{(2)}$. (These sign flips are the action of the Weyl group $\mathbb Z_2$ of $SU(2)$
on the $a$ parameters.) Secondly, these linking factors approach 1 if we send any of the parameters
$x_0$, $a$, $a^{(1)}$, or $a^{(2)}$ to zero.

\subsection{Some remarks on fusion}

In this subsection we comment on 
some aspects of the fusion of topological operators of the bulk SymTFT 
described by \eqref{eq_BF}.

\subsubsection{Wilson line operators}
For any  $d$, the bulk $(d+1)$-dimensional
BF theory \eqref{eq_BF} admits simple topological Wilson line operators labeled by irreducible representations of the group $G$, which is a compact Lie group.  They close under fusion. Indeed,
they fuse according to the tensor product of the representations that label them.
In other words, the  
bulk $(d+1)$-dimensional topological theory
contains a set of topological line operators described by the category ${\rm Rep}(G)$.

The fusion of Wilson lines can be argued as follows. As recalled around \eqref{eq_Wilson_loop}, a Wilson line corresponds to an insertion in the path integral of the trace in the representation $\mathbf R$ of the holonomy $h$ of the connection around a loop. Thus, the path integral insertion is a character
$\chi_{\mathbf R}(h)$
labeled by $\mathbf R$.
Characters are known to be multiplicative under tensor product ($\chi_{\mathbf R_1 \otimes \mathbf R_2}(h) = 
\chi_{\mathbf R_1}(h) \chi_{\mathbf R_2}(h)$)
and additive under direct sum
($\chi_{\mathbf R_1 \oplus \mathbf R_2}(h) = 
\chi_{\mathbf R_1}(h) +\chi_{\mathbf R_2}(h)$).
As a result, when a Wilson line in the irrep $\mathbf R_1$ is put on top of a Wilson line in the irrep   
$\mathbf R_2$, the net path integral insertion takes the form
$\chi_{\mathbf R_1}(h) \chi_{\mathbf R_2}(h)= 
\chi_{\mathbf R_1 \otimes \mathbf R_2}(h) = 
\chi_{\oplus_{j}\mathbf R'_j}  (h)
= \sum_j \chi_{\mathbf R'_j}(h)$,
where $\mathbf R_1 \otimes \mathbf R_2 = \oplus_j \mathbf  R_j'$
is the decomposition of the tensor product of 
$\mathbf R_1$ and $\mathbf R_2$ into irreps.
The above argument is semiclassical.
In a general quantum field theory, additional subtleties related to regularization may emerge. In our topological quantum field theory, however, this problem does not arise.
This is related to the fact that our theory has no non-trivial gauge-invariant local operators.
We refer the reader to 
\cite[sec.~2.2]{Heidenreich:2021xpr} for further comments on fusion of Wilson lines in general.

\subsubsection{Observations on the fusion of $B$-defects}

In this work we do not address systematically the fusion of $B$-defects
in the bulk $(d+1)$-dimensional
BF theory \eqref{eq_BF}. 
To gain some preliminary intuition, however,
we can first revisit the case of $G$ a finite 
non-Abelian group.

\paragraph{Finite non-Abelian group.}
Let us  consider a $(d+1)$-dimensional Dijkgraaf-Witten theory \cite{Dijkgraaf:1989pz} with gauge group $G$.
For simplicity, we restrict ourselves to the case of trivial cocycle.
The theory contains topological
codimension-two operators
labeled by a conjugacy class $C$ in $G$, which we denote $D_{d-1}^{(C)}$. They can be thought of as Gukov-Witten operators, in the sense that inserting 
$D_{d-1}^{(C)}$ on $\Sigma^{d-1}$ amounts to
imposing a holonomy for the (flat) discrete $G$ gauge field
around $\Sigma^{d-1}$, given up to conjugation by the class $C$. We regard the operators $D_{d-1}^{(C)}$ as the finite-group analog of the $B$-defects studied earlier in this work.

Let us recall some properties of the  fusion of the topological operators $D_{d-1}^{(C)}$,
following \cite{Dijkgraaf:1989hb}.
The operator $D_{d-1}^{(C_3)}$ is an allowed fusion channel in 
$D_{d-1}^{(C_1)} \otimes D_{d-1}^{(C_2)}$ if
there exist $g_1 \in C_1$,
$g_2 \in C_2$ such that
$g_1 g_2 \in C_3$. More precisely, we can write
\be \label{eq_finite_GW_fusion}
D_{d-1}^{(C_1)} \otimes 
D_{d-1}^{(C_2)} = \sum_{C_3} N_{C_1 C_2}{}^{C_3} D_{d-1}^{(C_3)} + \dots 
\ee 
The fusion coefficient
$N_{C_1 C_2}{}^{C_3}$ is the number of $G$-orbits 
in the set
\be \label{eq_set_of_triples}
\{ (g_1, g_2, g_3) \in C_1 \times C_2 \times C_3 |    g_1 g_2= g_3
\} \ ,
\ee 
where $G$ acts on triples $(g_1, g_2, g_3)$ by simultaneous conjugation.

The ellipses on the RHS of \eqref{eq_finite_GW_fusion} indicate that generically we expect additional contributions in the fusion rule. 
To clarify this point we can focus on the case $d=2$. 
The 3d bulk topological theory admits simple line operators labeled by pairs $(C, \pi)$,
where $C$ is a conjugacy class of $G$ and $\pi$ is an  irreducible representation of the centralizer subgroup $N_{g}$ of   $g\in C$.
(Different choices of representatives $g\in C$ is give isomorphic centralizer subgroups.)
Heuristically,
we can think of the topological line labeled $(C,\pi)$ as a ``pure'' Gukov-Witten operator 
with label $C$
stacked with a Wilson line in the representation $\pi$
of  the ``unbroken gauge group'' $N_g \subset G$
living on the worldvolume of the Gukov-Witten operator.
The lines with label $(C = \{e\},\pi)$, namely with trivial conjugacy class, are topological Wilson line operators for the full gauge group $G$ in the 3d bulk.

Let us denote the topological line with label $(C,\pi)$
as $D_1^{(C,\pi)}$.
The  fusion algebra of the operators 
$D_1^{(C,\pi)}$ 
takes the form 
\be 
D_1^{(C_1,\pi_1)}
\otimes 
D_1^{(C_2,\pi_2)}
= \sum_{(C_3, \pi_3)}
N_{(C_1, \pi_1) (C_2,\pi_2)}{}^{(C_3, \pi_3)}
D_1^{(C_3,\pi_3)} \ ,
\ee 
where the fusion coefficients 
are known in closed form
\cite{Dijkgraaf:1989hb,Roche:1990hs}.
For brevity, we refrain from writing down their expressions. Let us emphasize, however, two salient features of the fusion coefficients.
\begin{itemize}
    \item If we consider the case in which $\pi_1$, $\pi_2$, $\pi_3$ are the trivial representation (denoted 
    $\mathbf 1$), then the fusion coefficient $N_{(C_1, \mathbf 1) (C_2,\mathbf 1)}{}^{(C_3, \mathbf 1)}$ equals the fusion coefficient $N_{C_1 C_2}{}^{C_3}$ described in \eqref{eq_finite_GW_fusion}, \eqref{eq_set_of_triples}. 
    \item If $\pi_1$ and $\pi_2$ are trivial, but $\pi_3$ is non-trivial,
    the fusion coefficient 
    $N_{(C_1, \mathbf 1) (C_2,\mathbf 1)}{}^{(C_3, \pi_3)}$ can be different from zero. This can be checked, for example, in the case $G =S_3$.
\end{itemize}
The second point shows that the subset of ``pure'' Gukov-Witten operators
with labels $(C,  \pi = \mathbf 1)$ does not generically close under fusion. This substantiates the claim made earlier that
additional terms are present in \eqref{eq_finite_GW_fusion}, besides those written down explicitly.
In contrast, one can 
verify using the expressions of the fusion coefficients  $N_{(C_1, \pi_1) (C_2,\pi_2)}{}^{(C_3, \pi_3)}$ that 
the lines with labels
$(C=\{e\}, \pi)$ (i.e.~Wilson lines)
do close under fusion,
in accordance with our general claim above.

In categorical language, 
the topological line operators of 
the 3d Dijkgraaf-Witten theory with gauge group $G$ form a category, which is the 
Drinfeld center $\mathcal Z({\rm Vec}_G)$ of ${\rm Vec}_G$. The discussion of the previous paragraphs is an informal account of some properties of $\mathcal Z({\rm Vec}_G)$, see e.g.~\cite{lusztig1984characters,lusztig1987leading}, \cite[Chapter 3]{bakalov2001lectures}, \cite[ex.~8.5.4]{etingof2015tensor}. 
In particular, it is known that \cite{willerton2008twisted}
\be \label{eq_ZVec_and_Rep}
\mathcal Z({\rm Vec}_G) 
\cong \bigoplus_{[g]} {\rm Rep}(N_g) \ , 
\ee 
corresponding to the fact that simple lines in the Drinfeld center are labeled by a conjugacy class $[g]$ and an irreducible representation of the centralizer of $g$.
(An analogous statement holds in the twisted case, see e.g.~\cite{willerton2008twisted}.)

An analog of \eqref{eq_ZVec_and_Rep} is available in the case $d=3$, namely for a 4d Dijkgraaf-Witten theory. 
In this case the topological surface operators in the 4d bulk form a 2-category,
which is the Drinfeld center
$\mathcal Z(2{\rm Vec}_G)$
of the 2-category $2{\rm Vec}_G$. One has \cite{kong2020center} 
\be 
\mathcal Z( 2{\rm Vec}_G) \cong 
\bigoplus_{[g]} 2 {\rm Rep}(N_g) \ .
\ee 
(Once again, for simplicity we restrict to the untwisted case, but analogous results are available with twist.)
This encodes the fact that simple topological surface operators in the 4d bulk are labeled by a conjugacy class $[g]$ and an irreducible 2-representation  of the centralizer subgroup $N_g$. 
Now, an irreducible 
2-representation  of  $N_g$
is labeled by: (i) a subgroup $H$ of $N_g$; (ii) a group cohomology class $\psi \in H^2(H,\mathbb C^\times)$ \cite{ostrik2002module,kong2020center}.
These   data also label
irreducible 2d TQFTs with $N_g$ non-anomalous 0-form symmetry, see e.g.~\cite{Bhardwaj:2023ayw}.
Thus, heuristically,
we can think of the topological surface operators in the 4d bulk of the Dijkgraaf-Witten theory as ``pure'' Gukov-Witten operator labeled by the conjugacy class $[g]$ stacked with a 2d TQFT
whose global symmetry equals the ``unbroken gauge group'' on the worldvolume of the Gukov-Witten operator.\footnote{\,This is the spirit of theta defects \cite{Bhardwaj:2022lsg,Bhardwaj:2022kot}.}

The natural generalization of 
\eqref{eq_ZVec_and_Rep} for  $d\ge 4$,
\be \label{eq_higher_Drinfeld}
\mathcal Z ((d-1){\rm Vec}_G) \cong 
\bigoplus_{[g]} (d-1){\rm Rep}(N_g) \ ,
\ee 
has also been put forward in the mathematical literature \cite{kong2020center} and motivated by 
physical arguments \cite{Bhardwaj:2023ayw}.
Heuristically, we interpret 
\eqref{eq_higher_Drinfeld} as the statement that simple codimension-two topological operators in the bulk of the $(d+1)$-dimesional
Dijkgraaf-Witten theory
with gauge group $G$ are realized by starting from a ``pure'' Gukov-Witten operator labeled by a conjugacy class $[g]$ and stacking it with a simple $(d-1)$-dimensional TQFT with global symmetry $N_g$.

Building on the case $d=2$,
we have the following expectations regarding the fusion of codimension-two operators in the  $(d+1)$-dimesional
Dijkgraaf-Witten theory.
\begin{itemize}
    \item The fusion of ``pure'' Gukov-Witten operators follows the pattern \eqref{eq_finite_GW_fusion} with the specified fusion coefficients $N_{C_1 C_2}{}^{C_3}$.
    \item The set of ``pure'' Gukov-Witten operators may in general not close under fusion.  
\end{itemize}
It would be interesting to test these claims in the case $d=3$, using the explicit fusion rules given in \cite{kong2020center}, but this  goes beyond the scope of this paper.

\paragraph{Continuous non-Abelian group.}
The intuition coming from the case of finite non-Abelian group leads us to the following expectations for the fusion of $B$-defects \eqref{eq_def_B_op} in the case of 
a compact Lie group $G$.

Recall that the $B$-defect
$\mathbf Q_{X_0}(\Sigma^{d-1})$
is a Gukov-Witten operator
inducing a holonomy labeled by the conjugacy class $[e^{2\pi X_0}]$. In analogy with 
\eqref{eq_finite_GW_fusion}, \eqref{eq_set_of_triples},
we expect that the operator
$\mathbf Q_{X_0^{(3)}}(\Sigma^{d-1})$
is an allowed fusion channel
in $\mathbf Q_{X_0^{(1)}}(\Sigma^{d-1}) \otimes \mathbf Q_{X_0^{(2)}}(\Sigma^{d-1})$
if there exist $g, \ell \in G$ such that
\be \label{eq_allowed_channel}
e^{2\pi X_0^{(1)}}   g e^{2\pi X_0^{(2)}} g^{-1} = 
\ell e^{2\pi X_0^{(3)}} \ell^{-1} \ .
\ee 
In particular,
for  generic $X_0^{(1)}$, $X_0^{(2)}$
we have a continuum of allowed fusion channels~$X_0^{(3)}$.

For example, let $G= SU(2)$. As described in Section \ref{sec_example_SUN}, without loss of generality
the parameters $X_0^{(1)}$, $X_0^{(2)}$ take the form
\be 
X_0^{(1)} = i\, {\rm diag}(a^{(1)} ,-a^{(1)})  \ ,
\qquad 
- \tfrac 12 \le a^{(1)} \le 0  \ , 
\ee 
namely we can take $a^{(1)}$ in the fundamental Weyl alcove,
and similarly for $X_0^{(2)}$.
A generic element $g \in SU(2)$ can be parametrized as
\be 
g = e^{i \frac 12 \psi \sigma_3}
e^{i \frac 12 \theta \sigma_1}
e^{i \frac 12 \phi \sigma_3} \ , \qquad 
\theta\in [0,\pi] \ ,  \quad \phi \in [0,2 \pi)
\ , \quad \psi \in [0,4 \pi) \ ,
\ee 
where $\sigma$'s denote the standard Pauli matrices.
A direct computation shows that 
\be 
e^{2\pi X_0^{(1)}} g e^{2\pi X_0^{(2)}} g^{-1} \sim {\rm diag}(e^{2\pi i a^{(3)}} ,e^{-2\pi i a^{(3)}}  ) \ , 
\ee 
where $\sim$ denotes equality up to conjugation in $SU(2)$
and $a^{(3)}$ satisfies
\be 
\cos (2\pi a^{(3)})
=  
\cos (2\pi a^{(1)}) \cos (2\pi a^{(2)})
- \cos \theta \sin (2\pi a^{(1)}) \sin (2\pi a^{(2)}) \ . 
\ee 
As $\theta$ varies, we run over the possible fusion channels. The allowed values for $a_3$
in the fundamental Weyl alcove are 
\be 
- \tfrac 12 \le \left|
a^{(1)} + a^{(2)} + \tfrac 12
\right| - \tfrac 12
\le a^{(3)} \le -|a^{(1)} - a^{(2)}| \le 0 \ . 
\ee 
This proposal for the fusion of $B$-defects differs from analogous proposals, such as the fusion rule of averaged operators in \cite{Cordova:2022rer}.
According to that paper,
for $G=SU(2)$ one finds only two fusion channels on the RHS, as opposed to a continuum of possible values. 
We hope to return to this point in the future to clarify it further. 

Let us point out that
trivalent junctions among 
 surface operators labeled by conjugacy classes of $G$ are studied in \cite[sec.~3.5]{Chun:2015gda}.
The knowledge of allowed trivalent junctions amounts to knowledge of allowed fusion channels. The authors of 
\cite{Chun:2015gda} show
how to find trivalent junctions 
by solving the multiplicative Horn problem:
given conjugacy classes $\cC$, $\cC'$, $\cC''$, the problem is to determine whether $\cC$ is contained in $\cC' \cdot \cC''$.
This fits with our claim  above
\eqref{eq_allowed_channel}
that 
$\mathbf Q_{X_0^{(3)}}(\Sigma^{d-1})$
is an allowed fusion channel
in $\mathbf Q_{X_0^{(1)}}(\Sigma^{d-1}) \otimes \mathbf Q_{X_0^{(2)}}(\Sigma^{d-1})$
if $e^{2 \pi X_0^{(3)}}$ appears
in the product of the conjugacy classes
of $e^{2 \pi X_0^{(1)}}$,  $e^{2 \pi X_0^{(2)}}$.
We refer the reader to \cite{Chun:2015gda} for a detailed account of solutions to the multiplicative Horn problem for 
$SU(N)$ and further remarks for general compact connected $G$.

As a final comment, the intuition from the case of finite $G$ suggests that the fusion of $B$-defects might not be closed. Additional topological operators might have to be included in the fusion algebra, which are the analogs of the operators in the Dijkgraaf-Witten theory with a non-trivial (higher) representation label. Conjecturally, 
they might be obtained from a ``pure'' $B$-defect as in \eqref{eq_def_B_op} by stacking
a $(d-1)$-dimensional TQFT with global symmetry ${\rm Stab}(X_0) \subset G$, which corresponds to the ``unbroken gauge group'' on the worldvolume of the $B$-defect,  and performing a (flat) gauging of the 
diagonal ${\rm Stab}(X_0)$.
Carrying out this construction, however, is beyond the scope of this note.

\subsection{Topological operators in the sandwich construction}

The $B$-operators and the Wilson loops discussed in the previous section can be formulated
without making assumptions on the spacetime
$X^{d+1}$. In this section we discuss some additional operators that can be defined when
we take $X^{d+1}$ to be the product of
a $d$-manifold $M^d$ with an interval,
$X^{d+1} = M^d \times [z_0, z_1]$.
The boundary component of $X^{d+1}$ at $z=z_0$ is identified with the symmetry boundary of the sandwich construction. On this component,
we impose the Dirichlet boundary condition
\be 
A \Big|_{M^d \times \{z_0\}} = 0    \  . 
\ee 
Correspondingly, we
restrict the parameter $g$ of bulk gauge transformations \eqref{eq_BF_gauge_trans} to be trivial
at the symmetry boundary.
The boundary component of $X^{d+1}$ at $z=z_1$ 
 is identified with the physical boundary.

\subsubsection{Stretched Wilson line operators}

A simple class of operators that can be defined on $X^{d+1} = M^d \times [z_0, z_1]$
are open Wilson line operators 
labeled by an irreducible representation
$\mathbf R$ and
stretching
along the interval direction $z$.
This operator is constructed
with the  parallel transport $\Gamma(\gamma)_0^1$
along a path $\gamma$ connecting a point $Q$ on the physical boundary to a point $P$ on the
symmetry boundary,
\be 
\gamma:[0,1]\rightarrow X^{d+1} \ , \qquad 
\gamma(0) =Q \in M^d\times \{z_1\} \ , \qquad 
\gamma(1) =P \in M^d\times \{z_0\} \ ,
\ee 
\be 
\text{(open Wilson line in irrep $\mathbf R$)}= 
\mathbf W_{\mathbf R}^o(\gamma)=
D_{\mathbf R}(\Gamma(\gamma)_0^1) \ . 
\ee 
Here the symbol $D_{\mathbf R}(g)$
denotes the linear map that represents the abstract
group element $g\in G$ in the irrep $\mathbf R$.
The operator 
$\mathbf W_{\mathbf R}^o(\gamma)$ 
(where `o' stands for open)
is matrix-valued, whereas its closed loop counterpart $\mathbf W_{\mathbf R}(\gamma)$
is real-valued.
Under a gauge transformation,
\be 
\mathbf W_{\mathbf R}^o(\gamma) \mapsto
\mathbf W_{\mathbf R}^o(\gamma)
  D_{\mathbf R} ( g(Q)^{-1} ) \ , 
\ee 
where we have used $g(P)=1$,
hence $D_{\mathbf R} ( g(P) )= \mathbb I$,
because $P$ lies on the symmetry boundary.
The term $D_{\mathbf R}(g(Q)^{-1})$ in the gauge variation of $\mathbf W_{\mathbf R}^o(\gamma)$ is compensated by the non-topological
endpoint on the physical boundary.
Alternatively, we may consider a quiche-like
configuration, in which the physical boundary
at $z_1$ is pushed to infinity and  
$g(Q)=1$.

Since the connection $A$ is flat
in the bulk (away from operator insertions),
we already know that
$\mathbf W_{\mathbf R}^o(\gamma)$
is invariant under small deformations of the path $\gamma$ that leave the endpoints $P$, $Q$ fixed.
Moreover, $\mathbf W_{\mathbf R}^o(\gamma)$ is invariant if we move the point $P$ on the symmetry boundary.
More precisely, let us consider a different path $\gamma'$ connecting 
the same $Q$ on the physical boundary
to some $P'$ on the symmetry boundary. We can smoothly deform $\gamma'$ to a path $\gamma''$ that first connects 
$Q$ to $P$ along the original path $\gamma$,
and then 
$P$ to $P'$
along the symmetry boundary.
The only potential difference in the final result can originate from the
portion of the path $\gamma''$ connecting
$P$ to $P'$. But on the physical boundary the connection is identically zero due to the Dirichlet boundary condition, ensuring that
parallel transport along any path inside the symmetry boundary is trivial.
We then conclude that 
$\mathbf W_{\mathbf R}^o(\gamma)$ is invariant
if we deform $\gamma$ and/or move the point $P$.
In contrast, the operator is not invariant if we 
move the point $Q$ on the physical boundary.
Indeed, upon closing the sandwich,
$\mathbf W_{\mathbf R}^o(\gamma)$ yields a non-topological local operator in the
QFT $\mathcal T$.

\subsubsection{Non-genuine $B$-operators}
\label{sec_non_genuine_B}
\begin{figure}
        \centering
        \includegraphics[width=11cm]{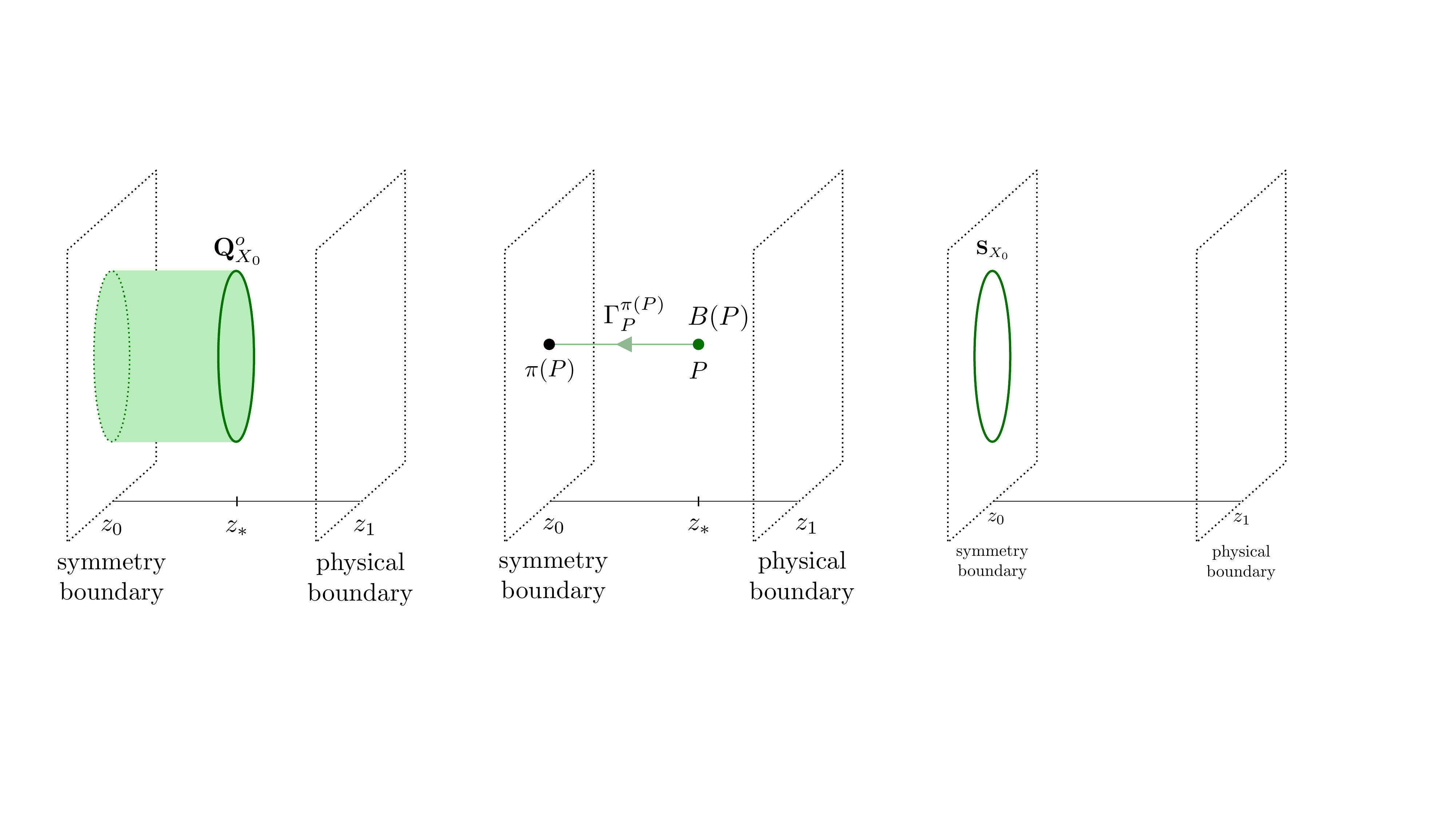}
        \caption{On the left: The non-genuine $B$-operator
        $\mathbf Q^o_{X_0}$
        is supported on the cylinder $\Sigma^{d-1} \times [z_0, z_*]$. 
        It is constructed by integrating 
        the dressed $\mathcal B$ operator \eqref{eq_mathcalB_def}.
        On the right: the dressed operator $\mathcal B$ is defined combining the field $B$ at the point $P$ with the parallel transport operator along a segment connecting $P$ to its projection $\pi(P)$ onto the symmetry boundary.
        }
        \label{fig_non_genuine}
\end{figure}

Let us now discuss a different version of $B$-operators, that are non-genuine.\footnote{These 
correspond to case 2 in the terminology of section 2 of \cite{Kapustin:2014gua}:
the $B$-operator is attached to a $d$-dimensional surface, and the dependence of the $B$-operator on the surface is topological. 
} More precisely, 
they are defined on 
$d$-dimensional cylinder
$\Sigma^{d-1} \times [z_0, z_*]$,
where $\Sigma^{d-1}$ is a cycle in
$M^d$ and $z_* \in (z_0, z_1)$ is a 
fixed value along the interval direction, 
see Figure \ref{fig_non_genuine}.

We start with a preliminary definition.
Let $P$ be a point in the interior of
$X^{d+1} = M^d \times [z_0, z_1]$,
sitting at a value $z_*$ of the interval coordinate
and a value $x$ for coordinates on $M^d$.
Consider the path connecting 
the point $P$ to its projection $\pi(P)$ onto
the symmetry boundary, which in coordinates has the same value $x$ as $P$, but $z=z_0$.
We multiply the field $B$ at the point $P$
by the parallel transport operator 
from $P$ to $\pi(P)$
along a straight path  stretching 
in the interval direction, see Figure \ref{fig_non_genuine}.
In this way we define a dressed version of $B$, denoted $\mathcal B$,
\be \label{eq_mathcalB_def}
\mathcal \cB  (P)
= \Gamma_P^{\pi(P)} B(P)  (\Gamma_P^{\pi(P)})^{-1} \ . 
\ee 
We are using the compact notation 
$\Gamma_P^{\pi(P)}$ for the parallel transport
from $P$ to $\pi(P)$.
Under a bulk gauge transformation \eqref{eq_BF_gauge_trans},
\be 
A \mapsto g(d+A)g^{-1} \ , \qquad 
B \mapsto g(B-d_A \tau) g^{-1} \ , \qquad 
\Gamma_{P}^{\pi (P)} \mapsto 
g(\pi(P))
\Gamma_{P}^{\pi (P)}
g(P)^{-1} \ .
\ee 
Notice, however, that $g(\pi(P))=1$
because $\pi  (P)$ lies on the symmetry boundary.
As a result, we have
\be \label{eq_mathcalB_trans}
\mathcal \cB  (P) \mapsto 
\mathcal \cB  (P)  - 
\Gamma_P^{\pi(P)} (d_A \tau)(P)  (\Gamma_P^{\pi(P)})^{-1} \ . 
\ee 

Next, we use the dressed operator $\cB$ to construct a non-genuine version of
a $B$-operator. This is achieved by integrating
$\cB$ on $\Sigma^{d-1} \times \{ z_* \}$.
Each point $P$ on $\Sigma^{d-1} \times \{ z_* \}$
comes with its  interval attaching it to $\pi(P)$, so that result of the integration is not a genuine operator on $\Sigma^{d-1} \times \{ z_* \}$,
but rather an operator on the cylinder
$\Sigma^{d-1} \times [z_0, z_* ]$, as anticipated.
We can write this operator as
\be \label{eq_def_B_op_open}
\mathbf Q^o_{X_0} (\Sigma^{d-1} \times [z_0, z_* ]) \;: \;\; 
 \exp \bigg \{
\frac{i}{2\pi} \int_{P \in \Sigma^{d-1} \times \{z_* \}} {\rm Tr}
( 
\cB(P)   X_0  
)
\bigg \} \ . 
\ee 
We may contrast this operator
with the genuine $B$-operator
in \eqref{eq_def_B_op}.
We notice that the definition
of $\mathbf Q^o_{X_0}$
does not require localized fields,
in contrast to the fields $U$, $\beta$ in 
\eqref{eq_def_B_op}.
Indeed, the expression in 
\eqref{eq_def_B_op_open} is gauge invariant without any additional compensating local fields.\footnote{One could ask whether $d$-dimensional counter-terms might be added to the action for the non-genuine $B$-operator. We stress that \eqref{eq_def_B_op_open} is the minimal definition compatible with gauge invariance.  If the   bulk Lagrangian is the standard BF theory \eqref{eq_BF}, topological counter-terms are not needed. For a continuous SymTFT with a more complicated bulk action, topological counter-terms might be required. Here we do not explore this possibility and use \eqref{eq_def_B_op_open} as the definition of the non-genuine $B$-operator.} This follows from 
\eqref{eq_mathcalB_trans} and the identity 
\be \label{eq_identity_for_mathcalB}
\Gamma_P^{\pi(P)} (d_A \tau)(P)  (\Gamma_P^{\pi(P)})^{-1} 
= d \Big[ 
\Gamma_P^{\pi(P)}   \tau (P)  (\Gamma_P^{\pi(P)})^{-1} 
] \ . 
\ee 
This identity ensures that the $d_A \tau$ shift
in \eqref{eq_mathcalB_trans} yields a total derivative in the integrand of \eqref{eq_def_B_op_open}. 
It can be derived following steps analogous to those leading to the non-Abelian Stokes' formula
\eqref{eq_nonAb_Stokes}, using flatness of the connection.

We stress that $\mathbf Q^o_{X_0}$ depends on
$X_0 \in \mathfrak g$,
while its genuine counterpart
$\mathbf Q_{X_0}$ depends on $X_0 \in \mathfrak g$ only up to the adjoint action of $G$ on $\mathfrak g$.

\paragraph{Operators restricted on the boundary.}
Finally, we may also consider the limit in which
the position $z_*$ along the interval direction approaches the value $z_0$ of the symmetry boundary. In this limit, the operator
$\mathbf Q^o_{X_0}$ yields a genuine operator, albeit one that is restricted to lie inside the symmetry boundary. We denote this operator as
\be \label{eq_S_as_a_limit}
D_{X_0}(\Sigma^{d-1} \times \{ z_0 \}) =  \lim_{z_* \rightarrow z_0} \mathbf Q^o_{X_0}(\Sigma^{d-1} \times [z_0 , z_*]) \ .
\ee 
Comparing with \eqref{eq_def_B_op_open}, we see that this operator can be described by the exponentiated action
\be 
\exp \bigg \{
\frac{i}{2\pi} \int_{P \in \Sigma^{d-1} \times \{z_0 \}} {\rm Tr}
( 
B   X_0  
)
\bigg \} \ . 
\ee
The presentation \eqref{eq_S_as_a_limit} 
will be useful below in discussing  linking
with Wilson lines.

\paragraph{Linking with Wilson lines.}

\begin{figure}
    \centering
    \includegraphics[width=7.5cm]{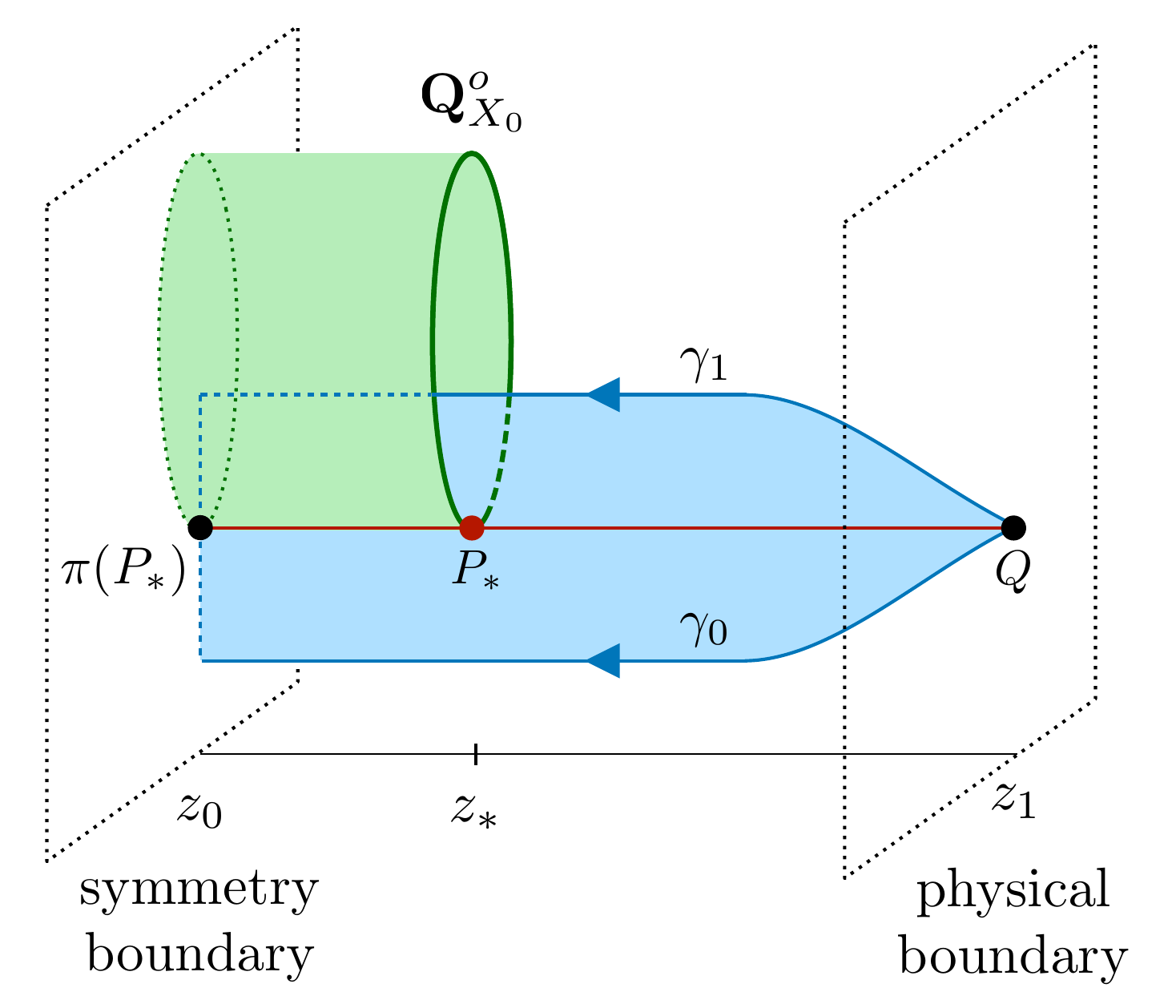}
    \caption{A one-parameter family of Wilson paths sweeps a surface intersecting a non-genuine $B$-operator
    on $\Sigma^{d-1} \times [z_0, z_*]$. The point $P_*$ is the intersection of the surface swept by the one-parameter family of paths and 
    $\Sigma^{d-1} \times \{ z_* \}$.}
    \label{fig_nongenuine_link}
\end{figure}

Next, let us consider the linking factor between
a non-genuine $B$-operator
and a stretched Wilson line operator.
We can analyze this linking by deforming the path of the Wilson line and applying the non-Abelian Stokes' formula,
see Figure \ref{fig_nongenuine_link}.
More precisely, we consider a one-parameter family $\gamma_t$ of paths with
\be 
t\in [0,1] \ , \quad 
s\in [0,1] \ , \quad 
\gamma_t(0) = Q \ , \qquad 
\gamma_t(1) = P_t \ , 
\ee 
where $Q$ is on the physical boundary,
$P_t$ is on the symmetry boundary.
The surface swept by this 
one-parameter family of paths intersects the support of the non-genuine $B$-operator
at the point $P_*$, corresponding to
values $t_*$, $s_*$ of the $t$, $s$ parameters.
Without loss of generality, we can arrange
the one-parameter family in such a way that
the projection $\pi(P_*)$ of $P_*$ onto the
symmetry boundary is the same as the endpoint
$P_{t_*}$ of the path $\gamma_{t_*}$.

After these preliminaries, we apply the non-Abelian Stokes' formula
\eqref{eq_nonAb_Stokes}.
(We get no contribution from the variation of the endpoint $P_t$ on the symmetry boundary because the connection is zero there.)
We get a non-zero contribution from
a localized field strength,
\be \label{eq_F_source_non_genuine}
F_A + (\Gamma_{P_*}^{\pi(P_*)})^{-1}
X_0 \Gamma_{P_*}^{\pi(P_*)} \delta_1(\Sigma^{d-1} \subset M^d) \delta(z-z_*) dz =0 \ ,
\ee 
as follows from the variation of the total action
\eqref{eq_BF} plus \eqref{eq_def_B_op_open}
with respect to $B$.
Plugging this field strength
   into \eqref{eq_nonAb_Stokes} we get the equation
\be  \label{eq_derivation_nongenuine_link}
\ba  
\Big( \Gamma(\gamma_t)_0^1 \Big)^{-1} \frac{\partial  \Gamma(\gamma_t)_0^1}{\partial t}
 =  \delta(t-t_*)
\Big( \Gamma(\gamma_{t_*})_0^{s_*} \Big)^{-1}
(\Gamma_{P_*}^{\pi(P_*)})^{-1}
X_0 \Gamma_{P_*}^{\pi(P_*)}
 \Gamma(\gamma_{t_*})_0^{s_*}   \ .
 \ea
\ee 
We observe that the product
$\Gamma_{P_*}^{\pi(P_*)}
 \Gamma(\gamma_{t_*})_0^{s_*}$
amounts to parallel transport 
from the starting point $Q$ at $s=0$,
to the contact point $P_*$ at $s=s_*$,
followed by transport from $P_*$ to $\pi(P_*)$.
All in all, this is equal to 
$\Gamma(\gamma_{t_*})_0^1$.
(Here we are using  the fact 
the connection is flat away from $P_*$
and hence parallel transport depends 
on the initial and final points, but not on the
specific shape of the path connecting them.
Moreover, we recall that the endpoint $\gamma_{t_*}(1)$ 
is the same as $\pi(P_*)$.)
The above remarks shows that we can recast 
\eqref{eq_derivation_nongenuine_link} in the form
\be  
\Big( \Gamma(\gamma_t)_0^1 \Big)^{-1} \frac{\partial  \Gamma(\gamma_t)_0^1}{\partial t}
 =  \delta(t-t_*)
\Big( \Gamma(\gamma_{t_*})_0^1 \Big)^{-1}
X_0 \Gamma(\gamma_{t_*})_0^1  \ ,
\ee 
or equivalently 
\be \label{eq_derivation_nongenuine_linkBIS}
\bigg[  \Gamma(\gamma_t)_0^1 
\Big( \Gamma(\gamma_{t_*})_0^1 
\Big)^{-1}
\bigg]^{-1}
\frac{\partial  }{\partial t}\bigg[  \Gamma(\gamma_t)_0^1 
\Big( \Gamma(\gamma_{t_*})_0^1 
\Big)^{-1}
\bigg]
 =  \delta(t-t_*)
X_0  \ .
\ee 
Since the RHS is of the form
$\delta(t-t_*)$ times a constant element of $\mathfrak g$,
we adopt the parametrization
\be 
\Gamma(\gamma_t)_0^1 
\Big( \Gamma(\gamma_{t_*})_0^1 
\Big)^{-1} = \exp (f(t) X_0) \ , 
\ee 
where $f(t)$ is a scalar function of $t$.
We see that, if we choose
$f(t) = \theta(t-t_*)$,
we satisfy
\eqref{eq_derivation_nongenuine_linkBIS},
as well as the consistency requirement
$\Gamma(\gamma_t)_0^1 
\Big( \Gamma(\gamma_{t_*})_0^1 
\Big)^{-1} = 1$ for $t=t_*$.
In conclusion,
\be 
\ba 
\Gamma(\gamma_t)_0^1  & = \Gamma(\gamma_{t_*})_0^1  \ , & 
&t\le t_* \ , \\ 
\Gamma(\gamma_t)_0^1  & = e^{X_0} \Gamma(\gamma_{t_*})_0^1  \ , &
&t > t_* \ . 
\ea 
\ee 
In other words, upon sweeping the path
across the $B$-surface, the value of parallel transport is multiplied on the left by $e^X_0 \in G$.
With this result we can immediately write down the effect of  
passing a non-genuine $B$-operator
past a stretched Wilson line,
\be \label{eq_Qo_acts_on_stretched_Wilson}
\text{swipe $\mathbf Q^o_{X_0}$ through $\mathbf W_{\mathbf R}^o$:}
\qquad 
\mathbf W_{\mathbf R}^o(\gamma) \mapsto
 D_{\mathbf R} (  
e^{X_0}  
)
\mathbf W_{\mathbf R}^o(\gamma)
  \ .
\ee 
Recall that $D_{\mathbf R}$ is the map
that associates to an element of $G$
the matrix that represents it in the irrep 
$\mathbf R$.

It is straightforward to specialize the above
results to the limit $z_* \rightarrow z_0$,
in which the non-genuine operator
$\mathbf Q_{X_0}(\Sigma^{d-1} \times [z_0, z_*])$
yields the genuine operator
$D_{X_0}(\Sigma^{d-1} \times \{ z_0 \})$
stuck at the symmetry boundary.
In particular, \eqref{eq_Qo_acts_on_stretched_Wilson} implies that 
passing an  $D_{X_0}$ operator
past a stretched Wilson line
has the   effect 
\be \label{eq_S_acts_on_stretched_Wilson}
\text{swipe $D_{X_0}$ through $\mathbf W_{\mathbf R}^o$:}
\qquad 
\mathbf W_{\mathbf R}^o(\gamma) \mapsto
 D_{\mathbf R} (  
e^{X_0}  
)
\mathbf W_{\mathbf R}^o(\gamma)
  \ .
\ee

The operator 
$D_{X_0}(\Sigma^{d-1} \times \{ z_0 \})$
provides an explicit
realization of a class
of operators discussed  in 
\cite{Bonetti:2024cjk}, see also
\cite{Jia:2025jmn}.
These operators are stuck on the symmetry boundary and
are labeled by a group element (here $e^{X_0}$),
as opposed to a conjugacy class.
They act on stretched Wilson lines 
as in \eqref{eq_S_acts_on_stretched_Wilson}.

\subsection{Proposal to accommodate  non-flat Dirichlet boundary conditions}
\label{sec_nonflat_BF}

According to the proposal of \cite{Brennan:2024fgj,Antinucci:2024zjp,Bonetti:2024cjk}
discussed in Section \ref{sec_SymTFT_review},
the bulk-boundary description of a continuous
global 0-form symmetry $G$
makes use of the topological BF theory 
\eqref{eq_BF} with gauge group $G$.
This is equipped with Dirichlet boundary conditions for the gauge field $A$ on the symmetry boundary.
Due to the bulk equations of motion, the boundary value of $A$ must necessarily be a flat connection.
In this section we propose a generalization of the above framework, which allows us to impose Dirichlet boundary conditions with non-flat boundary values for $A$.
To achieve this goal, we consider a family of
bulk theories on $X^{d+1}$, all based on the same gauge group $G$.
A member of the family is 
determined by a choice of background $G$-connection $\cA$
on $X^{d+1}$.  
The proposed bulk action does not give rise to any propagating degrees of freedom
(in contrast, say, to a standard Yang-Mills action).

\paragraph{Action, gauge symmetries, equations of motion.}

The dynamical fields on $X^{d+1}$ are:
a $G$-connection $A$;
a $\mathfrak g$-valued $(d-1)$-form $B$;
a  $\mathfrak g$-valued $(d-2)$-form $\beta$; a $G$-valued 0-form $U$.
We also need to choose a non-dynamical
background $G$-connection $\cA$.
The action reads 
\be \label{eq_BFvariation_action}
S  = \frac{1}{2\pi} \int_{X^{d+1}}
{\rm Tr} \bigg[ 
(B + d_A \beta) \wedge (F_A - U \cF U^{-1} )
\bigg]  \ . 
\ee 
This action is written in terms of the quantities
\be
F_A = dA + \tfrac 12[A,A] \ , \qquad 
d_A \beta = d\beta + [A,\beta] \ , \qquad 
\cF = d\cA + \tfrac 12 [\cA, \cA]  \ .
\ee 
The action \eqref{eq_BFvariation_action} is invariant under 
\be \label{eq_BFvariation_gauge}
\ba 
A &\mapsto g(d+A) g^{-1} \ , & 
B &\mapsto g(B - d_A \tau) g^{-1} \ , & 
\beta &\mapsto g(\beta + \tau) g^{-1} \ , \\ 
\cA &\mapsto \overline g(d+\cA) \overline g^{-1} \ , &
U &\mapsto g U \overline g^{-1}  \ .
\ea 
\ee 
Here $g$, $\overline g$ are a $G$-valued 0-forms, with $g$ the parameter of a gauge transformation of the dynamical $A$, and $\overline g$ for $\cA$.
As usual, $\tau$ is a $\mathfrak g$-valued $(d-2)$-form.

The equations of motion
for $A$, $B$, $U$, $\beta$ read 
\be \label{eq_BFvariation_EOM}
\ba 
0&=d_A (B  +d_A \beta)
+ [\beta ,F_A - U \cF U^{-1} ]
\ , \\ 
0 & = F_A - U \cF U^{-1} \ , \\
0 & = [B + d_A \beta , U \cF U^{-1}] \ , \\
0 & = d_A(U \cF U^{-1}) \ ,
\ea 
\ee 
respectively. We have made use of the Bianchi identity $d_A F_A = 0$ and of  
$d_A d_A \beta = [F_A, \beta]$.
We also observe that the equations of motion for $U$ and $\beta$ (third and fourth equation)
can be derived from those of $A$, $B$
(first and second equation) by acting with $d_A$ and using $d_A F_A = 0$,
$d_A d_A \beta = [F_A,\beta] =  [U \cF U^{-1},\beta]$.

The bulk action \eqref{eq_BFvariation_action} reduces to the standard BF bulk action \eqref{eq_BF} if   $\cF = 0$. (The bulk term $d_A \beta \wedge F_A$
becomes purely a boundary term by integration by parts and $d_A F_A = 0$.)
For any $\cF$, the equation of motion of $B$ fixes the field strength $F_A$ of $A$ to be equal to $\cF$ up to conjugation.
Thus, for non-zero $\cF$, the modified action
\eqref{eq_BFvariation_action} can describe non-flat $G$-connections. On the other hand, 
we have no propagating degrees of freedom, because $F_A$ is completely fixed
in terms of the $c$-number $\cF$
up to gauge transformations.

\paragraph{Dirichlet boundary conditions.}

We are interested in considering a theory of the form
\eqref{eq_BFvariation_action} on a spacetime of the form
$X^{d+1}  = M^d \times [z_0, z_1]$.
The boundary component $M^d \times \{z_0\}$ is interpreted as the symmetry boundary
where we wish to impose
the Dirichlet boundary condition.

We proceed as follows.
First we select a background $G$-connection 
$\overline \cA$ on $M^d \times \{z_0\}$.
Next, we extend $\overline \cA$ to a background $G$-connection $\cA$ on $X^{d+1}$. 
This can be done because the interval direction is topologically trivial (contractible). 
We can now consider
the BF theory \eqref{eq_BFvariation_action}
written in terms of this 
background $G$-connection $\cA$. 
We equip the bulk action 
\eqref{eq_BFvariation_action} with the following 
 boundary condition at $z_0$, 
\be \label{eq_Dirichlet_nonflat}
A \Big|_{M^d \times \{ z_0 \}} = \overline \cA   \ , \qquad 
U \Big|_{M^d \times \{ z_0 \}} = 1 \ . 
\ee 
Notice that we 
have to use a bulk action
that is tailored to the chosen value boundary value
 $\overline \cA$. This is in contrast to the standard
SymTFT paradigm, in which the bulk is fixed and one scans over topological boundary conditions of the bulk theory at $B^d_{\rm sym}$. 

We remark that, 
without loss of generality,
we can work with an extension $\cA$ of $\overline \cA$ such that 
the field strength
$\cF$  has no legs along the interval direction $z$.
If we use $\mu,\nu$ for spacetime indices along the $M^d$ directions,
\be \label{eq_no_z_legs}
\cF_{\mu z} = 0  \ , \qquad 
\cD_z \cF_{\mu\nu} = 0 \ . 
\ee 
In the second equation, $\cD_z = \partial_z + [\cA_z,\cdot]$ is the covariant derivative constructed with $\cA$. Note that the second relation in \eqref{eq_no_z_legs} follows form the first and the Bianchi identity.\footnote{Given  $\overline \cA(x)$ on $M^d$ (where $x$ are local coordinates),
we can extend it to a connection $\cA$ on $M^d \times [z_0,z_1]$ satisfying
(up to gauge transformations)
\be 
\cA_\mu (x,z) = \overline \cA_\mu(x) \ , \qquad 
\cA_z(x,z)=0 \ . \nonumber
\ee 
Here, 
the components of $\cA$ along the $M^d$ directions are denoted $\cA_\mu(x,z)$,
and the component along the $z$ direction
is denoted $\cA_z(x,z)$.
Then, 
the covariant 
relations in~\eqref{eq_no_z_legs} follow.  
}

The considerations of the previous paragraphs apply also to the Abelian setting
$G = U(1)$. In this case, the 2-form $\cF$
is a gauge-invariant closed 2-form with integral periods.
We do not need to introduce the fields $U$, $\beta$.

\paragraph{Heuristic picture: from \eqref{eq_BF} to \eqref{eq_BFvariation_action} in terms of $B$-defects.}
Let us reconsider the standard BF action
\eqref{eq_BF}, and insert one $B$-defect \eqref{eq_def_B_op}
along a submanifold $\Sigma^{d-1}$ of $X^{d+1}$.
The total action can be written as
\be 
S = \frac{1}{2\pi}\int_{X^{d+1}} {\rm Tr}\bigg[  B \wedge  F_A
+ 
(B + d_A \beta) \wedge U \Big( 2\pi X_0 \delta_2(\Sigma^{d-1} \subset X^{d+1}) \Big)  U^{-1}
\bigg] \ . 
\ee 
Suppose that $\Sigma^{d-1}$ is of the form
$\Sigma^{d-1} = \widehat \Sigma^{d-2} \times [z_0,z_1]$, namely, that the $B$-defect
extends along the interval direction.
Then, the 2-form $\delta_2(\Sigma^{d-1} \subset X^{d+1})$ has no leg along the interval direction, and can be taken to be independent of the interval coordinate $z$.
Next, we can imagine inserting a smeared collection of $B$-defects, all extending along the interval direction.
Heuristically, the net result is to replace the delta-function localized $\mathfrak g$-valued quantity $X_0 \delta_2(\Sigma^{d-1} \subset X^{d+1})$ with a smooth $\mathfrak g$-valued 2-form $\cF$,
which has no legs along the interval direction,
and is independent of the interval coordinate.
At the same time, the fields $\beta$, $U$,
which are originally only living on the worldvolume of the $B$-defect,
can now be regarded as a bulk fields.
All in all, we get a structure like the proposed modified bulk action 
\eqref{eq_BFvariation_action}.
(To get a precise match, we need to add a term of the form $F_A d_A \beta$, which is a total derivative.)
The idea of using $B$-defects stretching along the interval direction to
implement non-flat connections
has been advanced in \cite{Brennan:2024fgj} in the Abelian setting.

\subsubsection{Comments on topological operators}

Let us comment briefly on some aspects of topological operators in the BF theory \eqref{eq_BFvariation_action} with non-trivial $\cF$.
We focus on two classes of operators that are arguably the most important in the sandwich construction: Wilson lines that stretch in the interval direction,
and $B$-operators living inside  the Dirichlet boundary.

\paragraph{Wilson lines.}
The theory with action \eqref{eq_BFvariation_action} admits standard
Wilson line operators, defined in terms of the path ordered exponential of the dynamical connection $A$. For generic background 2-form $\cF$ and generic support, however,
these Wilson lines are not topological, since the dynamical connection $A$ is not flat on-shell.

Our main application of Wilson line operators is
in the sandwich construction, 
where we take them to stretch from one boundary to the other.
Upon closing the sandwich, this configuration in general corresponds to a non-topological local operator.
We can consider what happens if we deform the support of the stretched Wilson line in the bulk, keeping its endpoints fixed. We  argue that, if \eqref{eq_no_z_legs} holds, the Wilson line is actually invariant under such deformations, even if $\cF$ is nonzero. This can be seen as follows.
Let us consider a Wilson line stretched along a straight line, parametrized by $z$ and located at some fixed $x(z) = x_*$ in the $M^d$ directions.
We can imagine to deform the support of the Wilson line, introducing a small ``bump''
pointing, say, in the $x^1$ direction. In other words, the new support of the Wilson line is described by the equations $x^1(z) = x^1_* + f(z)$,
$x^\mu(z) = x_*^\mu$, $\mu = 2,\dots,d$, where $f(z)$
is supported on a small interval centered around some value of $z$, and zero outside.
The change in the Wilson line due to the change in its support is related to the integral
of the component $\cF_{1z}$
of $\cF$, which however is zero, thanks to \eqref{eq_no_z_legs}.
A more detailed version of the above argument can be made using the non-Abelian Stokes' formula, leading to the same conclusions.

\paragraph{$B$-defects inside the Dirichlet boundary.}
In the usual setting with flat Dirichlet boundary conditions, we know that the $G$ symmetry is implemented by $(d-1)$-dimensional topological operators inside
$B^d_{\rm sym}$. They were denoted
$D_{X_0}$ in Section \ref{sec_non_genuine_B},
where they were defined as limits of non-genuine $B$-defects, see 
\eqref{eq_S_as_a_limit}. 

We can repeat a similar analysis in the present setting. 
The non-genuine $B$-defect 
extends along $[z_0, z_*]$ in the interval direction. It 
has the same form as
\eqref{eq_def_B_op_open}. The quantity $\mathcal B$ is constructed 
as in \eqref{eq_mathcalB_def}, but using the  
combination $B + d_A \beta$ in place of $B$ only. When we consider the limit
$z_* \rightarrow z_0$, we get an operator stuck on the Dirichlet boundary, which can be described by the action
\be \label{eq_stuck_operator_nonflat}
\frac{1}{2\pi} \int_{\Sigma^{d-1} \times \{z_0\}} {\rm Tr} \Big[ 
(B + d_{\overline \cA} \beta) X_0
\Big] \ .
\ee 
We have used the fact that $A = \overline \cA$ at the Dirichlet boundary.
This operator should be topological under deformations of its support
that keep it inside the Dirichlet boundary.
This leads us to consider the exterior derivative of the integrand in \eqref{eq_stuck_operator_nonflat},
\be 
d{\rm Tr} \Big[ 
(B + d_{\overline \cA} \beta) X_0
\Big]  
= (-1)^{d-1} {\rm Tr}
(B + d_{\overline \cA} \beta) d_{\overline 
\cA}X_0 \ .
\ee 
We have used $d_{\overline \cA}(B + d_{\overline \cA}
\beta)=0$, which follows from the bulk equations of motion \eqref{eq_BFvariation_EOM} restricted to the Dirichlet boundary.
We are led to demand
that the constant parameter $X_0$
is covariantly constant with respect to the background connection,
\be 
d_{\overline \cA} X_0 = 0 \ .
\ee 
Thanks to this relation, we can actually drop the $d_{\overline \cA}\beta$ term in 
\eqref{eq_stuck_operator_nonflat}.

\paragraph{Field theory comments.}
The fact that the parameter $X_0$ should be convariantly constant with respect to $\overline \cA$ 
can also be
seen purely in field theory.
Suppose we have a system with a non-anomalous $G$ symmetry. 
The associated current $*J$ is a $(d-1)$-form valued in the adjoint representation of $G$. Let us couple the system to a
background $G$-connection $\overline \cA$.
In the presence of $\overline \cA$,
the conservation equation for $*J$ reads
\be 
d_{\overline \cA} *J = 0 \ . 
\ee 
Our goal is to construct a $(d-1)$-dimensional topological operator to implement $G$ transformations.
Our strategy is to build it by exponentiating a quantity constructed from $*J$. Since $*J$ is in the adjoint,
we consider
\be 
*j = {\rm Tr} (X_0 *J) \ , 
\ee 
where $X_0$ is a $\mathfrak g$-valued parameter. We require $*j$ to be closed,
\be 
d*j = {\rm Tr} (d_{\overline \cA} X_0 \wedge  *J
+ X_0 \wedge d_{\overline \cA} *J)
= {\rm Tr} (d_{\overline \cA} X_0 \wedge  *J) \ .
\ee 
We see that the parameter $X_0$ must be covariantly constant with respect to $\overline \cA$. If this is the case,
$*j$ is closed and can be exponentiated
to give a topological operators
(modulo  caveats associated to  regularization, see   \cite{Bah:2024ucp}).

\section{SymTFT and non-linearly realized symmetries}\label{sec:nonlinearly}

This section is devoted to the study of bulk-boundary realizations of QFTs admitting a global symmetry that is realized non-linearly. We stress that the standard sandwich construction, based on a bulk BF theory and Dirichlet boundary conditions at $B^d_{\rm sym}$, does not capture non-linear realizations.
Indeed, charged operators in the QFT are realized by open Wilson lines, stretched from $B^d_{\rm sym}$ to $B^d_{\rm phys}$,
and the endpoint of the Wilson line on 
$B^d_{\rm sym}$
transforms linearly upon linking with the topological  operators on $B^d_{\rm sym}$ that implement the global $G$ symmetry \cite{Bonetti:2024cjk}.

Our analysis builds on \cite{Benini:2022hzx,Antinucci:2024bcm,Argurio:2024ewp,Paznokas:2025auw}.
The rest of this section is structured as follows. First, we consider an Abelian setting,
with a non-linearly
realized $p$-form symmetry,
discussing general $p$ and dimension $d$.
Next, we turn to the discussion of a non-linearly realized non-Abelian 0-form symmetry $G$. In general, a subgroup $H\subset  G$ may be linearly realized.
We start by studying the case in which $H$ is trivial. After that, we 
analyze the case of non-trivial~$H$.
We proceed by exhibiting a club sandwich construction \cite{Bhardwaj:2023bbf}
that engineers 
the Callan-Coleman-Wess-Zumino
action 
associated to a non-Abelian group $G$ and
an Abelian subgroup $H$ of $G$. We conclude by studying some examples of 
symmetry structures that are beyond ordinary groups:
 2-groups, and $\mathbb Q/\mathbb Z$ symmetry.

\subsection{Non-linearly realized $U(1)$ $p$-form symmetry}

\subsubsection{Field theory preliminaries}

Let us consider a QFT in $d$ dimensions with a non-linearly realized $U(1)$ $p$-form symmetry.
By this we mean that the QFT contains a dynamical field $\varphi_p$ that shifts under
the action of the global symmetry $U(1)^{(p)}$.
More precisely, 
let $A_{p+1}$
be a background gauge field for 
$U(1)^{(p)}$.
The variations of $A_{p+1}$ and $\varphi_p$  under 
background gauge transformations read
\be 
A_{p+1} \rightarrow A_{p+1} + d\Lambda_p \ , \qquad 
\varphi_p \rightarrow \varphi_p - \Lambda_p \ . 
\ee 
Both the dynamical field $\varphi_p$ and the gauge parameter $\Lambda_p$ are $U(1)$
$p$-form gauge fields.
In the presence of the background gauge field $A_{p+1}$,
the field strength
of $\varphi_p$ is
\be 
D\varphi_p = d\varphi_p + A_{p+1} \ . 
\ee 

Let us define 
\be 
*_dJ_{d-p-1} = \frac{D\varphi_p}{2\pi}  \ . 
\ee 
The quantity 
$*_dJ_{d-p-1}$
satisfies
\be 
d*_dJ_{d-p-1} = \frac{dA_{p+1}}{2\pi} \ . 
\ee 
This is interpreted as follows. If we turn off the background gauge field $A_{p+1}$ for $U(1)^{(p)}$, $J_{d-p-1}$ is a conserved current. It signals a global $U(1)^{(d-p-2)}$  symmetry of the system.
In the presence of a generic background field $A_{p+1}$, the conservation of the current $J_{d-p-1}$
is violated by a c-number proportional to $dA_{p+1}$.
This means that the two global symmetries
$U(1)^{(p)}$ and $U(1)^{d-p-2}$ have a mixed 't Hooft anomaly. This anomaly can be captured by descent via the anomaly polynomial
\be 
\cI_{d+2} = \frac{1}{(2\pi)^2} 
d\widetilde A_{d-p-1}\wedge A_{p+1}   \ ,
\ee 
where $\widetilde A_{d-p-1}$ is the background gauge field
for $U(1)^{(d-p-2)}$.
Equivalently, the anomaly is captured via inflow
by the 5d topological action
\be \label{eq_emergent_mixed}
\frac{1}{2\pi} \int_{X^{d+1}} \widetilde A_{d-p-1} \wedge dA_{p+1} \ ,
\ee 
where $X^{d+1}$
bounds   physical spacetime.

\subsubsection{SymTFT description}
Building on \cite{Argurio:2024ewp,Paznokas:2025auw},
we describe a sandwich construction for the system described above. 
As we close the sandwich, we will reproduce the action 
for $\varphi_p$, including 
 background fields for both $U(1)^{(p)}$
and $U(1)^{d-p-2}$.
We also comment on electromagnetic duality for
$d=2p+2$.

\paragraph{Bulk TQFT.}
As we have seen, the physical theory
has a global $U(1)^{(p)} \times U(1)^{(d-p-2)}$ global symmetry with the mixed 't Hooft anomaly \eqref{eq_emergent_mixed}.
To capture these global symmetry structures we can use a bulk $(d+1)$-dimensional TQFT 
consisting of
two mixed $\mathbb R$-$U(1)$ BF theories, coupled via an additional term in the action,
\be 
S_{\rm bulk} = \frac{1}{2\pi} 
\int_{X^{d+1}} \bigg[ 
B_{d-p-1} \wedge dA_{p+1}
+ C_{p+1} \wedge d\widetilde A_{d-p-1}
+ \widetilde A_{d-p-1} \wedge dA_{p+1}
\bigg]  \ . 
\ee 
Here $B_{d-p-1}$,
$C_{p+1}$ are $\mathbb R$ gauge fields, while
$A_{p+1}$, $\widetilde A_{d-p-1}$
are $U(1)$ gauge fields.
This TQFT admits an equivalent description in terms of a single BF term, where both field are $\mathbb R$ gauge fields \cite{Antinucci:2024zjp},
\be \label{eq_RR_bulk_action}
S_{\rm bulk} = \frac{1}{2\pi} \int_{X^{d+1}} 
B_{d-p-1} \wedge dC_{p+1} \ . 
\ee 
The gauge transformations are
\be \label{eq_RR_gauge}
B_{d-p-1} \mapsto B_{d-p-1} + d\omega_{d-p-2} \ , \qquad 
C_{p+1} \mapsto C_{p+1} + d \mu_p \ , 
\ee 
where $\omega_{d-p-2}$, $\mu_p$
are globally defined forms.

\paragraph{Topological boundary condition on $B^d_{\rm sym}$.}
We consider the following topological boundary condition on $B^d_{\rm sym}$, labeled by a real parameter $R$,
\be \label{eq_RR_top}
\ba 
S_{\rm sym} &= \frac{1}{2\pi} \int_{B^d_{\rm sym}}
\bigg\{  
\Big[
B_{d-p-1} -  R^{-1} (d\widetilde \varphi_{d-p-2} + \widetilde \cA_{d-p-1}) 
\Big] \wedge \widetilde \lambda_{p+1}
\\
&+ \lambda_{d-p-1} \wedge 
\Big[ 
C_{p+1} -  R (d\varphi_p  + \cA_{p+1})
\Big]
+ (-1)^{d-p} \widetilde \cA_{d-p-1} \wedge (d\varphi_p + \cA_{p+1})
\bigg\} \ . 
\ea 
\ee 
We have introduced the dynamical fields
$\widetilde \lambda_{p+1}$, $\lambda_{d-p-1}$,
$\varphi_p$, and $\widetilde \varphi_{d-p-2}$ on $B^d_{\rm sym}$.
The Lagrange multipliers $\widetilde \lambda_{p+1}$, $\lambda_{d-p-1}$ are forms, while $\varphi_p$,  $\widetilde \varphi_{d-p-2}$
are $U(1)$ gauge fields.
The gauge transformations of the dynamical fields on $B^d_{\rm sym}$ are taken to be 
\be 
\ba 
\varphi_p &\mapsto \varphi_p + d \alpha_{p-1} + R^{-1} \mu_p \ , & 
\widetilde \varphi_{d-p-2} &\mapsto \widetilde \varphi_{d-p-2}
+ d \widetilde \alpha_{d-p-3} + R \omega_{d-p-2}  \ , \\ 
\lambda_{d-p-1} & \mapsto \lambda_{d-p-1} + (-1)^{d-p} d\omega_{d-p-2} \ , & 
 \widetilde \lambda_{p+1} & \mapsto  \widetilde \lambda_{p+1}  \ . 
\ea 
\ee 
The quantities $\cA_{p+1}$, $\widetilde \cA_{d-p-1}$
are non-dynamical $U(1)$ gauge fields on $B^d_{\rm sym}$, which we take to be flat,
\be
d\cA_{p+1} = 0 \ , \qquad 
d \widetilde \cA_{d-p-1} = 0  \ . 
\ee 
In our approach, they do not transform under gauge transformations of the dynamical fields in the bulk and on $B^d_{\rm sym}$.

The gauge variation of the bulk action
\eqref{eq_RR_bulk_action} produces a boundary term.
Combined with the gauge variation of the
action \eqref{eq_RR_top}, we get in total
(we suppress form degrees and wedge products for brevity)
\be 
\ba 
\Delta S_{\rm bulk} + \Delta S_{\rm sym}
&= \frac{1}{2\pi} \int_{Y^d} \bigg[ 
(-1)^{d-p-1} R d\omega d\varphi 
+ (-1)^{d-p-1} R d\omega \cA
+(-1)^{d-p} R^{-1} \widetilde \cA d\mu
\bigg] \ . 
\ea 
\ee 
We notice the cancellation of all terms with $C$. The remaining terms can be collected into a total derivative.
(In each term we have at least one $\mu$ or $\omega$, which are globally defined forms.)
We have thus verified that we have a gauge invariant combined action
$S_{\rm bulk} + S_{\rm sym}$.

Next, let us consider the variations of 
$S_{\rm bulk}$,
$S_{\rm sym}$ under arbitrary variations of the fields.
In particular, we observe that the variation of
$S_{\rm bulk}$ gives rise to boundary terms,
\be \label{eq_RR_bulk_delta_gives_bdy}
\delta S_{\rm bulk} = \frac{1}{2\pi} \int_{\partial X^{d+1}}
(-1)^{d-p-1} B_{d-p-1} \wedge \delta C_{p+1} \ . 
\ee 
We use the orientation convention
$\partial X^{d+1} = B^d_{\rm sym} - B^d_{\rm phys}$.
We collect all terms on $B^d_{\rm sym}$, and we get
\be 
\ba 
& \delta S_{\rm bulk} + \delta S_{\rm sym}   = \frac{1}{2\pi} \int_{Y^d} \bigg\{ 
\Big[ B - R^{-1}(d\widetilde \varphi  + \widetilde \cA )\Big] \delta \widetilde \lambda
+ \delta \lambda \Big[ 
C - R (d\varphi + \cA)
\Big]
\\
& + \delta B  \widetilde \lambda 
+ \Big[\lambda + (-1)^{d-p-1} B \Big] \delta C
+ (-1)^{d-p} R^{-1} \delta \widetilde \varphi d\widetilde \lambda 
+ (-1)^{d-p-1} d(R \lambda - (-1)^{d-p} \widetilde  \cA) \delta \varphi
\bigg\} \ . 
\ea 
\ee 
From the terms with $\delta B$, $\delta C$, $\delta \lambda$,
$\delta \widetilde \lambda$ we infer 
\be \label{eq_RR_EOMs_on_sym}
\ba 
&\text{on $B^d_{\rm sym}$:}   &
B_{d-p-1} &= R^{-1}(d\widetilde \varphi_{d-p-2} 
+ \widetilde \cA_{d-p-1} ) \ , &
C_{p+1} &= R(d \varphi_p + \cA_{p+1}) \ , \\
&& 
\lambda_{d-p-1} &= (-1)^{d-p} B_{d-p-1} \ , &
\widetilde \lambda_{p+1} &= 0 \ .
\ea 
\ee 
We get no new information from the terms with
$\delta \varphi$, $\delta \widetilde \varphi$.
However, the sum over the fluxes of $d\varphi$,
$d\widetilde \varphi$ in the boundary action
\eqref{eq_RR_top} imposes that 
\be \label{eq_RR_sym_integrality}
\int_{\Sigma^{p+1}} R^{-1} \widetilde \lambda_{p+1}\in 2\pi \mathbb Z \ , \qquad 
\int_{\Sigma^{d-p-1}} \Big[ R \lambda_{d-p-1} - (-1)^{d-p} \widetilde \cA_{d-p-1} \Big] \in 2\pi \mathbb Z \ ,
\ee 
where $\Sigma^{p+1}$, $\Sigma^{d-p-1}$ are cycles in $B^d_{\rm sym}$.
Let us consider the integrands in \eqref{eq_RR_sym_integrality}
and use \eqref{eq_RR_EOMs_on_sym}. We get the expressions
\be \label{eq_RR_these_are_the_integrands}
R^{-1} \widetilde \lambda_{p+1} = 0 \ , \qquad 
R \lambda_{d-p-1} - (-1)^{d-p} \widetilde \cA_{d-p-1} = (-1)^{d-p} d\widetilde \varphi_{d-p-2} \ . 
\ee 
In the second equality we have a cancellation of the terms with $\widetilde \cA$. Indeed, this is how we have fixed the coefficient
of the term $\widetilde \cA (d\varphi + \cA)$ in \eqref{eq_RR_top}.
Thanks to \eqref{eq_RR_these_are_the_integrands}, we see that \eqref{eq_RR_sym_integrality} are automatically satisfied, without imposing any new constraints.

The choice of topological boundary action
\eqref{eq_RR_top} implies that:
\begin{itemize}
    \item an open operator $e^{i n R \int B_{d-p-1}}$ with $n\in \mathbb Z$ can end perpendicularly on $B^d_{\rm sym}$;
    \item a  closed operator $e^{i n R \int B_{d-p-1}}$ with $n\in \mathbb Z$ is trivialized if projected parallel onto  $B^d_{\rm sym}$;
    \item an open operator $e^{i m R^{-1} \int C_{p+1}}$ with $m\in \mathbb Z$ can end perpendicularly on $B^d_{\rm sym}$;
    \item a  closed operator $e^{i m R^{-1} \int C_{p+1}}$ with $m\in \mathbb Z$ is trivialized if projected parallel onto~$B^d_{\rm sym}$.
\end{itemize}
We can see this as follows. On $B^d_{\rm sym}$ we can construct
 operators of the form
\be \label{eq_RR_ops_on_sym_bdy}
e^{i n \int  \widetilde \varphi_{d-p-2}} \ , \qquad 
e^{i m \int   \varphi_{p}}  \ , \qquad 
n, m \in \mathbb Z \ , 
\ee 
where the integrality requirement on $n$, $m$ stems from the fact
that $\varphi_p$, $\widetilde \varphi_{d-p-2}$ are $U(1)$ gauge fields.
From the gauge transformations of $\varphi_p$, $\widetilde \varphi_{d-p-2}$, we see that the operators in \eqref{eq_RR_ops_on_sym_bdy}
can be at the endpoint of open operators 
$e^{i \alpha \int B_{d-p-1}}$,
$e^{i \beta \int C_{p+1}}$ in the bulk, but only provided that
$\alpha = n R$, $\beta = m R^{-1}$.
By a similar token, let us consider
a closed $e^{i \beta \int C_{p+1}}$ operator, and project it parallel onto $B^d_{\rm sym}$. There, we can make use of \eqref{eq_RR_EOMs_on_sym}. We get 
\be 
e^{i \beta \int C_{p+1}} = e^{i \beta R \int \cA_{p+1}} e^{i \beta R \int d\varphi_p} \ . 
\ee 
The first factor is a c-number,
well-defined for $\beta R\in \mathbb Z$ because $\cA_{p+1}$ is a $U(1)$ gauge field.
The second factor is trivialized if
$\beta R\in \mathbb Z$, because $\int d\varphi_p \in 2\pi \mathbb Z$.
The same argument applies to $e^{i \alpha \int B_{d-p-1}}$ operators.
We close by noticing that the operators
\be \label{eq_RR_Lagrangian}
e^{i n R \int B_{d-p-1}} \ , \qquad 
e^{i m R^{-1} \int C_{p+1}} \ , \qquad 
n,m \in \mathbb Z \ ,
\ee 
have trivial mutual braiding in the bulk,
as can be verified from the bulk action \eqref{eq_RR_bulk_action}.
In fact, they constitute a Lagrangian algebra, labeled by $R$ \cite{Antinucci:2024zjp}.
The sign of $R$ can be flipped by a flipping the sign of $\varphi_p$. Without loss of generality we can take $R > 0$.
Then, different values of $R$ 
correspond to distinct topological boundary conditions for the bulk TQFT.
The limits $R\rightarrow 0$,
$R \rightarrow +\infty$ correspond to
pure Dirichlet boundary conditions
for $C_{p+1}$, $B_{d-p-1}$.

\paragraph{Physical boundary conditions on $B^d_{\rm phys}$.}
On the physical boundary we impose a non-topological boundary condition. To do so, we assume that $B^d_{\rm phys}$ is equipped with a metric.
We do not include any localized degrees of freedom on $B^d_{\rm phys}$. The action is written in terms of the bulk field $C_{p+1}$ and reads
\be \label{eq_RR_phys_boundary}
S_{\rm phys} = \frac{1}{2\pi} \int_{B^d_{\rm phys}}
\bigg[ - \frac 12 
C_{p+1} \wedge *_d C_{p+1}
\bigg]  \ . 
\ee 
Recall that $B_{d-p-1}$ and $C_{p+1}$ can be rescaled at will.
We have implicitly used this freedom to fix the prefactors of the bulk action
\eqref{eq_RR_bulk_action} and of the $B^d_{\rm phys}$ action \eqref{eq_RR_phys_boundary}.
The action $S_{\rm phys}$ is not invariant under the bulk gauge transformations
\eqref{eq_RR_gauge}. Hence, the bulk gauge parameter $\mu_p$ must be restricted to be trivial on $B^d_{\rm phys}$.

The variation of $S_{\rm phys}$ 
with respect to $C_{p+1}$
combines with a $B \delta C$ contribution from $\partial X^{d+1}$, see \eqref{eq_RR_bulk_delta_gives_bdy}.
In total, we infer that 
\be \label{eq_RR_phys_EOM}
\text{on $B^d_{\rm phys}$:} \qquad 
*_dC_{p+1} + (-1)^{pd} B_{d-p-1} = 0 \ .
\ee

\paragraph{Closing the sandwich.}
Upon closing the sandwich, we obtain the sought-for physical action on $M^d$. It receives contributions from the action \eqref{eq_RR_top} on $B^d_{\rm sym}$, and from the action \eqref{eq_RR_phys_boundary} on $B^d_{\rm phys}$.
We evaluate these contributions
making use of \eqref{eq_RR_EOMs_on_sym}.
The total action upon closing the sandwich reads
\be \label{eq_RR_closed_sandwich}
S = \frac{1}{2\pi} \int_{M^d} \bigg[  - \frac 12 R^2 ( d\varphi_p + \cA_{p+1}) \wedge *_d ( d\varphi_p + \cA_{p+1}) 
+ (-1)^{d-p} \widetilde \cA_{d-p-1} \wedge (d\varphi_p + \cA_{p+1})
\bigg] \ . 
\ee 
This is the expected action for the mode $\varphi_p$,
coupled to background fields $\cA_{p+1}$, $\widetilde \cA_{d-p-1}$
for $U(1)^{(p)}$ and $U(1)^{d-p-2}$ (see e.g.~\cite{Brennan:2022tyl}).
We can also combine
\eqref{eq_RR_phys_EOM} and \eqref{eq_RR_EOMs_on_sym} to see that
$\widetilde \varphi_{d-p-2}$ is the electromagnetic dual of $\varphi_p$ in $d$ dimensions,
\be 
d\widetilde \varphi_{d-p-2}
+ \widetilde \cA_{d-p-1} = (-1)^{pd}
R^{-2} *_d(d\varphi_p + \cA_p) \ . 
\ee

\paragraph{Comments on electromagnetic duality.}
Let us specialize to the case $d = 2p+2$, in which
$B$ and $C$ both have form degree $p+1$.
The bulk action in invariant under a discrete 0-form symmetry exchanging $B$ and $C$. More precisely, let us consider the transformation
\be \label{eq_RR_exchange}
B \rightarrow C \ , \qquad 
C \rightarrow (-1)^p B \ . 
\ee 
It squares to $(-1)^p$ times the identity.
Under this transformation,
\be \label{eq_RR_exchange_bulk}
S_{\rm bulk} \rightarrow S_{\rm bulk } + \frac{1}{2\pi} \int_{\partial X^{d+1}} (-1)^p B_{p+1} \wedge C_{p+1}  \ .
\ee 
Let us now consider the action on $B^d_{\rm sym}$.
To find the new action, we have to implement
\eqref{eq_RR_exchange}, and take into account the additional term from
\eqref{eq_RR_exchange_bulk}.
The result reads
\be \label{eq_RR_top_after}
\ba
S_{\rm sym} &= \frac{1}{2\pi} \int_{B^d_{\rm sym}} \bigg\{ 
\Big[ B_{p+1} - R'^{-1} (d\widetilde \varphi_p' + \widetilde \cA_{p+1}')
\Big] \wedge \widetilde \lambda' _{p+1}
\\
& + \lambda'_{p+1} \wedge \Big[ 
C_{p+1} - R' (d\varphi'_p + \cA'_{p+1})
\Big] 
+ \cA'_{p+1} \wedge (d\widetilde \varphi_p' + \widetilde \cA_{p+1}')
+(-1)^p B_{p+1} \wedge C_{p+1}
\bigg\} \ .
\ea
\ee 
We have introduced the notation
\be \label{eq_RR_primed_fields}
\ba 
R^{-1}  & = R' \ , &
\widetilde \lambda_{p+1} &= (-1)^{p+1} \lambda_{p+1} ' \ , & 
\lambda_{p+1} &= \widetilde \lambda_{p+1} ' \ , & 
\widetilde \varphi_p &= \varphi'_p \ , & 
\varphi _p &= (-1)^p \widetilde \varphi'_p \ , \\ 
\widetilde \cA_{p+1} &= \cA'_{p+1} \ , & 
\cA _{p+1} &= (-1)^p \widetilde \cA'_{p+1} \ . 
\ea 
\ee 
The new action on the symmetry boundary can be analyzed
as we have done above for the original action.
We find that it leads to the equations of motion
\be   \label{eq_RR_sym_boundary_EOMs_after}
\ba 
&\text{on $B^d_{\rm sym}$:}   &
B_{p+1}  &= R'^{-1}(d\widetilde \varphi'_{p} + \widetilde \cA'_{p+1} ) \ , &
C_{p+1} &= R'(d \varphi'_p + \cA'_{p+1}) \ , \\
&& 
\lambda'_{p+1} &= 0 \ , &
\widetilde \lambda'_{p+1} &= (-1)^{p+1} C_{p+1} \ .
\ea 
\ee 
We also verify that the sum over fluxes of
$d\varphi'$, $d\widetilde\varphi'$ does not lead to any new constraint.
Even though the functional form
of the new boundary action \eqref{eq_RR_top_after}
is not identical to that of \eqref{eq_RR_top},
we see that both actions implement the same class
of boundary conditions for $B$ and $C$.
The parameter $R$ of the old boundary action
is mapped to the new parameter $R' = R^{-1}$.

Let us now briefly discuss the effect of the transformation
\eqref{eq_RR_exchange} on the action on the physical boundary.
We apply \eqref{eq_RR_exchange} to \eqref{eq_RR_phys_boundary}, and keep track of the boundary additional term in \eqref{eq_RR_exchange_bulk} (with a minus sign due to orientation).
We get the new action
\be \label{eq_RR_phys_boundary_BIS}
S_{\rm phys} = \int_{B^d_{\rm phys}} \bigg[ 
- \frac 12 B_{p+1} \wedge *_d B_{p+1} 
- (-1)^p B_{p+1} \wedge C_{p+1}
\bigg]  \ . 
\ee 
The presence of the second term is important to have a well-defined
variational problem. Indeed, taking into account the terms from integration by parts in the variation of the bulk action,
we have 
\be 
\delta S_{\rm bulk} + \delta S_{\rm phys} \supset \frac{1}{2\pi} \int_{B^d_{\rm phys}} -\delta B_{p+1} \wedge (  *_d B_{p+1} + (-1)^p C_{p+1}) \ ,
\ee 
where the terms with $\delta C$ have canceled.
The terms with $\delta B$ give the relation
$*_d B_{p+1} + (-1)^p C_{p+1}=0$, which is actually equivalent to
\eqref{eq_RR_phys_EOM}. In other words, even though the functional forms
of \eqref{eq_RR_phys_boundary} and \eqref{eq_RR_phys_boundary_BIS} are not identical,
they impose the same relation \eqref{eq_RR_phys_EOM} on $B^d_{\rm phys}$. This relation is invariant under the exchange
\eqref{eq_RR_exchange} on $B$ and $C$.

In summary, our explicit formulae confirm that
the exchange \eqref{eq_RR_exchange} 
maps a symmetry boundary condition with parameter $R$ to one with parameter $R^{-1}$, and leaves the physical boundary condition invariant.
The bulk 0-form symmetry  \eqref{eq_RR_exchange}
can be implemented by codimension-one topological operators in $X^{d+1}$, constructed by condensation of holonomies of $B$ and $C$. By considering an open version of these condensation defects, one can construct
a topological operator implementing the duality between
\eqref{eq_RR_closed_sandwich} and the same theory with $R\mapsto R^{-1}$.
This program has been carried out for $p=0$, $d=2$ 
(T-duality of the compact boson) \cite{Argurio:2024ewp}
and for $p=1$, $d=4$ (S-duality in Maxwell theory) \cite{Paznokas:2025auw} -- see also  \cite{Arbalestrier:2024oqg,Hasan:2024aow}.

\subsection{Review on non-linear realizations of non-Abelian 0-form symmetries}
\label{sec_nonlinear_review}

Let us consider an effective field theory in $d$ dimensions on which the global 0-form symmetry group $G$ acts non-linearly, with the Lie subgroup $H$ of $G$ realized linearly. It is proven in \cite{Coleman:1969sm,Callan:1969sn} that, up to field redefinitions,
the action of $G$ can 
always be described in terms of a set of scalar fields
$\phi$ parameterizing 
the coset $G/H$,
together with a collection of fields $\Psi$ that transform in a linear representation of $H$, as reviewed below in some detail.

We take $G$ to be a compact, connected, semisimple Lie group. The Lie algebra $\mathfrak g$ of $G$ can be decomposed as
\be \label{eq_Lie_alg_decomposition}
\mathfrak g \cong \mathfrak h \oplus \mathfrak h^\perp \ , 
\ee 
where $\mathfrak h$ is the Lie algebra of the subgroup $H$ and $\mathfrak h^\perp$ is the orthogonal complement of $\mathfrak h$ with respect to the Cartan-Killing metric on $
\mathfrak g$.
Notice that
$[\mathfrak h,\mathfrak h] \subset \mathfrak h$,
$[\mathfrak h,\mathfrak h^\perp] \subset \mathfrak h^\perp$,
while in general
$[\mathfrak h^\perp,\mathfrak h^\perp] \subset \mathfrak g$.
(The special case $[\mathfrak h^\perp,\mathfrak h^\perp] \subset \mathfrak h$
corresponds to $G/H$ being a symmetric space.)

Let us consider the coset $G/H$, i.e.~the set of equivalence classes of the equivalence relation $g\sim gh$, $g\in G$, $h\in H$.
Any element of $G$ can be written uniquely in the form
$\cV(\phi) h$,
where $\phi$ are coordinates on the coset space $G/H$,
$\cV(\phi) \in G$ is the standard 
representative for the coset 
element with coordinates $\phi$,
and $h$ is an element of $H$.
We sometimes refer to $\cV(\phi)$ as 
the scalar vielbein on the coset space $G/H$.

The group $G$ acts on $G/H$ from the left.
More precisely, a $G$ transformation with constant parameter $g_0$ takes the form
\be \label{eq_G_left}
g_0 \cV(\phi) = \cV(\phi') h(\phi; g_0) \ .
\ee 
The above equation defines  the transformed coordinate $\phi'$ on $G/H$
as well as the element
$h(\phi; g_0)\in H$, which plays the role of a compensating $H$ transformation.
We observe that it is always possible to choose the coset representative $\cV(\phi)$ in such a way that\footnote{To see this, we may choose
$\cV(\phi) = \exp(\phi^a x_a)$,
where $\{x_a\}_{a=1}^{\dim G-\dim H}$ is a basis of $\mathfrak h^\perp$.
From $[\mathfrak h, \mathfrak h^\perp] \subset \mathfrak h^\perp$
it follows that,  for $g_0 \in H$,
$g_0 \cV(\phi)= \cV(\phi') g_0$ where
$\phi'^a x_a = g_0 (\phi^a x_a) g_0^{-1}$.
} 
\be 
h(\phi ; g_0) = g_0 \qquad 
\text{for $g_0 \in H \subset G$}  \ ,
\ee 
namely, the compensating $H$ transformation is independent of the fields $\phi$ for transformations in the subgroup $H$.

From the coset representative $\cV(\phi)$ we construct the 1-forms $P$, $Q$ as follows,
\be \label{eq_P_and_Q_are_def}
\cV(\phi)^{-1} d\cV(\phi)
= P+Q \ , \qquad 
P \in \mathfrak h^\perp \  , \qquad 
Q \in \mathfrak h \ . 
\ee 
Under the $G$ action \eqref{eq_G_left} on $\phi$,
$P$ and $Q$ transform as
\be 
P' = h(\phi; g_0)
P h(\phi; g_0)^{-1} \ , \qquad 
Q' = h(\phi; g_0)
(d+Q) h(\phi; g_0)^{-1} \ .
\ee 
In particular, $Q$ transforms as a connection for the group $H$ and can be used to construct covariant derivatives and curvatures.
For instance, the covariant derivative of $P$ and the curvature of $Q$ are given respectively as
\be 
d_Q P = dP + [Q,P] = dP + Q\wedge P + P\wedge Q  \ , \qquad 
F_Q = dQ + \tfrac 12 [Q,Q] = dQ + Q\wedge Q \ .
\ee 
Under \eqref{eq_G_left} they
transform as
\be 
(d_Q P)' = h(\phi; g_0)
d_Q P h(\phi; g_0)^{-1} \ , \qquad 
(F_Q)' = 
h(\phi; g_0)
F_Q h(\phi; g_0)^{-1} \ . 
\ee

Next, suppose $\Psi$ is a field or a collection of fields transforming in a linear (unitary, not necessarily irreducible) representation
$D$
of $H$. Using the scalar fields $\phi$ we can define an action of the full group $G$ on $\Psi$:
under  
\eqref{eq_G_left}, $\Psi$
transforms as
\be 
\Psi' = D(h(\phi; g_0)) \Psi \ .
\ee 
The covariant derivative of 
$\psi$ is defined as
\be 
d_Q \Psi = d\Psi + D(Q)\Psi \ , 
\ee 
where, by abuse of notation, we have used the same symbol $D$ to denote the representation of $\mathfrak h$ induced by the representation $D$ of $H$.

After these preliminaries, we may now state the main results of \cite{Coleman:1969sm,Callan:1969sn}.
Firstly, 
any effective action which is constructed  from $P$, $F_Q$, $\Psi$
and their covariant derivatives $d_Q$,
and which is 
invariant under the 
subgroup $H$, is automatically invariant under the action 
\eqref{eq_G_left} of the full group $G$.
Secondly, any non-linear realization of $G$, with $H$ realized linearly, arises in this way.
If the effective action describes the IR dynamics of a system that exhibits spontaneous symmetry breaking of $G$ to $H$,
the scalars $\phi$ are identified with the massless Goldstone modes.

\subsection{Non-linearly realized non-Abelian 0-form symmetry}

We now discuss a sandwich construction for a non-linearly realized non-Abelian symmetry $G$. For ease of exposition, we treat first the case in which
the  subgroup $H \subset G$ of linearly realized
symmetries is trivial.
We discuss the most general case in the second part of this section.

\subsubsection{The case of trivial $H$}
\label{sec_H_trivial}

\paragraph{Bulk TQFT.}
For non-Abelian 0-form symmetries there is no obvious analog of the global symmetry $U(1)^{(d-p-2)}$ or the mixed anomaly in the Abelian discussion. 
For these reasons, 
in the SymTFT description,
we use the standard non-Abelian BF theory
\eqref{eq_BF}
as bulk TQFT, repeated here for convenience,
\be \label{eq_nonlin_nonAb_bulk}
S_{\rm bulk} = \frac{1}{2\pi} \int_{X^{d+1}} {\rm Tr}(B \wedge F_A) \ .
\ee 
We also repeat the bulk gauge transformations,
\be \label{eq_BF_gauge_yet_again}
A \mapsto g(d+A)g^{-1}  \ , \qquad 
B \mapsto g(B - d_A \tau) g^{-1} \ , 
\ee 
where the gauge parameter $g$ is a $G$-valued 0-form,
while $\tau$ is a $\mathfrak g$-valued $(d-2)$-form.

\paragraph{Topological boundary condition on $B^d_{\rm sym}$.}
We consider the action
\be   \label{eq_nonlinG_top}
S_{\rm sym}  = \frac{1}{2\pi} \int_{B^d_{\rm sym}} {\rm Tr} \bigg( 
(\cV^{-1} d\cV + \cV^{-1} \cA \cV -A) \wedge (B + d_A \beta)
+ \beta \wedge F_A
\bigg) \ . 
\ee 
The fields $\cV$ and $\beta$ are localized on $B^d_{\rm sym}$, with $\cV$ a $G$-valued 0-form
and $\beta$ a $\mathfrak g$-valued $(d-2)$-form.
(In Section \ref{sec_bulk_top_operators},
we have used the  symbol $\beta$ for the localized field on a $B$-defect.
It should be clear from context if we are referring to a $B$-defect or to the symmetry boundary $B^d_{\rm sym}$.)
Their gauge transformations are 
\be 
\cV \mapsto \cV g^{-1}  \ , \qquad 
\beta \mapsto g (\beta + \tau) g^{-1} \ . 
\ee 
The quantity $\cA$ is a non-dynamical $G$ gauge field,
which we take to be flat,
\be \label{eq_nonAb_nonline_cA_flat}
d\cA + \tfrac 12 [\cA,\cA] = 0\ .
\ee 
In our approach, $\cA$ does not transform under the gauge transformations of the dynamical fields.
The action
$S_{\rm bulk} + S_{\rm sym}$
is invariant under gauge transformations with arbitrary parameters $g$, $\tau$.
This is checked in Appendix \ref{app_nonAb_boundary_conditions}, where we also study the variation  of $S_{\rm bulk} + S_{\rm sym}$ under arbitrary variations of the fields.
The variational problem is well-posed and leads to 
\be \label{eq_nonlinear_nonAb_eom}
\text{bulk:} \quad 
F_A = 0 \ , \quad 
d_A B = 0 \ , \qquad 
\text{on $B^d_{\rm sym}:$} \quad 
A = \cV^{-1} d\cV  + \cV^{-1} \cA \cV\ .
\ee 
The bulk equations of motion are familiar.
We observe that the boundary relation
$A = \cV^{-1} d\cV + \cV^{-1} \cA \cV$ is compatible with $F_A = 0$ by virtue of the Maurer-Cartan equation
\be 
d(\cV^{-1} d\cV) + \tfrac 12 [\cV^{-1} d\cV,\cV^{-1} d\cV] = 0 \ ,
\ee 
and of flatness of $\cA$, \eqref{eq_nonAb_nonline_cA_flat}.

From the point of view of the topological Wilson line operators in the bulk, the boundary condition
\eqref{eq_nonlinG_top} acts as a Dirichlet boundary condition, in the sense that: 
\begin{itemize}
    \item closed bulk Wilson loops 
projected parallel onto $B^d_{\rm sym}$ give trivial operators;
\item open bulk Wilson lines can end perpendicularly on $B^d_{\rm sym}$.
\end{itemize}
This can be seen as follows.
We know that 
$A = \cV^{-1} d\cV + \cV^{-1} \cA \cV$
on $B^d_{\rm sym}$.
This means that $A$ is a gauge transform of the c-number $\cA$. It follows that a bulk Wilson loop, which is  invariant under gauge transformations,  becomes a c-number if projected parallel onto $B^d_{\rm sym}$.
Next, let us consider an  the open Wilson operator in the irrep $\mathbf R$ of $G$. It is a matrix-valued operator.
If the Wilson line connects points $P_0$ and $P_1$, its gauge transformation is 
\be 
\mathbf W_{\mathbf R}^{(P_1 \leftarrow P_0)}\mapsto D_\mathbf R(g(P_1)) \mathbf W_{\mathbf R}^{(P_1 \leftarrow P_0)}
D_\mathbf R(g(P_0))^{-1} \ . 
\ee 
The function $D_\mathbf R$ assigns to an abstract group element the matrix representing in the irrep~$\mathbf R$.
For definiteness, suppose $P_0$ is on $B^d_{\rm sym}$.
Then, we insert at $P_0$ the local operator
$D_{\mathbf R}(\cV^{-1})$ to compensate the gauge transformation at $P_0$,
\be 
\bigg( \mathbf W_{\mathbf R}^{(P_1 \leftarrow P_0)}
 D_{\mathbf R}(\cV^{-1}(P_0))
 \bigg)
\mapsto D_\mathbf R(g(P_1)) 
\bigg( \mathbf W_{\mathbf R}^{(P_1 \leftarrow P_0)}
 D_{\mathbf R}(\cV^{-1}(P_0)) 
 \bigg) \ . 
\ee 
As usual, the other endpoint of the Wilson line is either pushed to infinity (in the quiche picture), or ends on the other boundary (sandwich construction).

\paragraph{Physical boundary conditions on $B^d_{\rm phys}$.}
On the physical boundary we impose a non-topological boundary condition. To do so, we assume that $B^d_{\rm phys}$ is equipped with a metric.
We do not include any localized degrees of freedom on $B^d_{\rm phys}$. The action is written in terms of the bulk field $A$
restricted to $B^d_{\rm phys}$
and reads
\be \label{eq_nonAB_phys_boundary}
S_{\rm phys} = \frac{1}{2\pi} \int_{B^d_{\rm phys}}
\bigg[ - \frac 12 f^2 
{\rm Tr} (A  \wedge *_d A)
\bigg]  \ , 
\ee 
where $f$ is a real parameter.
The action $S_{\rm phys}$ is not invariant under  bulk gauge transformations of $A$. Hence, the bulk gauge parameter $g$ is restricted to be trivial on $B^d_{\rm phys}$.
In principle we could consider a more general action $S^{\rm phys}$,
including higher powers of $A$ and/or derivatives of $A$. In this section we study simply \eqref{eq_nonAB_phys_boundary}, and we comment on this issue in the next section for general $H$.

We observe that the variation of bulk action
$S_{\rm bulk}$ includes a boundary term 
\be \label{eq_bulk_variation_gives_boundary}
\delta S_{\rm bulk} \supset \frac{1}{2\pi} \int_{\partial X^{d+1}} {\rm Tr} (\delta A \wedge B) \ , 
\ee 
which originates from an integration by parts.
The variation of $S_{\rm phys}$ 
with respect to $A$
combines with this term. 
In total, we infer that 
\be 
\text{on $B^d_{\rm phys}$:} \qquad 
B + f^2 *_dA = 0 \ .
\ee

\paragraph{Closing the sandwich.}
If we collapse the interval direction,
the dynamical $G$-valued 0-form $\cV$, which was originally localized on $B^d_{\rm sym}$,
becomes a field in physical spacetime~$M^d$.
Its action is determined by \eqref{eq_nonAB_phys_boundary}, plugging in the expression for $A$ in terms of $\cV$ from~\eqref{eq_nonlinear_nonAb_eom},
\be 
S = \frac{1}{2\pi} \int_{M^d}   
- \frac 12 f^2 
{\rm Tr} \bigg[  \Big( \cV^{-1} d\cV + \cV^{-1} \cA \cV  \Big) \wedge *_d 
\Big( \cV^{-1} d\cV + \cV^{-1} \cA \cV  \Big)
\bigg]  \ . 
\ee 
This is the standard kinetic term for a $G$-valued scalar $\cV$, including the coupling to a background gauge field $\cA$.

\subsubsection{The case of non-trivial $H$}
\label{sec_H_nontrivial}

We repeat the analysis of the previous subsection in the case of non-trivial subgroup $H$.
We still have a bulk-boundary realization
in terms of a TQFT on the slab $X^{d+1}$,
the symmetry boundary $B^d_{\rm sym}$,
and the physical boundary $B^d_{\rm phys}$.

\paragraph{Bulk and symmetry boundary.}
These are exactly as in the case of trivial $H$,
see \eqref{eq_nonlin_nonAb_bulk} and \eqref{eq_nonlinG_top}. 
We already know that $S_\text{bulk}+ S_{\rm sym}$
is gauge-invariant and that it gives rise to a well-defined variational problem, with the result that on-shell
\be
\text{bulk $X^{d+1}$:} \quad 
F_A = 0 \ , \quad 
d_A B = 0 \ , \qquad 
\text{on $B^d_{\rm sym}:$} \quad 
A = \cV^{-1} d\cV  + \cV^{-1} \cA \cV \ .
\ee

\paragraph{Physical boundary.}
The action on $B^d_{\rm phys}$ is non-topological. 
We do not require it 
to be invariant under a generic bulk gauge transformation
\eqref{eq_BF_gauge_yet_again}
on  $X^{d+1}$.
We do require, however,
that $B^d_{\rm phys}$ be invariant under
transformations 
\eqref{eq_BF_gauge_yet_again} in which the gauge parameter $g$ is restricted to take values in $H \subset G$.

It is convenient to
think about the action on $B^d_{\rm phys}$ as the sum of  universal terms (whose form depend only on the choice of $G$ and $H \subset G$),
and non-universal terms (which depend on the specific theory with non-linearly realized $G$ symmetry).
We discuss them in turn.

The universal terms on 
$B^d_{\rm phys}$ are not written in terms of dynamical fields on $B^d_{\rm phys}$. Rather, they are written entirely in terms
of the restriction of the bulk field $A$ to 
$B^d_{\rm phys}$.
To discuss them, 
we    decompose the $\mathfrak g$-valued field $A$ into its $\mathfrak h$ and $\mathfrak h^\perp$ components,
\be \label{eq_A_Lie_decom}
A = a + \mathscr A  \ , \qquad 
a\in \mathfrak h \ , \qquad 
\mathscr A \in \mathfrak h^\perp \ . 
\ee 
Under a gauge transformation
\eqref{eq_BF_gauge_yet_again} in which the gauge parameter $g$  takes values in $H$,
 $a$ and $\mathscr A$ do not mix. Indeed, we have 
\be \label{eq_a_and_K_gauge}
\text{$H$-valued $g$:}  \qquad 
a \mapsto g (a+d)g^{-1} \ , \qquad 
\mathscr A \mapsto g \mathscr A g^{-1} \ . 
\ee 
We see that the component
$\mathscr A$ transforms homogeneously as a field in the adjoint representation of $H$, while the component $a$ 
transforms inhomogeneously
as an $H$ connection. 
We are led to define the quantities
\be 
d_a \mathscr A = d \mathscr A + [a, \mathscr A]  \ , \qquad 
f_a = da + \tfrac 12 [a,a] \ .
\ee 
They transform homogeneously
in the adjoint representation of $H$,
\be \label{eq_fa_and_daK_gauge}
\text{$H$-valued $g$:}
\qquad 
d_a \mathscr A \mapsto  
g  d_a \mathscr A g^{-1} \ , \qquad 
f_a \mapsto  
g  f_a g^{-1}
\  .
\ee 
With this notation,
the universal terms in the
action on $B^d_{\rm phys}$ are
constructed from a Lagrangian $\cL_{\rm univ}$,
\be \label{eq_withH_univ}
S_{\rm phys,univ} = \frac{1}{2\pi} \int_{B^d_{\rm phys}} \cL_{\rm univ}[\mathscr A , d_a \mathscr A , f_a] \ .
\ee 
Here $\cL_{\rm univ}$ is a local functional to $\mathscr A $, $d_a \mathscr A $, $f_a$ (and   higher $d_a$ covariant derivatives thereof) restricted from  $X^{d+1}$ onto $B^d_{\rm phys}$.
We also require that
$\cL_{\rm univ}$ be invariant under the $H$ transformations of $\mathscr A $, $d_a \mathscr A $, $f_a$, see \eqref{eq_a_and_K_gauge}, \eqref{eq_fa_and_daK_gauge}.
We regard $\cL_{\rm univ}$ as an effective
Lagrangian, organized as a derivative expansion.
To lowest order in derivatives we have the term
\be \label{eq_leading_goldstone}
\cL_{\rm univ}
= -\frac{f^2}{2} {\rm Tr} \Big( \mathscr A
 \wedge *_d
\mathscr A  \Big) \ ,
\ee  
where $*_d$ is the Hodge star operator
of the metric on $B^d_{\rm phys}$
and $f$ is a   constant.

Let us now turn to non-universal terms.
They can be written down if $B^d_{\rm phys}$ supports additional localized degrees of freedom, denoted collectively $\Psi$. We assume that the fields $\Psi$ transform  linearly under  $H$.
This allows us to define a standard covariant derivative $d_a\Phi$ using $a$ as connection, schematically
$d_a\Psi = d\Psi + D(a)\Psi$, if $D$ is the representation of $\mathfrak h$ induced by the representation of $H$ in which $\Psi$ transforms.
Then, we consider non-universal terms of the form
\be 
S_\text{phys,non-univ} = \frac{1}{2\pi} \int_{B^d_{\rm phys}} \cL_\text{non-univ}[\Psi;\mathscr A , d_a \mathscr A , f_a] \ .
\ee 
Here $\cL_\text{non-univ}$ is the most general local Lagrangian that is invariant under $H$ and is constructed with
$\Psi$, $\mathscr A$, $d_a \mathscr A$, $f_a$ (and   $d_a$ covariant derivatives thereof).
As before, $\cL_\text{non-univ}$ is understood as an effective action in a derivative expansion.

Let us now comment on the 
variational problem from
$S_\text{bulk} + S_{\rm phys}$,
with $S_{\rm phys}$ the total action
on $B^d_{\rm phys}$.
Recall from \eqref{eq_bulk_variation_gives_boundary}
that the variation
$\delta S_{\rm bulk}$ induces a
boundary term of the   form $\delta A B$.
To write this term
it is convenient to split the field $B$ into its $\mathfrak h$ and $\mathfrak h^\perp$ components,
\be 
B = b + \mathscr B \ , \qquad 
b \in \mathfrak h \ , \qquad 
\mathscr B \in 
\mathfrak h^\perp \ . 
\ee 
With this notation,
the term from $\delta S_{\rm bulk}$ on $B^d_{\rm phys}$ reads
\be 
\delta S_{\rm bulk} \supset  -\frac{1}{2\pi} \int_{I^d_{\rm int}} 
{\rm Tr}(\delta a \wedge b 
+ \delta \mathscr A \wedge \mathscr B) \ . 
\ee 
We have recalled that the decomposition $\mathfrak g = \mathfrak h \oplus \mathfrak h^\perp$ is orthogonal with respect to Tr.
On the other hand, we can parametrize the variation of $S_{\rm phys}$ with respect to $a$, $\mathscr A$ as
\be 
\delta S_{\rm phys} \supset \frac{1}{2\pi} \int_{B^d_{\rm phys}} {\rm Tr}\bigg[ 
\delta a \wedge  \frac{\delta S_{\rm phys}}{\delta a}
+ \delta \mathscr A \wedge  \frac{\delta S_{\rm phys}}{\delta \mathscr A}
\bigg]  \ . 
\ee 
We see that the variational problem for 
$S_\text{bulk} + S_{\rm phys}$ gives us the conditions
\be 
\text{on $B^d_{\rm phys}$:} \qquad 
b -  \frac{\delta S_{\rm phys}}{\delta a} = 0 \ , \qquad 
\mathscr B -  \frac{\delta S_{\rm phys}}{\delta \mathscr A} = 0 \ . 
\ee

\paragraph{Closing the sandwich.}

We start with some preliminary remarks.
We know that $A = \cV^{-1} d\cV + \cV^{-1} \cA \cV$ on $B^d_{\rm sym}$.
For any $G$-valued 0-form 
$\cV$, the 1-form $\cV^{-1} d\cV + \cV^{-1} \cA \cV$ can be decomposed into
its $\mathfrak h$ and $\mathfrak h^\perp$ pieces,
\be 
\cV^{-1} d\cV + \cV^{-1} \cA \cV = P + Q  \ , \qquad 
P \in \mathfrak h^\perp \ , \qquad 
Q \in \mathfrak h \ .
\ee 
This is the same as 
 \eqref{eq_P_and_Q_are_def}, 
with the inclusion of a background field $\cA$. 
We have a similar decomposition for $A$, see \eqref{eq_A_Lie_decom}. We conclude that
\be 
\text{on $B^d_{\rm sym}$:} \qquad 
Q = a  \ , \qquad 
P = \mathscr A  \ . 
\ee 

Let us now close the sandwich.
The resulting system living on physical spacetime $M^d$ is captured by the action
\be 
\frac{1}{2\pi} \int_{M^d} \bigg[ 
\cL_{\rm univ}[P , d_Q P , F_Q]
+ \cL_\text{non-univ}[\Psi;P , d_Q P, F_Q]
\bigg] \ . 
\ee 
This comes from the action on the physical 
boundary, in which the bulk gauge field components $a$, $\mathscr A$ are replaced 
by their expressions in terms of the scalar vielbein $\cV$.
Our bulk-boundary system reproduces the most general 
effective action with non-linearly realized symmetry $G$ of the 
Callan-Coleman-Wess-Zumino form
\cite{Coleman:1969sm,Callan:1969sn},
including the coupling to the background gauge field
$\cA$.

\subsection{Callan-Coleman-Wess-Zumino and a club sandwich construction}
\label{sec_club_sandwich}

In this subsection we consider the following problem.
Suppose we are given a $d$-dimensional QFT $\cT_d$
with a linearly realized global 0-form symmetry $H$.
Now, we fix a Lie group embedding $H \subset G$.
Our goal is to construct 
an extension $\cT'_d$
of the given QFT $\cT_d$, 
such that 
the full symmetry group $G$ acts
on $\cT'_d$. The $G$-action is non-linear, with the $H$ subgroup realized linearly. 
Moreover, $\cT'_d$ should be minimal, in the sense that
its degrees of freedom consist of those of $\cT_d$, plus 
the scalar vielbein $\cV$ 
that parametrizes the coset space $G/H$,
see review in Section \ref{sec_nonlinear_review}.

Our goal is to exhibit a  bulk-boundary description of 
the relationship between $\cT_d$ and $\cT'_d$.
For technical reasons, which should become clear in due course, we restrict ourselves to the case in which the subgroup $H$ is Abelian.

Let us start with the original theory $\cT_d$. It can be engineered by  a sandwich construction of the form
\be \label{eq_initial_H_sandwich}
\begin{tikzpicture}
\draw [LightBlue2, fill=LightBlue2] (0,0) -- (0,1.5) -- (3,1.5) -- (3,0) -- (0,0) ;
\draw [very thick] (0,0) -- (0,1.5) ; 
\draw [very thick] (3,0) -- (3,1.5) ; 
\node at (1.5,0.75) {$H$ Maxwell} ;
\node at (3.75,0.75) {$\cong$} ;
\draw [very thick] (4.5,0) -- (4.5,1.5) ; 
\node[below] at (0,0) {$B^d_{\rm Dir}$}; 
\node[below] at (3,0) {$B^d_{\rm phys}$};
\node[below] at (1.5,0.05) {$\widehat X^{d+1}$}; 
\node[below] at (4.5,0) {$\cT_d$}; 
\node at (5.5,0.75) {.} ;
\end{tikzpicture}
\ee 
In the spirit of the SymTh proposal \cite{Apruzzi:2024htg}, we  choose a bulk-boundary realization 
in which the bulk is not topological, but 
rather the free Maxwell theory in $(d+1)$ dimensions with Abelian gauge group $H$.
This choice is motivated by the fact that
we need a construction that can support non-flat $H$-connections in the slab, for reasons explained below.

Using \eqref{eq_initial_H_sandwich} as starting point, 
we propose a bulk-boundary construction of $\cT'_d$,
in the spirit of the ``club sandwich'' construction of \cite{Bhardwaj:2023bbf}.
The proposed realization of $\cT'_d$ is
\be \label{eq_new_H_sandwich}
\begin{tikzpicture}
\draw [thick,-stealth](-3.3,0.6) .. controls (-3.7,0.3) and (-3.7,1.1) .. (-3.3,0.9);
\node at (-4,0.75) {$G$};
\draw [DarkSeaGreen3, fill=DarkSeaGreen3] (-3,0) -- (-3,1.5) -- (0,1.5) -- (0,0) -- (-3,0) ;
\draw [very thick] (-3,0) -- (-3,1.5) ; 
\node at (-3+1.5,0.73) {$G$ BF-theory} ;
\node[below] at (-3,0) {$B^d_{\rm sym}$};
\draw [LightBlue2, fill=LightBlue2] (0,0) -- (0,1.5) -- (3,1.5) -- (3,0) -- (0,0) ;
\draw [very thick] (0,0) -- (0,1.5) ; 
\draw [very thick] (3,0) -- (3,1.5) ; 
\node at (1.5,0.75) {$H$ Maxwell} ;
\node at (3.75,0.75) {$\cong$} ;
\draw [very thick] (4.5,0) -- (4.5,1.5) ; 
\node[below] at (-1.5,0) {$X^{d+1}$}; 
\node[below] at (0,0) {$I^d_{\rm int}$}; 
\node[below] at (3,0) {$B^d_{\rm phys}$};
\node[below] at (1.5,0.05) {$\widehat X^{d+1}$}; 
\node[below] at (4.5,0) {$\cT'_d$}; 
\node at (5.5,0.75) {.} ;
\end{tikzpicture}
\ee 
We will describe all ingredients of the above diagram in detail below.
We  collect here some 
general comments on the construction.
\begin{itemize}
    \item The right-most slab ($H$ Maxwell) and $B^d_{\rm phys}$ are the same as in \eqref{eq_initial_H_sandwich}.
    \item The left-most slab supports a topological BF theory with gauge group $G$.
    \item The boundary condition at $B^d_{\rm sym}$ is topological and supports a $G$-valued 0-form $\cV$.  Together with 
    the left-most slab, it describes a ``quiche'' that captures a global 0-form symmetry $G$.
    \item The interface $I^d_{\rm int}$ is not topological. It does not support localized fields. The action on $I^d_{\rm int}$ is universal, in the sense its functional form only depends on the group-theory data $G$ and $H \subset G$.\footnote{The action on $I^d_{\rm int}$ is understood as an effective action in a derivative expansion, see below. Its form is determined by symmetry considerations, but the specific coefficients in the Lagrangian are model-dependent.}
\end{itemize}
We can imagine collapsing the left-most slab in \eqref{eq_new_H_sandwich}. If we do so, we obtain 
\be 
\begin{tikzpicture}
\draw [thick,-stealth](-.3,0.6) .. controls (-.7,0.3) and (-.7,1.1) .. (-.3,0.9);
\node at (-1,0.75) {$G$};
\draw [LightBlue2, fill=LightBlue2] (0,0) -- (0,1.5) -- (3,1.5) -- (3,0) -- (0,0) ;
\draw [very thick] (0,0) -- (0,1.5) ; 
\draw [very thick] (3,0) -- (3,1.5) ; 
\node at (1.5,0.75) {$H$ Maxwell} ;
\node at (3.75,0.75) {$\cong$} ;
\draw [very thick] (4.5,0) -- (4.5,1.5) ; 
\node[below] at (0,0) {$B^d_\text{non-lin}$}; 
\node[below] at (3,0) {$B^d_{\rm phys}$};
\node[below] at (1.5,0.05) {$\widehat X^{d+1}$}; 
\node[below] at (4.5,0) {$\cT_d'$}; 
\node at (5.5,0.75) {.} ;
\end{tikzpicture}
\ee 
Here $B^d_\text{non-lin}$ is a non-topological boundary condition. It originates from
the collapse of $(B^d_{\rm sym}|\text{$G$ BF on $X^{d+1}$}|I^d_{\rm int})$. The $G$-valued 0-form $\cV$, which originally lived on $B^d_{\rm sym}$, now lives on $B^d_\text{non-lin}$. It combines with the non-topological terms on $I^d_{\rm int}$.
As a result, $\cV$ is now interpreted as the scalar vielbein for the coset $G/H$.
The boundary $B^d_\text{non-lin}$ admits an action of the full group $G$,
which is non-linear, with $H$ realized linearly.

We now give a more detailed account of the elements of the  diagram \eqref{eq_new_H_sandwich}.
The analysis bears many technical similarities with Section \ref{sec_H_nontrivial}, so we will be brief.
Explicit expressions for the example
$G=SU(2)$, $H=U(1)$ are reported in Appendix
\ref{app_GH_example}.

\paragraph{First slab and symmetry boundary.}
The first (left-most) slab 
is described by the standard non-Abelian BF theory. The symmetry boundary $B^d_{\rm sym}$
is as in Sections \ref{sec_H_trivial}, \ref{sec_H_nontrivial}.
For convenience, we repeat them here,
\be 
\ba 
S_\text{slab 1} &= \frac{1}{2\pi} \int_{X^{d+1}} {\rm Tr}(B \wedge F_A) \ , \\ 
S_{\rm sym} &= \frac{1}{2\pi} \int_{B^d_{\rm sym}} {\rm Tr} \bigg( 
(\cV^{-1} d\cV + \cV^{-1} \cA \cV -A) \wedge (B + d_A \beta)
+ \beta \wedge F_A
\bigg)  \ . 
\ea 
\ee 
The total action $S_\text{slab 1}+ S_{\rm sym}$
is gauge invariant and its equations of motion imply
\be
\text{first slab $X^{d+1}$:} \quad 
F_A = 0 \ , \quad 
d_A B = 0 \ , \qquad 
\text{on $B^d_{\rm sym}:$} \quad 
A = \cV^{-1} d\cV + \cV^{-1} \cA \cV\ .
\ee

\paragraph{Interface between the slabs.}
The action on $I^d_{\rm int}$ is non-topological. We assume that 
$I^d_{\rm int}$ is equipped with a metric.
We do not include any localized fields on $I^d_{\rm int}$. The action is written only in terms of the bulk field $A$ of the first slab $X^{d+1}$.
To write the action
we 
use the same notation as in Section \ref{sec_H_nontrivial}.
We decompose $A$ as
\be \label{eq_A_Lie_decom2}
A = a + \mathscr A  \ , \qquad 
a\in \mathfrak h \ , \qquad 
\mathscr A \in \mathfrak h^\perp \ . 
\ee 
We do not require the action on $I^d_{\rm int}$ 
to be invariant under a generic bulk gauge transformation
\eqref{eq_BF_gauge_yet_again}
on the first slab $X^{d+1}$.
We do require, however,
that $I^d_{\rm int}$ be invariant under
transformations 
\eqref{eq_BF_gauge_yet_again} in which the gauge parameter $g$ is restricted to take values in $H \subset G$.
As we have already discussed in Section \ref{sec_H_nontrivial},
$a$ and $\mathscr A$ are not mixed by these transformations. $\mathscr A$ transforms linearly in a representation of $H$,
while $a$ transforms as an $H$-connection.
As in Section \ref{sec_H_nontrivial},
we make use of the quantities 
\be 
d_a \mathscr A = d \mathscr A + [a, \mathscr A]  \ , \qquad 
f_a = da   \ .
\ee 
In the first equation, the commutator is in $\mathfrak g$. We notice the absence of the commutator term in $f_a$, due to the assumption of Abelian $H$.

The action on the interface is completely analogous to \eqref{eq_withH_univ},
\be 
S_{\rm int} = \frac{1}{2\pi} \int_{I^d_{\rm int}} \cL_{\rm univ}[\mathscr A , d_a \mathscr A , f_a] \ .
\ee 
The Lagrangian $\cL_{\rm univ}$ is a  generic $H$-invariant functional of
$\mathscr A , d_a \mathscr A , f_a$
(and their $d_a$ derivatives), understood as a derivative expansion. The leading term is as in \eqref{eq_leading_goldstone}.

The field $B$ is also decomposed 
in its $\mathfrak h$ and $\mathfrak h^\perp$ parts,
\be 
B = b + \mathscr B \ , \qquad 
b \in \mathfrak h \ , \qquad 
\mathscr B \in 
\mathfrak h^\perp \ . 
\ee 
The variation of $S_\text{slab 1} + S_{\rm int}$
contains terms on $I^d_{\rm int}$ of the form
\be \label{eq_variation_slab1_and_int}
\ba 
\delta S_\text{slab 1} + \delta S_{\rm int}
&= \frac{1}{2\pi} \int_{I^d_{\rm int}} {\rm Tr} \bigg[ 
- \delta a \wedge b - \delta \mathscr A \wedge \mathscr B 
 + \delta a \wedge  \frac{\delta S_{\rm int}}{\delta a}
+ \delta \mathscr A \wedge  \frac{\delta S_{\rm int}}{\delta \mathscr A}
\bigg]  \ , 
\ea 
\ee 
where the first two terms 
originate from the variation of the bulk action, see \eqref{eq_bulk_variation_gives_boundary}, while the last two terms  originate from the interface action.
To deal with the terms $\delta \mathscr A$,
we impose
\be 
\text{on $I^d_{\rm int}$:} \quad 
\mathscr B +  \frac{\delta S_{\rm phys}}{\delta \mathscr A} = 0 \ .
\ee 
This is as in Section \ref{sec_H_nontrivial}.
In contrast, we now treat differently 
the  terms with $\delta a$, exploiting the
presence of the second slab in the club sandwich construction.

\paragraph{Action in the second slab.} 
 Our proposed action for the second (right-most) slab
$\widehat X^{d+1}$ in 
\eqref{eq_new_H_sandwich} 
is Maxwell theory with gauge group $H$.
We denote the $H$ gauge field as $\widehat A$.
If $H \cong U(1)^r$, the action is of the form 
\be 
S_\text{slab 2} =\frac{1}{2\pi} \int_{\widehat X^{d+1}} \bigg[ 
- \frac{1}{2} \tau_{ij} \widehat F^i \wedge *_{d+1} \widehat F^j 
\bigg] \ , 
\ee 
where $i,j=1,\dots,r$, $\widehat F^i = d \widehat A^i$ are the components of the 
$H \cong U(1)^r$ field strength,
and $\tau_{ij}$ are the gauge coupling constants.

\paragraph{Matching conditions
across $I^d_{\rm int}$.}
Let us now consider the variational problem for
$S_\text{slab 1} + S_{\rm int} + S_\text{slab 2}$.
The $\delta a$ terms in \eqref{eq_variation_slab1_and_int}
are yet uncanceled. We  write them as
\be \label{eq_uncanceled}
- \frac{1}{2\pi} \int_{I^d_{\rm int}} C_{ij} \delta a^i \wedge \widetilde b^j \ . 
\ee 
Some comments on our notation are in order.
We have introduced the quantity
\be 
\widetilde b := b -  \frac{\delta S_{\rm int}}{\delta a} \ . 
\ee 
We have also chosen a basis $T_i$ of $\mathfrak h$ and used 
\be 
a = a^i T_i \ , \qquad 
\widetilde b = \widetilde b^i T_i  \ , \qquad 
{\rm Tr}(T_i T_j) =: C_{ij} \ . 
\ee 
The variation of the Maxwell action in the second slab is another source of terms on $I^d_{\rm int}$, via integration by parts
(with a relative minus sign due to orientation),
\be \label{eq_boundary_from_second_slab}
 \frac{1}{2\pi} \int_{I^d_{\rm int}} \tau_{ij} \delta \widehat A^i \wedge \widetilde b^j \ . 
\ee 
Comparison of \eqref{eq_uncanceled} and \eqref{eq_boundary_from_second_slab} suggest to impose the following matching conditions on the interface,
\be \label{eq_matching_1}
\text{on $I^d_{\rm int}$:} \qquad 
a^i = \widehat A^i \ , \qquad 
C_{ij} \widetilde b^j = \tau_{ij} *_{d+1} \widehat F^j \ . 
\ee 
In these relations, the LHSs are bulk fields on the first slab restricted to $I^d_{\rm int}$,
and the RHSs are bulk fields on the second slab restricted to $I^d_{\rm int}$.
Due to the gluing condition \eqref{eq_matching_1}, we also have a match the gauge parameters.
On the side of the first slab, we have $g$,
which was originally $G$-valued, but then restricted to be valued in $H \subset G$.
On the side of the second slab, we have
the standard Abelian gauge transformations of Maxwell theory with gauge group $H$.
The gauge parameters from the two sides must agree on the interface.

A simple corollary of the gluing condition
\eqref{eq_matching_1} is 
\be 
\text{on $I^d_{\rm int}$:} \qquad
f_a = \widehat F^i T_i \ . 
\ee 
As explained in the next paragraph, $f_a$ is non-zero. Hence, on the second slab we must accommodate for non-flat $H$-gauge field $\widehat A^i$. This is the reason why we have selected Maxwell theory as filling for the second slab of the sandwich.

\paragraph{Comment on non-vanishing of $f_a$.}
In the first slab, we know that $F_A = 0$. In terms of the decomposition \eqref{eq_A_Lie_decom} of $A$ into $a$ and $\mathscr A$,
this condition reads
\be \label{eq_flatness_after_decom}
d_a \mathscr A + \tfrac 12 [\mathscr A , \mathscr A] + f_a = 0 \ .
\ee 
This is a $\mathfrak g$-valued equation. While $d_a \mathscr A$ takes values in $\mathfrak h^\perp$ and $f_a$ takes values in $\mathfrak h$, in general the term $[\mathscr A, \mathscr A]$ has components both in $\mathfrak h$ and $\mathfrak h^\perp$. Thus, if we separate the $\mathfrak h$ and $\mathfrak h^\perp$ components of \eqref{eq_flatness_after_decom} we get
\be \label{eq_total_flatness_pieces}
d_a \mathscr A + \tfrac 12 {\rm proj}_{\mathfrak h^\perp} [\mathscr A, \mathscr A] = 0 \ , \qquad 
f_a   + \tfrac 12 {\rm proj}_{\mathfrak h} [\mathscr A, \mathscr A] = 0 \ .
\ee 
In particular, we see that
$f_a$ is not set to zero
by the dynamics in the first slab. 
Demanding $f_a = 0$ by hand would impose a condition on $\mathscr A$ that is too restrictive. (This is exemplified for $G=SU(2)$, $H=U(1)$ in Appendix \ref{app_GH_example}.)

\paragraph{Physical boundary.}
The physical boundary $B^d_{\rm phys}$ can support any QFT with a global $H$ symmetry, linearly realized. This corresponds to the fields $\Psi$ in the review of Section \ref{sec_nonlinear_review}.
Since we assume that the theory on $B^d_{\rm phys}$ has an $H$ symmetry, we can couple it to a background $H$ connection.
This gives us a prescription to couple $B^d_{\rm phys}$ to the second slab:
we take the $H$ gauge field
$\widehat A$ on the second slab, we restrict it to the physical boundary $B^d_{\rm sym}$, and we regard it as a background $H$-connection.
Let us write schematically
\be 
S_{\rm phys} = \frac{1}{2\pi} \int_{B^d_{\rm phys}} \cL_{\rm phys}[\Psi ; \widehat A] \ . 
\ee

\paragraph{Closing the sandwich.}

We know that $A = \cV^{-1} d\cV + \cV^{-1} \cA \cV$ on $B^d_{\rm sym}$.
As above, we write
\be 
\cV^{-1} d\cV + \cV^{-1} \cA \cV= P + Q  \ , \qquad 
P \in \mathfrak h^\perp \ , \qquad 
Q \in \mathfrak h \ .
\ee 
As a result, we have 
\be 
\text{on $B^d_{\rm sym}$:} \qquad 
Q = a  \ , \qquad 
P = \mathscr A  \ . 
\ee 
Let us now collapse the first
slab of the sandwich.
We get a new boundary for the second slab,
which we denote $B^d_\text{non-lin}$.
The action that describes it is 
\be 
S  = \frac{1}{2\pi} \int_{B^d_\text{non-lin}} \cL_{\rm univ}[P , d_Q P , F_Q] \ .
\ee 
This is the interface action,
in which we plug in $P$ and $Q$ constructed from the field $\cV$ that was originally living on $B^d_{\rm sym}$.
The fields $P$, $Q$ now live
on $B^d_\text{non-lin}$. They are matched to the bulk fields of the second slab,
in such a way that
\be \label{eq_corollary_matching}
Q = \widehat A^i T_i \ , \qquad \text{hence} \quad 
F_Q = \widehat F^i T_i  \ ,
\ee 
which follow from \eqref{eq_matching_1}.

The name $B^d_\text{non-lin}$
is motivated by the fact that this is a boundary for $H$ Maxwell theory, but it encodes a non-linear realization of the full group $G$, thanks to the scalar $\cV$ and the 1-forms $P$, $Q$ constructed out of $\cV$.

Finally, let us collapse the second slab. 
We get a $d$ dimensional system formulated on the physical spacetime $M^d$.
The total action that describes the resulting system is 
\be 
S  = \frac{1}{2\pi} \int_{M^d}
\bigg[ 
\cL_{\rm univ}[P , d_Q P , F_Q]
+ \cL_{\rm phys}[\Psi ;  Q]
\bigg] \ .
\ee 
Here we have used the fact that 
$\widehat A$ collapses to $Q$
because of \eqref{eq_corollary_matching}.
We have constructed an
effective action with non-linearly realized symmetry $G$ of the 
Callan-Coleman-Wess-Zumino form
\cite{Coleman:1969sm,Callan:1969sn}.

\paragraph{Comments on non-Abelian $H$.}
We close this section with some speculative remarks on how to repeat the above construction if the subgroup $H$ is non-Abelian.
The point that requires care is the choice of a suitable theory to fill the second slab.
Crucially, the theory must be able to support non-flat connections.
A first candidate could be Yang-Mills theory with group $H$. Yet, this is not a free theory,
and its use in a sandwich construction
appears to be subtle and to depend in crucial ways on the dimension~$d$.
Another approach, which might be better-behaved, is to use 
a semi-Abelian gauge theory in place of Yang-Mills theory. More precisely, the semi-Abelian gauge theory of interest has gauge group
$U(1)^{{\rm rank}(H)} \rtimes \cW_{\mathfrak h}$, with ${\rm rank}(H)$ the rank of $H$ and
$\cW_{\mathfrak h}$ the Weyl group of $\mathfrak h$. This semi-Abelian theory
has a rich spectrum of topological operators
with non-trivial fusion rules \cite{Antinucci:2022eat}.
Its potential application to a club sandwich construction lies beyond the scope of this work and is left for the future.

\subsection{Non-linearly realized Abelian 2-group  symmetry}

We proceed now with applications of the bulk BF TQFT towards non-linearly realized generalized symmetries organized according to a higher group. While our methods apply generally, for definiteness in this section we focus on a four-dimensional QFT with a $U(1)$ 0-form symmetry and a $U(1)$ 1-form symmetry
participating in a 2-group. In the following section we will consider the case of a 2-group with a non-Abelian $0$-form symmetry.

\paragraph{Bulk TQFT.}
The SymTFT that captures this symmetry structure is \cite{Antinucci:2024zjp}
\be \label{eq:bulk2group}
S_{\rm bulk} = \frac{1}{2\pi} \int_{X^5} \bigg[ 
B_3 \wedge dA_1 + B_2 \wedge dA_2 + \kappa B_2 \wedge A_1 \wedge dA_1
\bigg] \ , 
\ee 
where $\kappa = k/(2\pi)$, $k$ integer.
The fields $B_3$, $B_2$ are $\mathbb R$ gauge fields,
while $A_1$, $A_2$ are $U(1)$ gauge fields.
The gauge transformations of the bulk fields read
\be \label{eq:2groupgaugetransf}
\ba 
A_1 & \mapsto A_1 + d\Lambda_0 \ , \\ 
A_2 & \mapsto A_2 + d\Lambda_1  + \kappa d\Lambda_0 \wedge A_1 \ , \\ 
B_2 & \mapsto B_2 + d\Omega_1 \ , \\ 
B_3 & \mapsto B_3 + d\Omega_2 - \kappa d\Omega_1 \wedge A_1
\ , 
\ea 
\ee 
where the gauge parameters $\Lambda_0$, $\Lambda_1$
are $U(1)$ gauge fields, while $\Omega_1$, $\Omega_2$
are globally defined forms.
Under the above gauge transformation, the bulk action varies into a boundary term,
\be 
\Delta S_{\rm bulk} = \frac{1}{2\pi} \int_{\partial X^5}
\bigg[ 
\Omega_1 \wedge dA_2 + \Omega_2 \wedge dA_1
\bigg]  \ . 
\ee 
If we consider a variation of the bulk action under arbitrary variations of the fields, we infer the following bulk equations of motion,
\be 
dA_1 = 0 \ , \qquad 
dA_2 + \kappa A_1 \wedge dA_1= 0 \ , \qquad 
dB_2 = 0 \ , \qquad 
dB_3 = 0 \ . 
\ee 
Integration by parts generate the following boundary terms,
\be \label{eq_Ab2group_bulk_varies_into_boundary}
\delta S_{\rm bulk} \supset \frac{1}{2\pi} \int_{\partial X^5} \bigg[
\delta A_1 \wedge B_3 + \delta A_2 \wedge B_2
+ \kappa \delta A_1 \wedge B_2 \wedge A_1
\bigg]  \ . 
\ee 
It is also useful to observe that the sum over the fluxes of $A_2$ in the bulk action imposes that the closed form $B_2$ has integral periods.
As a result, we can write it on-shell as the field strength of a $U(1)$ gauge field $v_1$,
\be \label{eq_Ab2group_B_onshell}
\text{on-shell:} \qquad  B_2 = dv_1 \ . 
\ee 

\paragraph{Topological boundary conditions on $B^4_{\rm sym}$.} We consider the following   action,
\be \label{eq_Ab2group_sym}
\ba
S_{\rm sym} &= \frac{1}{2\pi} \int_{B^4_{\rm sym}} \bigg[ 
- B_2 \wedge (A_2 - d\varphi_1 - \cA_2)
+ B_3 \wedge (A_1 - d\varphi_0 - \cA_1)
\\
&
- \kappa B_2 \wedge A_1 \wedge (d\varphi_0 + \cA_1)
+  r \kappa B_2 \wedge  \cA_1 \wedge  d \varphi_0 
\bigg]  \ . 
\ea
\ee 
The quantities $\varphi_0$, $\varphi_1$ are dynamical
$U(1)$ gauge fields on $B^4_{\rm sym}$.
The quantities $\cA_1$, $\cA_2$ are non-dynamical
$U(1)$ gauge fields on $B^4_{\rm sym}$, which we take to be flat,
\be 
d\cA_1 = 0 \ , \qquad 
d\cA_2 = 0 \ . 
\ee 
The parameter $r$ is a constant, which will be determined below. The gauge transformations of $\varphi_0$, $\varphi_1$ read
\be 
\varphi_0 \mapsto \varphi_0 + \Lambda_0 \ , \qquad 
\varphi_1 \mapsto \varphi_1 + d\sigma_0 
+ \Lambda_1  + 
\kappa \Lambda_0 (d\varphi_0 + \cA_1)
+ r \kappa \cA_1 \Lambda_0 \ . 
\ee 
We have verified that the combined $S_{\rm bulk} + S_{\rm sym}$ is gauge invariant, up to total derivatives.
This holds for any value of $r$.

Next, let us consider the variational problem
from $S_{\rm bulk} + S_{\rm sym}$.
We keep  \eqref{eq_Ab2group_bulk_varies_into_boundary}
into account.
The terms with $\delta B_2$, $\delta B_3$ impose the relations
\be \label{eq_Ab2group_sym_boundary_EOMs}
\ba 
&\text{on $B^4_{\rm sym}$:} & 
A_1 &= d\varphi_0 + \cA_1 \ ,  & 
A_2 & = d\varphi_1 + \cA_2 - r \kappa d\varphi_0 \wedge \cA_1 \ .
\ea 
\ee 
All other variations do not 
give any new information and/or confirm the bulk equations of motion. Let us consider, however,
the sum over fluxes $d\varphi_0$, $d\varphi_1$.
These impose 
\be 
\int_{\Sigma^2}B_2 \in 2\pi \mathbb Z \ , \qquad 
\int_{\Sigma^3}( B_3 + \kappa B_2 \wedge A_1 - r \kappa B_2 \wedge \cA_1)\in 2\pi \mathbb Z  \ , 
\ee 
where $\Sigma^1$, $\Sigma^3$ are cycles in $B^4_{\rm sym}$.
The first relation is already true on-shell in the bulk.
To analyze the second relation, we plug the expression for
$A_1$ in the integrand, 
\be 
B_3 + \kappa B_2 \wedge A_1 - r \kappa B_2 \wedge \cA_1
= B_3 + \kappa B_2 \wedge d\varphi_0 +(1- r) \kappa B_2 \wedge \cA_1 \ .
\ee 
We already know that $B_2$ and $d\varphi_0$
have periods in $2\pi \mathbb Z$. We want to get a relation for the field $B_3$ without constraining $\cA_1$. This
selects the value of the parameter $r$ to be
\be 
r = 1 \ .
\ee

\paragraph{Physical boundary condition on $B^4_{\rm phys}$.}
We consider the following action,
\be \label{eq_Ab2group_phys}
S_{\rm phys} = \frac{1}{2\pi} \int_{B^4_{\rm phys}}
\bigg[ 
- \frac 12 f^2 A_1 \wedge *_4 A_1 
- \frac 12 g^2 B_2 \wedge *_4 B_2 
+  A_2 \wedge B_2
\bigg] \ . 
\ee 
To motivate the last term,
we consider the total variation of
$S_{\rm bulk} + S_{\rm phys}$.
The terms \eqref{eq_Ab2group_bulk_varies_into_boundary} enter with a minus sign, due to orientation. In total, we find 
\be 
\frac{1}{2\pi} \int_{B^4_{\rm phys}} \bigg[ 
- \delta A_1 \wedge \Big(f^2 *_4 A_1 + B_3  + \kappa B_2 \wedge A_1 \Big)
-  \delta B_2 \wedge \Big( g^2 *_4 B_2  -A_2 \Big)
\bigg] \ , 
\ee 
while the terms with $\delta A_2$ cancel.
We have a well-posed variational problem,
which gives the relations
\be
\text{on $B^4_{\rm phys}$:} \qquad 
f^2 *_4 A_1 + B_3  + \kappa B_2 \wedge A_1 =0  \ , \qquad 
g^2 *_4 B_2  -A_2  =0   \ . 
\ee 

\paragraph{Closing the sandwich.}
We want to determine the total action on physical spacetime
$M^4$ upon closing the sandwich. 
Potential contributions originate from the  actions \eqref{eq_Ab2group_sym} and \eqref{eq_Ab2group_phys}. We evaluate them using
\eqref{eq_Ab2group_sym_boundary_EOMs}.
The action \eqref{eq_Ab2group_sym} on the symmetry boundary gives a vanishing contribution. From the action \eqref{eq_Ab2group_phys} on the physical boundary we obtain instead 
\be 
\ba 
S &= \frac{1}{2\pi} \int_{M^4} \bigg[ 
- \frac 12 f^2 (d\varphi_0 + \cA_1) \wedge *_4 (d\varphi_0 + \cA_1)
- \frac 12 g^2 B_2 \wedge *B_2
\\
&
+ (d\varphi_1 + \cA_2 - \kappa d\varphi_0 \wedge \cA_1)\wedge B_2
\bigg]  \ .
\ea 
\ee 
We are effectively integrating out to bulk, so we put the bulk field $B_2$ on-shell and use \eqref{eq_Ab2group_B_onshell}.
The topological terms on the second line become
\be 
\frac{1}{2\pi} \int_{M^4} \bigg[ 
d\varphi_1 \wedge dv_1 + \cA_2 \wedge dv_1 
- \kappa d\varphi_0 \wedge \cA_1 \wedge dv_1 
\bigg] 
\ . 
\ee 
The first term can be dropped, because it has an integer coefficient and both $d\varphi_1$ and $dv_1$ have periods in $2\pi \mathbb Z$. (The term with $\cA_2$ is not a total derivative because $v_1$ is not a globally defined form.)
In total, we arrive at the action
\be \label{eq_Goldstone_Maxwell_action}
\ba 
S &= \frac{1}{2\pi} \int_{M^4} \bigg[ 
- \frac 12 f^2 (d\varphi_0 + \cA_1) \wedge *_4 (d\varphi_0 + \cA_1)
- \frac 12 g^2 dv_1 \wedge * dv_1
\\
&
+ \cA_2 \wedge dv_1 
- \kappa d\varphi_0 \wedge \cA_1 \wedge dv_1
\bigg]  \ .
\ea 
\ee 
This describes the compact scalar $\varphi_0$ coupled
to the 0-form symmetry background field $\cA_1$,
as well as a photon $v_1$ whose magnetic 1-form symmetry
is coupled to $\cA_2$. The non-trivial 2-group structure is encoded in the last term, which can be regarded as a non-standard coupling between $\varphi_0$ and $v_1$ that is turned on in the presence of the background $\cA_1$.
The action \eqref{eq_Goldstone_Maxwell_action} was studied in \cite{Cordova:2018cvg} in reference to spontaneously broken continuous 2-group symmetry. It was derived in a holographic context in \cite{Antinucci:2024bcm}.

\subsection{Non-linearly realized non-Abelian 2-group symmetry}

In this section we study a four-dimensional QFT that enjoys a non-Abelian 0-form symmetry and a $U(1)$ 1-form symmetry that participate in a 2-group.
For definiteness, we consider the case $G = SU(N)$.

\paragraph{Bulk TQFT.}
The SymTFT that captures the non-Abelian 2-group
was derived in \cite{Antinucci:2024bcm} using the methods of \cite{Antinucci:2024zjp}. The bulk action reads
\be \label{eq_nonAb_2gr_bulk}
S_{\rm bulk} = \frac{1}{2\pi} \int_{X^5} \bigg[ 
b_2 \wedge da_2 
+ {\rm Tr}(B \wedge F_A)
+ \tfrac 12 \kappa b_2 \wedge {\rm Tr}(A\wedge dA + \tfrac 23 A^3)
\bigg]  \ ,
\ee 
where $\kappa = k/(2\pi)$, $k$ integer.
The dynamical fields in the bulk are:
a $U(1)$ gauge field $a_2$;
an $\mathbb R$ gauge field $b_2$;
a $G$-connection $A$;
a $\mathfrak g$-valued 3-form $B$.
The gauge transformation of $A$ is standard,
\be 
A \mapsto g(A+d) g^{-1} \ ,
\ee 
with $g$ a $G$-valued 0-form.
This implies 
\be 
{\rm Tr}(AdA + \tfrac 23 A^3) \mapsto 
{\rm Tr}(AdA + \tfrac 23 A^3) + d\Big[ 
\Theta_2  - {\rm Tr}(A g^{-1} dg)
\Big] \ . 
\ee 
The quantity $\Theta_2$ is a locally-defined 2-form with the property
\be 
d\Theta_2 = \tfrac 13 {\rm Tr}\Big[ (g^{-1} dg)^3  \Big]\ .
\ee 
(It exists because 
$d(g^{-1} dg) = -(g^{-1} dg)^2$
implies
$d{\rm Tr}[(g^{-1} dg)^3]=0$.)
With this notation, the gauge transformations of the remaining bulk fields are  
\be 
\ba 
b_2 & \mapsto b_2 + d\Omega_1 \ , \\ 
a_2 & \mapsto a_2 + d\Lambda_1 - \tfrac 12 \kappa 
\Big[ 
\Theta_2  - {\rm Tr}(A g^{-1} dg)
\Big] \ ,\\
B & \mapsto g \Big( B - d_A \tau 
- \tfrac 12 \kappa \Omega_1 \wedge F_A\Big)g^{-1} \ . 
\ea 
\ee 
The gauge parameters are:  a globally defined form $\Omega_1$;
 a $U(1)$ gauge field $\Lambda_1$;
  a $\mathfrak g$-valued 2-form $\tau$.
The gauge variation of the bulk action yields
a boundary term,
\be 
\Delta S_{\rm bulk} = \frac{1}{2\pi} \int_{\partial X^5} \bigg[ 
\Omega_1 da_2
- {\rm Tr}(\tau F_A)
+ \tfrac 12 \kappa \Omega_1 {\rm Tr}(AdA + \tfrac 23 A^3)
\bigg]  \ .
\ee 
The bulk equations of motion read
\be 
db_2 =0 \ , \qquad 
F_A = 0 \ , \qquad 
d_A B = 0 \ , \qquad 
da_2 + \tfrac 12 \kappa {\rm Tr}(A\wedge dA + \tfrac 23 A^3) = 0 \ .
\ee 
The variation of the bulk action generates also boundary terms,
\be \label{eq_nonAb_2gr_delta_bulk_gives}
\delta S_{\rm bulk} \supset \frac{1}{2\pi} \int_{\partial X^5} \bigg[ 
b_2 \delta a_2
+ {\rm Tr}(\delta A B)
- \tfrac 12\kappa b_2{\rm Tr}( A \delta A)
\bigg] \ .
\ee 
We also notice that, due to the sum over the fluxes of $da_2$ in the path integral, on-shell $b_2$ has periods in $2\pi \mathbb Z$. Hence, we can write $b_2$ as field strength of a $U(1)$ gauge field $v_1$,
\be \label{eq_nonAb_2gr_b_onshell}
\text{on-shell:} \qquad b_2 = dv_1 \ .
\ee 

\paragraph{Topological boundary conditions on $B^4_{\rm sym}$.}
We propose the following  action,
\be \label{eq_nonAb_2gr_sym_action}
\ba 
S_{\rm sym} & = \int_{B^4_{\rm sym}} \bigg\{
-b_2 \wedge (a_2 - \cA_2 - d \varphi_1)
 + {\rm Tr} \Big[ 
(\omega -A)\wedge (B + d_A \beta) + \beta \wedge F_A
\Big]
\\
&  - \tfrac 12 \kappa b_2\wedge  {\rm Tr}(A \omega)
- \tfrac 12  \kappa  b_2 \wedge {\rm Tr}( \cA d\cV \cV^{-1} )
+ \tfrac 12 \kappa  b_2 \wedge \Psi_2
\bigg\} \ . 
\ea 
\ee 
Some remarks on our notation are in order.
The dynamical fields on the boundary are:
a $G$-valued 0-form $\cV$;
a $\mathfrak g$-valued 2-form $\beta$;
a $U(1)$ gauge field $\varphi_1$.
We also have a non-dynamical $U(1)$ gauge field
$\cA_2$ and a non-dynamical $G$ gauge field $\cA$,
which are both flat,
\be 
d\cA_2 = 0 \ ,\qquad 
d\cA + \tfrac 12 [\cA, \cA] = 0 \ . 
\ee 
In writing \eqref{eq_nonAb_2gr_sym_action} we have adopted the shorthand notation
\be 
\omega := \cV^{-1} d\cV + \cV^{-1} \cA \cV \ . 
\ee 
This quantity satisfies the relation $d\omega = - \omega^2$ thanks to the flatness of $\cA$. 
We have also introduced the locally 
 defined 2-form $\Psi_2$ with the property
\be \label{eq_Psi2_def}
d\Psi_2 = \tfrac 13 {\rm Tr}\Big[(\cV^{-1} d\cV)^3\Big] \ .
\ee 
The gauge transformations of the dynamical boundary fields are
\be 
\cV \mapsto \cV g^{-1}  \ , \qquad 
\varphi_1 \mapsto \varphi_1 + d\sigma_0 + \Lambda_1 \ ,
\ee 
with $g$ and $\Lambda_1$ as in the bulk gauge transformations, and $\sigma_0$ a compact scalar on the boundary.
By studying the gauge transformation of the quantity ${\rm Tr}[(\cV^{-1} d \cV)^3]$, we can infer the gauge transformation of $\Psi_2$,
\be 
\ba 
\tfrac 13 {\rm Tr}[(\cV^{-1} d \cV)^3] &\rightarrow 
\tfrac 13 {\rm Tr}[(\cV^{-1} d \cV)^3]
- \tfrac 13 {\rm Tr} [(g^{-1} dg)^3]   
+   d {\rm Tr}(\cV^{-1} d \cV g^{-1} dg) \ , 
\\
\Psi_2 &\mapsto \Psi_2 -  \Theta_2 +  {\rm Tr}(
\cV^{-1} d \cV 
g^{-1} dg) \ .
\ea 
\ee 
(We are implicitly making a choice for $\Psi_2$, which is only defined modulo addition of closed forms.)
We have verified that the combined action
$S_{\rm bulk} + S_{\rm sym}$ is gauge invariant, 
provided that we restrict the gauge parameter $\Omega_1$ to be trivial on $B^4_{\rm sym}$.
This restriction still allows us to capture   the interesting symmetry structures related to the 2-group.

Next, let us consider the variation of the boundary action. 
We also have to take into account the terms originating from the bulk variation, see \eqref{eq_nonAb_2gr_delta_bulk_gives}.
The terms in the total variation with $\delta B$
and $\delta b_2$ impose the following relations,
\be \label{eq_nonAb_2gr_sym_eom}
\ba 
&\text{on $B^4_{\rm sym}$:} & 
A & = \cV^{-1} d\cV + \cV^{-1} \cA \cV \ , \\ 
&& a_2 & = d\varphi_1 + \cA_2 
- \tfrac 12 \kappa  {\rm Tr}(\cA d\cV \cV^{-1})
+ \tfrac 12  \kappa  \Psi_2 \ . 
\ea 
\ee 
The other variations do not give rise to any additional constraint.

\paragraph{Physical boundary condition on $B^4_{\rm phys}$.} We consider the non-topological action
\be \label{eq_nonAb_2gr_phys}
S_{\rm phys} = \frac{1}{2\pi} \int_{B^4_{\rm phys}} 
\bigg[ 
- \frac 12 f^2 {\rm Tr}(A \wedge *_4 A)
- \frac 12 g^2 b_2 \wedge *_4 b_2
+ a_2 \wedge b_2
\bigg]  \ .
\ee
The last term is motivated by the need to
have a cancellation of $\delta a_2 b_2$ terms in $\delta S_{\rm phys}$ against the similar term that
originates from the bulk, see \eqref{eq_nonAb_2gr_delta_bulk_gives}.
The variation of $S_{\rm bulk} + S_{\rm phys}$
gives the relations
\be
\text{on $B^4_{\rm phys}$:} \qquad 
f^2 *_4 A + B + \tfrac 12 \kappa b_2 \wedge A = 0 \ , \qquad 
g^2 *_4 b_2 - a_2 = 0 \ .
\ee

\paragraph{Closing the sandwich.}
Upon closing the sandwich, the total action 
on physical spacetime $M^4$
can a priori receive contributions from \eqref{eq_nonAb_2gr_sym_action} and \eqref{eq_nonAb_2gr_phys}. 
Moreover, we put the bulk field $b_2$ on shell, and then use
\eqref{eq_nonAb_2gr_b_onshell}.
We find that the action on the symmetry boundary gives a vanishing contribution.
The action on the physical boundary 
gives us the expected non-topological terms,
together with a non-trivial set of topological terms from $a_2 b_2$.
To analyze them, we need some integration by parts and 
the property \eqref{eq_Psi2_def}.
In the computation, we generate a term $\tfrac{1}{2\pi} \int dv_1 \wedge d\varphi_1$;
it can be dropped because 
 both $dv_1$ and $d\varphi_1$ have periods in $2\pi \mathbb Z$.
The final result for the action reads  
\be 
\ba 
S &= \frac{1}{2\pi}   \int_{M^4} \bigg\{ 
- \tfrac 12 f^2 {\rm Tr}\Big[ 
(\cV^{-1} d\cV + \cV^{-1} \cA \cV) \wedge  *_4 
(\cV^{-1} d\cV + \cV^{-1} \cA \cV)
\Big]
 - \tfrac 12 g^2 dv_1 \wedge *_4 dv_1 
\\
&
+ dv_1 \wedge \cA_2 
-   \tfrac 12 \kappa dv_1 \wedge {\rm Tr}(\cA d\cV \cV^{-1})
+  \tfrac 16  \kappa v_1 \wedge   {\rm Tr}[(\cV^{-1} d\cV)^3]
\bigg\}  \ . 
\ea 
\ee 
We have obtained the action for a $G$-valued 0-form $
\cV$, coupled to the $G$ background gauge field $\cA$,
together with a photon $v_1$, coupled to a background $\cA_2$ for its magnetic symmetry. 
Interestingly, we also have a Wess-Zumino-like coupling $v_1{\rm Tr}[(\cV^{-1} d\cV)^3]$.
This action was derived in \cite{Antinucci:2024bcm}
in a holographic context.

\subsection{Non-linearly realized $\mathbb Q/\mathbb Z$ symmetry}

In this section we discuss a $\mathbb Q/\mathbb Z$ 0-form symmetry in four dimensions.
A notable example of system exhibiting this symmetry is massless QED. 
At the classical level, the model has an axial $U(1)$ global symmetry, but at the quantum level it suffers from an ABJ anomaly. For rational values of the axial rotation angle, 
the symmetry can be recovered, albeit as a non-invertible symmetry \cite{Choi:2022jqy,Cordova:2022ieu}.\footnote{\, See  \cite{Karasik:2022kkq,GarciaEtxebarria:2022jky,Arbalestrier:2024oqg} for proposals to realize a non-invertible version of axial symmetry for irrational values of the angle.}

Growing evidence shows that the spontaneous breaking 
of $\mathbb Q/\mathbb Z$ 
symmetry leads to axion-Maxwell theory \cite{Cordova:2022ieu,Choi:2022fgx,Pace:2023mdo,Antinucci:2024bcm}. Below, we confirm this picture, by constructing an explicit sandwich realization, which is based on the bulk SymTFT appropriate for $\mathbb Q/\mathbb Z$ symmetry, and yields axion-Maxwell upon closing the sandwich. 

\paragraph{Bulk TQFT.} The SymTFT for 
$\mathbb Q/\mathbb Z$ symmetry was derived in \cite{Antinucci:2024zjp}. 
It is described by the action
\be \label{eq_QmodZ_bulk}
S_{\rm bulk} = \frac{1}{2\pi} \int_{X^5} \bigg[
B_3 \wedge dA_1 + B_2 \wedge dA_2 + \tfrac 12 \kappa A_1 \wedge B_2 \wedge B_2
\bigg] \ ,
\ee
where $\kappa = k/(2\pi)$ with integer $k$.
The fields $A_3$, $A_1$ are $U(1)$ gauge fields, while $B_3$, $B_2$ are $\mathbb R$ gauge fields.
The gauge transformations read
\be \label{eq_QmodZ_gauge}
\ba 
A_1 & \mapsto A_1 + d\Lambda_0 \ , \\
A_2 & \mapsto A_2 + d\Lambda_1
- \kappa \Omega_1 \wedge A_1
- \kappa B_2 \Lambda_0
- \kappa d\Omega_1 \Lambda_0 \ , \\
B_2 & \mapsto B_2 + d \Omega_1 \ , \\
B_3 & \mapsto B_3 + d\Omega_2 
- \kappa \Omega_1 \wedge B_2
- \tfrac 12 \kappa \Omega_1 \wedge d \Omega_1 \ . 
\ea 
\ee 
The bulk equations of motion are
\be 
dA_1 = 0 \ ,\qquad 
dB_2 = 0 \ ,\qquad 
dB_3 + \tfrac 12 \kappa B_2 \wedge B_2 = 0 \ , \qquad 
dA_2 + \kappa A_1 \wedge B_2 = 0 \ .
\ee 
It is  also be useful to observe that, due to the sum over
$dA_2$ fluxes in the bulk action 
\eqref{eq_QmodZ_bulk}, the closed 2-form $B_2$ has periods in $2 \pi \mathbb Z$.
As a result, on-shell we can write $B_2$ as the field strength of a $U(1)$ 1-form gauge field.
We write
\be \label{eq_QmodZ_B2_onshell}
\text{on-shell:} \qquad B_2 = dv_1 \ . 
\ee

The action 
\eqref{eq_QmodZ_bulk} gives the 
genuine bulk topological operators \cite{Antinucci:2024zjp}
\be 
\ba 
\mathbf W^{(n)} & = e^{i n \int_{\Sigma^1} A_1} \ , \\ 
\mathbf B^{(\alpha)} & = e^{i \alpha \int_{\Sigma^2} B_2} \ , \\ 
\widehat{ \mathbf W}^{(m)} & =  e^{i m \int_{\Sigma^2} A_2} \cT^m[\Sigma^2;A_1,B_2] \ , \\ 
\widehat{\mathbf B}^{(\frac{p}{kq})} & = e^{i \frac{p}{kq}  \int_{\Sigma^3}  B_3  }
\cA^{p,q}[\Sigma^3;B_2] \ , 
\ea 
\ee 
where $n,m\in \mathbb Z$, $\alpha \in \mathbb R/\mathbb Z$, $p$, $q$ are coprime integers, with $p \sim p + kq$.
Since $A_1$, $B_2$ are closed on-shell, their holonomies give rise to standard operators. 
In contrast, the non-closure of $A_2$ on-shell implies that the na\"ive operator 
$e^{i m \int A_2}$ must be dressed with a non-trivial TQFT. Here,
$\cT^{km}[\Sigma^2; A_1, B_2]$ denotes
the minimal 2d discrete $\mathbb Z_{km}$ gauge theory, coupled to $A_1$, $B_2$. 
The na\"ive operator
$e^{i \beta \int_{\Sigma^3} B_3}$
is treated
in a similar fashion. Upon restricting its parameter $\beta$ to be rational ($p/(qk)$),
we dress it with $\cA^{p,q}[\Sigma^3;B_2]$, which denotes the minimal TQFT $\cA^{p,q}$ of \cite{Hsin:2018vcg} coupled to $B_2$.\footnote{\, One may also consider irrational $\beta$ and enlarge the family of TQFTs used for dressing \cite{Arbalestrier:2024oqg}. The main conclusions of this section would still hold.}

For our discussion below, it will be useful to have a more explicit description of the dressed operator 
$\widehat{\mathbf W}^{(m)}$.
It can be described by the following total 
 2d topological  action
\be \label{eq_QmodZ_dressed_action}
\widehat{\mathbf W}^{(m)}= e^{i S} \ ,  \qquad
S=m \int_{\Sigma^2} \bigg[ 
A_2 + \kappa \Big( 
\psi_1 \wedge d \psi_0
+  B_2 \psi_0 
+ A_1 \wedge  \psi_1 
\Big)
\bigg] \ , 
\ee 
where $\psi_0$, $\psi_1$ are $U(1)$ gauge fields on $\Sigma^2$, with gauge transformations
\be \label{eq_QmodZ_psi_gauge}
\psi \mapsto \psi + d\lambda_0 - \Omega_1 \ , \qquad 
\psi_0 \mapsto \psi_0  + \Lambda_0 \ ,
\ee 
where $\lambda_0$ is a gauge parameter on $\Sigma^2$.
In fact, we can check that the gauge variation of the action $S$ in \eqref{eq_QmodZ_dressed_action} under
\eqref{eq_QmodZ_psi_gauge} and \eqref{eq_QmodZ_gauge}
reads
\be \label{eq_QmodZ_dressed_gauge_variation}
\Delta S = 
\int_{\Sigma_2} m \kappa \lambda_0 dA_1
+ m \int_{\partial \Sigma^2} 
\bigg[ 
\Lambda_1 
+\kappa \Omega_1 \psi_0 
- \kappa A_1 \lambda_0 
- \kappa \psi_0 d\lambda_0
\bigg]  \ . 
\ee 
If we use the bulk equation of motion $dA_1=0$, we see that the operator is gauge invariant on a closed $\Sigma^2$. We discuss the case of open $\Sigma^2$ below.

\paragraph{Symmetry boundary.}
We propose to consider the following topological action on the symmetry boundary,
\be \label{eq_QmodZ_symm_boundary}
S_{\rm sym} = \frac{1}{2\pi} \int_{B^4_{\rm sym}} \bigg[ 
B_3 \wedge (A_1  - d\varphi_0)
- B_2 \wedge (A_2- \cA_2 - d \varphi_1)
- \tfrac 12 \kappa \varphi_0 B_2 \wedge B_2
\bigg] \ . 
\ee 
The quantities $\varphi_0$, $\varphi_1$ are dynamical fields on $B^4_{\rm sym}$. They are $U(1)$ gauge fields ($\varphi_0$ is a compact scalar). The non-dynamical quantity $\cA_2$ is locally a 2-form, which we take to be closed,
\be 
d\cA_2 = 0 \ .
\ee 
The gauge transformations of the fields $\varphi_0$, $\varphi_1$ read
\be \label{eq_QmodZ_varphi_gauge}
\varphi_0 \mapsto  \varphi_0 
+ \Lambda_0
\ , \qquad 
\varphi_1 \mapsto \varphi_1 
+ d\sigma_{0}
+ \Lambda_1 
+ \kappa \varphi_0 \Omega_1  \ , 
\ee 
where $\sigma_0$ is a parameter on $B^4_{\rm sym}$, while $\Lambda_0$,
$\Lambda_1$, $\Omega_1$ are as in \eqref{eq_QmodZ_gauge}.
In Appendix \ref{app_QmodZ} we verify that the combined bulk and boundary action
is gauge invariant.
The equations of motion originating from the boundary action \eqref{eq_QmodZ_symm_boundary}
read
\be \label{eq_QmodZ_sym_EOM}
\text{on $B^4_{\rm sym}$:}
\qquad 
A_1 = d\varphi_0 \ , \qquad 
A_2 = \cA_2  + d\varphi_1 - \kappa \varphi_0 B_2 \ . 
\ee

With the choice \eqref{eq_QmodZ_symm_boundary} for the action on the symmetry boundary:
\begin{itemize}
    \item an open operator
    $\mathbf W^{(n)}$ can end perpendicularly on $B^4_{\rm sym}$;
   \item a closed operator
    $\mathbf W^{(n)}$ is trivialized if projected parallel onto   $B^4_{\rm sym}$;
    \item an open operator
    $\widehat{\mathbf W}^{(m)}$ can end perpendicularly on $B^4_{\rm sym}$;
   \item a closed operator
    $\widehat{\mathbf W}^{(m)}$ is trivialized if projected parallel onto   $B^4_{\rm sym}$.
\end{itemize}
Indeed, if we have an open operator
$\mathbf W^{(n)} = e^{i n \int A_1}$,
it is clear from the gauge transformation \eqref{eq_QmodZ_varphi_gauge} of $\varphi_0$ that it can end on $B^4_{\rm sym}$, on an operator of the form $e^{i n \varphi_0}$.
On the other hand, if we have a closed $\mathbf W^{(n)}$ and we project it onto $B^4_{\rm sym}$,
we get a trivial operator because
$\int d\varphi_0 \in 2\pi \mathbb Z$.
Let us now turn to the 
$\widehat {\mathbf W}^{(m)}$ operators.
If the support of
$\widehat {\mathbf W}^{(m)}$ is an open $\Sigma^2$, with $\partial \Sigma^2 \subset B^4_{\rm sym}$,
its gauge variation is obtained from
\eqref{eq_QmodZ_dressed_gauge_variation}, using $A_1 = d\varphi_0$,
\be 
\Delta S = 
m \int_{\partial \Sigma^2} 
\bigg[ 
\Lambda_1 
+\kappa \Omega_1 \psi_0 
- \kappa d\varphi_0 \lambda_0 
- \kappa \psi_0 d\lambda_0
\bigg] \ . 
\ee 
This gauge variation can be canceled
(up to total derivatives) by inserting the following operator on
$\partial \Sigma^2$,
\be 
m \int_{\partial \Sigma^2} 
\bigg[ 
- \varphi_1 + \kappa \psi_0 \psi_1 
- \kappa \varphi_0 \psi_1
\bigg] \ . 
\ee 
Thus, we have verified that an open
$\widehat {\mathbf W}^{(m)}$ operator can end perpendicularly  on $B^4_{\rm sym}$. Finally, we turn to a closed
$\widehat {\mathbf W}^{(m)}$ operator, projected onto $B^4_{\rm sym}$. We go back to the action 
\eqref{eq_QmodZ_dressed_action}, and we make use of  
of 
\eqref{eq_QmodZ_B2_onshell} and \eqref{eq_QmodZ_sym_EOM}.
After some integration by parts, we get
\be 
\int_{\Sigma^2} \bigg[ 
m \cA_2  
+ m d\varphi_1 
- m \kappa v_1 \wedge d\varphi_0 
+ m \kappa v_1 \wedge d\psi_0 
- m \kappa \psi_1 d\varphi_0
+ m \kappa \psi_1 \wedge d\psi_0
\bigg]  \ . 
\ee 
Finally, we perform a redefinition of the fields $\psi_0$, $\psi_1$ localized on $\Sigma^2$,
\be 
\psi_0 = \widetilde \psi_0 + \varphi_0 \ , \qquad 
\psi_1 = \widetilde \psi_1 - v_1 \ .
\ee 
The action becomes 
\be 
\int_{\Sigma^2} \bigg[ 
m \cA_2  
+ m d\varphi_1 
+ m \kappa \widetilde \psi_1 \wedge d\widetilde \psi_0
\bigg]  \ . 
\ee
The term $m \cA_2$ is a c-number.
The term $ m d\varphi_1 $ drops away because $\int_{\Sigma^2} d\varphi_1 \in 2\pi \mathbb Z$.
The term $m \kappa \widetilde \psi_1 \wedge d\widetilde \psi_0$ describes a completely decoupled discrete $\mathbb Z_{km}$ gauge theory.

We close with a comment about $\cA_2$.
One might wonder if it is possible to turn on an analogous quantity $\cA_1$ for $A_1$. The pair $(\cA_1, \cA_2)$ should satisfy 
\be 
d\cA_1 = 0 \ , \qquad 
d\cA_2 + \kappa \cA_1 \wedge B_2 = 0 \ . 
\ee 
On $B^4_{\rm sym}$, however, $B_2$ is fluctuating/summed over. Hence, it does not seem possible to have a non-zero $\cA_1$. This remark fits with the expectation that it should not be possible to turn on background fields for a non-invertible symmetry.

\paragraph{Physical boundary.}
We propose the following non-topological action on the physical boundary,
\be \label{eq_QmodZ_phys}
S_{\rm phys} = \frac{1}{2\pi} \int_{B^4_{\rm phys}} \bigg[ 
- \tfrac 12 f^2 A_1 \wedge * A_1
- \tfrac 12 g^2 B_2 \wedge * B_2
+ A_2 \wedge B_2
\bigg]  \ . 
\ee 
The quantities $f^2$, $g^2$ are constant parameters.
The last term $A_2 B_2$ is motivated as follows.
Upon varying the bulk action
\eqref{eq_QmodZ_bulk}, we get the boundary term
\be 
\delta S_{\rm bulk} \supset \frac{1}{2\pi} \int_{\partial X^5} 
\bigg[ 
\delta A_2 \wedge B_2 
+ \delta A_1 \wedge B_3 
\bigg]  \ . 
\ee 
We use the orientation convention
$\partial X^5 = B^4_{\rm sym} - B^4_{\rm phys}$. As a result, 
the combined variation of 
$S_{\rm bulk}$ and $S_{\rm phys}$ yields the following terms on $B^4_{\rm phys}$,
\be 
\frac{1}{2\pi} \int_{B^4_{\rm phys}} \bigg[ 
- f^2 \delta A_1 \wedge * A_1
- g^2 \delta B_2 \wedge * B_2
+ \delta B_2 \wedge A_2 
- \delta A_1 \wedge B_3
\bigg]  \ . 
\ee 
Thanks to the $A_2 B_2$ term in \eqref{eq_QmodZ_phys}, we have no terms with $\delta A_2$.
We have a well-posed variational problem, which gives
\be 
\text{on $B^4_{\rm phys}$:} \qquad 
B_3 + f^2 *A_1 = 0 \ , \qquad 
A_2 - g^2 *B_2 = 0  \ . 
\ee

\paragraph{Closing the sandwich.}
We now want to discuss the outcome of collapsing the sandwich constructed above. The final action
is obtained combing the contributions on the symmetry and physical boundaries. Moreover,
we put the bulk fields on-shell.
A first contribution to the final action on $M^4$ after closing the sandwich comes from \eqref{eq_QmodZ_symm_boundary}.
If we make use of \eqref{eq_QmodZ_sym_EOM} and \eqref{eq_QmodZ_B2_onshell}, we get
\be 
\frac{1}{2\pi} \int_{M^4} \tfrac 12 \kappa \varphi_0  dv_1 \wedge dv_1 \ . 
\ee 
A second contribution comes from \eqref{eq_QmodZ_phys}. Using \eqref{eq_QmodZ_sym_EOM} and \eqref{eq_QmodZ_B2_onshell}, it reads 
\be 
\frac{1}{2\pi} \int_{M^4} 
\bigg[ 
- \tfrac 12 f^2 d\varphi_0 \wedge * d\varphi_0 
- \tfrac 12 g^2 dv_1 \wedge * dv_1 
+ \cA_2 \wedge dv_1
+ d\varphi_1 \wedge dv_1
- \kappa \varphi_0 dv_1 \wedge dv_1
\bigg] 
\ . 
\ee 
The term 
 $d\varphi_1  dv_1$ can be dropped, since its integral is in $(2\pi)^2\mathbb Z$.
We keep the term  $\cA_2 \wedge dv_1$,
which is not a total derivative because $v_1$ is not a globally defined 1-form.
In conclusion, the final action
reads 
\be 
S = \frac{1}{2\pi} \int_{M^4} 
\bigg[ 
- \tfrac 12 f^2 d\varphi_0 \wedge * d\varphi_0 
- \tfrac 12 g^2 dv_1 \wedge * dv_1 
- \tfrac 12 \kappa \varphi_0 dv_1 \wedge dv_1
+ \cA_2 dv_1
\bigg] 
\ . 
\ee 
This is the expected axion-Maxwell system in four dimensions.
We also recover the coupling of the conserved current
$*J = \tfrac{1}{2\pi}dv_1$ of the magnetic 1-form symmetry to
the background~$\cA_2$.
In closing, we remark that axion-Maxwell enjoys 
an emergent electric 1-form symmetry, which is non-invertible \cite{Choi:2022fgx,Yokokura:2022alv}. 
This is not encoded in the bulk SymTFT \eqref{eq_QmodZ_bulk}.
It would be interesting to explore if
it can be captured by selecting a different TQFT in the bulk.

\section{Comments on spontaneous symmetry breaking}\label{sec:SSB}

In this Section we propose a scenario to realize spontaneous symmetry breaking of continuous symmetries from a topological symmetry theory perspective. Our setup is strongly inspired by  the construction of strip algebras for symmetry categories in two dimensions after \cite{Cordova:2024iti,Bhardwaj:2024igy,Choi:2024tri}. A related construction can be found in  \cite{Copetti:2024onh}.

\medskip

To have a spontaneously broken global symmetry it is necessary to study the field theory on a non-compact spacetime, so that allowed field configurations relax to energy-minimizing boundary conditions at infinity. To proceed with our analysis we focus on field theory formulated in Euclidean spacetimes, where we can still detect spontaneous symmetry breaking by considering the behavior of order parameters, charged field configurations which, if the symmetry is preserved, must all have zero 1-point functions. Of course the value of the 1-point functions in Euclidean spacetimes depends on choices of boundary conditions, which allows to detect spontaneous breaking in this setting. Our focus in this section is to give the SymTFT counterpart of this well-known phenomenon. 

\subsection{SSB and SymTFT with boundaries and corners: schematic description}
\label{sec_corners}

As we have reviewed in the section above, spontaneous breaking of symmetries occurs when an order parameter acquires a non-trivial one-point function. In this context therefore we obtain a family of QFT, parametrized by the choice of boundary conditions. We focus on a non-compact $d$-dimensional spacetime with boundary at infinity $\partial M^d \neq 0$. Therefore, in the SymTFT description the space $X^{d+1} = M^d \times [z_0,z_1]$ now is a space with boundaries and corners. This setup is close to the one considered in \cite{Cordova:2024iti,Choi:2024tri} in the construction of the strip algebras for two dimensional field theories. The same scenario is expected to generalize in other dimensions -- see e.g. \cite{Gagliano:2025gwr} for a recent application in four-dimensions. Let us review here briefly this idea. In presence of boundaries, the symmetry category $\mathcal C$ of the QFT of interest have an action on boundary conditions, which makes them into a $\mathcal C$-module, denoted $\mathcal M$. The precise structure of the module depends on the topology of $\partial M^d$.\footnote{\, In the examples discussed in the literature so far the choice of $M^d$ is Lorenzian, so one has a time direction and the space is chosen to have the form $[0,1] \times M^{d-2}$ so that one ends with a strip configuration, hence the name strip algebra.} The action of the symmetry category for field configurations in presence of boundary conditions is then encoded in a suitably generalized version of strip algebra depending on the topology of $M^d$. Boundary conditions along $\partial M^d$ form a module $\mathcal M$ of the latter, and the spontaneous breaking of the symmetry is understood in terms of the choice of $\mathcal M$. Charged field configurations transform according to $\mathcal C^*_{\mathcal M} = \text{Hom}_{\mathcal C}(\mathcal M,\mathcal M)$. In this context, the symmetry is fully broken if $\mathcal M \simeq \mathcal C$, while it is unbroken if $\mathcal M \simeq \text{Vec}$. Other choices of modules lead to partial symmetry breaking scenarios. Here we want to reproduce this logic in terms of our continuous SymTFT.

\medskip

\begin{figure}
    \centering
\includegraphics[width=14cm]{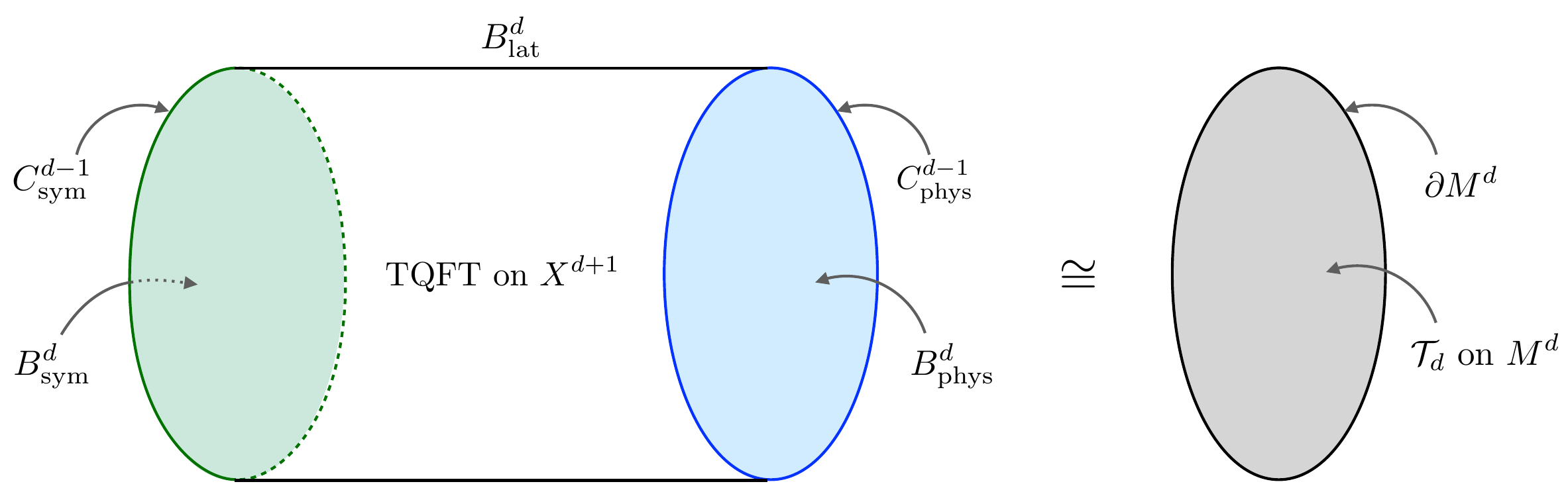}
    \caption{SymTFT realization of a  theory $\cT_d$ formulated on a spacetime $M^d$ with boundary. 
    The bulk TQFT is formulated on $X^{d+1} = M^d \times [z_0, z_1]$.
    The symmetry boundary $B^d_{\rm sym} = M^d \times \{ z_0\}$
    and the lateral boundary $B^d_{\rm lat} = \partial M^d \times [z_0, z_1]$
    support topological boundary conditions.
    They are joined along the topological interface  $C^{d-1}_{\rm sym} = \partial M^d \times \{z_0\}$.
    The physical boundary $B^d_{\rm phys} = M^d \times \{z_1\}$ supports a non-topological boundary condition in general.
    It is joined to $B^d_{\rm lat}$ along the interface $C^{d-1}_{\rm phys} = \partial M^d \times \{z_1 \}$, which is also non-topological in general.
    }
    \label{fig_corners}
\end{figure}

More precisely, we place the SymTFT on $X^{d+1} = M^d \times [z_0,z_1]$, where $M^d$ now has a non-trivial boundary $\partial M^d$. The resulting space then has three boundaries of codimension one 
\be
B^d_\text{lat} := \partial M^d \times [z_0,z_1]\,, \quad B^d_\text{sym} :=  M^d \times \{ z_0\}\,, \quad B^d_\text{phys} := M^d \times \{z_1\}
 \,,
\ee
and two codimension-two corners
\be
C^{d-1}_\text{sym} := B^d_\text{lat} \cap B^d_\text{sym} =  \partial M^d \times \{z_0\} \quad\text{and}\quad C^{d-1}_\text{phys} := B^d_\text{lat} \cap B^d_\text{phys} =  \partial M^d \times \{z_1\} \,.
\ee
See Figure \ref{fig_corners}.
In presence of boundaries of higher codimension, the collection of morphisms that are computed by the TFT in the bulk involves now also higher morphisms, related to the glueing conditions of $B^d_\text{lat}$, $B^d_{\text{sym}}$ and $B^d_\text{phys}$ along the codimension-two corners above. We obtain a family of theories because the system of interfaces $C^{d-1}_\text{sym}|B^d_\text{lat}|C^{d-1}_\text{phys}$ ends up being labeled by the choices boundary conditions for the physical theory -- see Figure \ref{fig_corners}.

\medskip

The SymTFT isomorphism is well defined only if the bulk of the SymTFT is completely topological. This implies that also the boundary conditions along $B^d_\text{lat}$ must be topological. Therefore we expect that $B^d_\text{sym}$ and $B^d_\text{lat}$ are each labeled by a choice of gapped boundary conditions for the bulk SymTFT, such that $B^d_\text{sym}$ encodes the symmetry of the theory, while the choice of $B^d_\text{lat}$ dictates which part of the symmetry is left unbroken. There are now two corners, of which only one is fully topological, namely $C^{d-1}_\text{sym}$. The latter is a topological interface of codimension two that intertwines the symmetry boundary condition with the unbroken part of the symmetry dictated by $B^d_\text{lat}$. This is the analogue of the choice of the module $\mathcal M$ for $\mathcal C$. The latter is encoded by the corner $C^{d-1}_\text{sym}$. The resulting algebra $\mathcal C^*_{\mathcal M}$ arises along $ B^d_\text{lat}$. On the other side of the SymTFT interval we have $C^{d-1}_\text{phys}$ which is a non-topological interface between $B^d_\text{lat}$ and the physical boundary. This corner connecting the physical boundary to $B^d_\text{lat}$ is then responsible for the explicit action of the symmetry on boundary conditions, encoding the spontaneous breaking on the physical theory from the bulk SymTFT description.

\medskip

Let us remark that the SymTFT setup with boundaries and corners described above and depicted in Figure
\ref{fig_corners} 
can be used to study a physical theory $\cT_d$
with a boundary that can be at infinity,
or at a finite distance. 
The difference between these two cases 
is captured by the metric on 
the physical boundary $B^d_{\rm phys}$
and physical corner $C^{d-1}_{\rm phys}$.
In contrast, the topological elements of the SymTFT construction -- namely
the bulk, $B^d_{\rm lat}$, $B^d_{\rm sym}$,
and $C^{d-1}_{\rm sym}$ --
do not discriminate between boundaries at infinity or at a finite distance.
Our main objective in this work is to study
spontaneous symmetry breaking. Therefore, 
we consider setups in which the metric on
$B^d_{\rm phys}$ is such that the boundary is
at infinity, which allows us to probe the 
long-distance infrared physics of the model, and hence the SSB.

\medskip

The above is meant to give a schematic description, let us make it concrete with some examples in detail.

\subsection{The case of an Abelian $p$-form symmetry}
\label{sec_corner_abelian_case}

For ease of exposition, here we focus
on the case of a (non-anomalous) $U(1)$ $p$-form symmetry in $d$ dimensions. We discuss  the non-Abelian case for $p=0$
in Section \ref{sec_corners_nonAb} below.

\subsubsection{Selected topological boundary conditions}

The bulk TQFT on $X^{d+1}$
is the BF theory
\be 
S_{\rm bulk} = \frac{1}{2\pi} \int_{X^{d+1}} B_{d-p-1} \wedge dA_{p+1} \ , 
\ee 
where $A_{p+1}$ is a $U(1)$ gauge field and $B_{d-p-1}$ is an $\mathbb R$ gauge field.
Our focus is on the following 
topological operators
of the bulk TQFT,
\be 
\mathbf W^{(n)}_{p+1} = \exp i n\int_{\Sigma^{p+1}} A_{p+1} \ , \qquad 
\mathbf Q^{(\alpha)}_{d-p-1} = \exp i \alpha \int_{\widehat \Sigma^{d-p-1}} B_{d-p-1} \ , 
\ee 
where $n\in \mathbb Z$, $\alpha \in \mathbb R$,
with $\alpha \sim \alpha +1$.

For concreteness, in this work we only consider two topological boundary conditions for the bulk TQFT,
which we denote $\cD_A$ and $\cD_B$ 
--- we refrain from attempting a classification
and we do not explore the option of dressing the boundary conditions and interfaces we find with additional TQFTs.
We now comment on some properties of $\cD_A$,
$\cD_B$  and on the relevant  topological interfaces between them.\footnote{See \cite{Bhardwaj:2024qiv,Bhardwaj:2024igy}  
for related discussions in the context of finite symmetries.
}
More details can be found in Appendix \ref{app_Dir_and_Neu}.

\paragraph{Topological boundary condition $\cD_A$.}
This is described by the action 
\be \label{eq_DA_action}
S_{\rm bdy}^{\cD_A} = \frac{1}{2\pi} \int_{Y^d}
(-1)^{d-p} B_{d-p-1} \wedge (A_{p+1} - 
\cA_{p+1} -
d\varphi_p) \ , 
\ee 
where $Y^d = \partial X^{d+1}$,  $\varphi_p$ is a dynamical $U(1)$ gauge field on $Y^d$,
$\cA_{p+1}$ is a non-dynamical $U(1)$ gauge field on $Y^d$, which we take to be flat,
\be 
d\cA_{p+1} = 0 \ . 
\ee 
The sum of the bulk action and \eqref{eq_DA_action}
is invariant under the gauge transformations
\be \label{eq_bulk_plus_DA_gauge}
\ba 
A_{p+1} & \mapsto A_{p+1} + d\Lambda_p \ , & 
B_{d-p-1} & \mapsto B_{d-p-1} + d\Omega_{d-p-2} \ , &
\varphi_p & \mapsto \varphi_p + d\sigma_{p-1} + \Lambda_p \ . 
\ea 
\ee 
The bulk gauge parameters $\Lambda_p$,
$\Omega_{d-p-2}$ are a $U(1)$ gauge field and a globally-defined form, respectively. The boundary gauge parameter $\sigma_{p-1}$ is a $U(1)$ gauge field.
The equations of motion of the boundary action imply
\be 
\text{on $Y^d$:} \qquad 
A_{p+1} = d\varphi_p + \cA_{p+1} \ . 
\ee 

The choice of boundary condition $\cD_A$ implies that:
\begin{itemize}
    \item An open bulk operator $\mathbf W_{p+1}^{(n)}$ can end perpendicularly on $Y^d$, on an operator of the form $e^{i n \int \varphi_p}$.
    \item A closed bulk operator $\mathbf W_{p+1}^{(n)}$ becomes a c-number if it is projected parallel onto $Y^d$.
    \item A closed bulk operator $\mathbf Q_{d-p-1}^{(\alpha)}$ remains a non-trivial operator if it is projected parallel onto $Y^d$. We write
    \be \label{eq_define_Dalpha}
\text{$\mathbf Q_{d-p-1}^{(\alpha)}$ projected onto $\cD_A$:} \quad D_{d-p-1}^{(\alpha)} \ .
\ee 
\end{itemize}
From the point of view of the bulk topological operators $\mathbf W_{p+1}^{(n)}$,
$\mathbf Q_{d-p-1}^{(\alpha)}$, the topological boundary condition $\cD_A$ is effectively the same as a Dirichlet boundary condition for $A_{p+1}$, and a free boundary condition for $B_{d-p-1}$.
The boundary condition $\cD_A$ differs from the usual Dirichlet boundary condition for $A_{p+1}$
because
it is formulated including the mode $\varphi_p$ on the boundary.
For our purposes, this
is useful to study interfaces (see below).

\paragraph{Topological boundary condition $\cD_B$.}
This is described by the action 
\be \label{eq_DB_action}
S_{\rm bdy}^{\cD_B} = \frac{1}{2\pi} \int_{Y^d}
(-1)^{d-p} (d\Phi_{d-p-2}  + \cB_{d-p-1})\wedge  A_{p+1} \ , 
\ee 
where $Y^d = \partial X^{d+1}$,  $\Phi_{d-p-2}$ is a dynamical $\mathbb R$ gauge field on $Y^d$,
$\cB_{d-p-1}$ is a non-dynamical form on $Y^d$, which we take to be closed,
\be 
d\cB_{d-p-1} = 0 \ . 
\ee 
The sum of the bulk action and \eqref{eq_DB_action}
is invariant under the same 
bulk gauge transformations as in \eqref{eq_bulk_plus_DA_gauge}, together with 
\be 
\Phi_{d-p-2} \mapsto \Phi_{d-p-2} + d \rho_{d-p-3} + \Omega_{d-p-2} \ , 
\ee 
where boundary gauge parameter $\rho_{d-p-3}$ is a globally defined form.
The equations of motion of the boundary action imply
\be 
\text{on $Y^d$:} \qquad 
B_{d-p-1} = d\Phi_{d-p-2} + \cB_{d-p-1} \ . 
\ee 

The choice of boundary condition $\cD_B$ implies that:
\begin{itemize}
    \item An open bulk operator $\mathbf Q_{d-p-1}^{(\alpha)}$ can end perpendicularly on $Y^d$, on an operator of the form $e^{i \alpha \int \Phi_{d-p-2}}$.
    \item A closed bulk operator $\mathbf Q_{d-p-1}^{(\alpha)}$ becomes a c-number if it is projected parallel onto~$Y^d$.
    \item A closed bulk operator $\mathbf W_{p+1}^{(n)}$ remains a non-trivial operator if it is projected parallel onto $Y^d$. We write
    \be \label{eq_define_tildeDn}
\text{$\mathbf W_{p+1}^{(n)}$ projected onto $\cD_B$:} \quad \widetilde D_{p+1}^{(n)} \ .
\ee 
\end{itemize}
From the point of view of the bulk topological operators $\mathbf W_{p+1}^{(n)}$,
$\mathbf Q_{d-p-1}^{(\alpha)}$, the topological boundary condition $\cD_B$ is effectively the same as a Dirichlet boundary condition for $B_{d-p-1}$, and a free boundary condition for $A_{p+1}$.
The boundary condition $\cD_B$ differs from the usual Dirichlet boundary condition 
for $B_{d-p-1}$ because
it is formulated including the mode $\Phi_{d-p-2}$ on the boundary.

\paragraph{$\cD_A$-$\cD_A$ interfaces.}
Let us partition the boundary $\partial X^{d+1}$ into two regions $Y^d_{1}$
and $Y^d_{2}$, 
and impose $\cD_A$ boundary conditions on both regions.
The two regions
meet at the interface
$Z^{d-1} = \partial Y^d_{1} = - \partial
Y^d_{2}$,
with the minus denoting orientation reversal.
We can construct a family of topological interfaces on $Z^{d-1}$. They are described by the action
\be \label{eq_interface_DA_DA}
S_{\rm int}^{\cD_A \,\cD_A} = \frac{1}{2\pi} \int_{Z^{d-1}}  B_{d-p-1} \wedge (\varphi_p^{(2)} - \varphi_p^{(1)}
- d\nu_{p-1}
- \xi_p  )   \ . 
\ee 
Each component $Y^d_i$ ($i=1,2$) comes with its
dynamical fields $\varphi_p^{(i)}$
(with gauge parameters $\sigma_{p-1}^{(i)}$ as in \eqref{eq_bulk_plus_DA_gauge}) and its non-dynamical $\cA_{p+1}^{(i)}$.
The quantity 
$\nu_{p-1}$ 
is a dynamical $U(1)$ gauge field localized on $Z^{d-1}$
with gauge transformation
\be 
\nu_{p-1} \mapsto \nu_{p-1} + du_{p-2} + \sigma^{(2)}_{p-1} - \sigma^{(1)}_{p-1} \ ,
\ee 
where the parameter $u_{p-2}$ is a $U(1)$ gauge field on $Z^{d-1}$.
The quantity $\xi_p$ in \eqref{eq_interface_DA_DA} 
is 
a non-dynamical form on $Z^{d-1}$, which we take such that
\be \label{eq_app_dxiBIS}
\text{on $Z^{d-1}$:}
\qquad 
d\xi_p = \cA^{(1)}_{p+1} - \cA^{(2)}_{p+1} \ . 
\ee 
We regard $\xi_p$ as a parameter that 
labels various possible $\cD_A$-$\cD_A$ interfaces. We give more comments on its physical interpretation in Section \ref{sec_Goldstone} below, where we connect it to the degeneracy of vacua expected for a theory with SSB of a continous symmetry.

An important property of 
$\cD_A$-$\cD_A$ interfaces of the form
\eqref{eq_interface_DA_DA} is the following.
On either side of the interface we can insert 
an operator $D^{(\alpha)}_{d-p-1}$
(see \eqref{eq_define_Dalpha} for the definition)
supported on a cycle $\Sigma^{d-p-1}$.
If we project $D^{(\alpha)}_{d-p-1}$ parallel to the interface,
the interface  absorbs it. In the process, the interface parameter $\xi_p$ shifts
\be \label{eq_app_DD_xi_shiftBIS}
\xi_p \rightarrow \xi'_p = \xi_p + 2\pi \alpha \delta_p(\Sigma^{d-p-1} \subset Z^{d-1}) \ ,
\ee 
where $\delta_p(\Sigma^{d-p-1} \subset Z^{d-1})$
is the Poincar\'e dual to $\Sigma^{d-p-1}$
in $Z^{d-1}$.
We refer the reader to Appendix \ref{app_Dir_and_Neu} for more details.

\paragraph{$\cD_A$-$\cD_B$ interface.}
Let us partition the boundary $\partial X^{d+1}$ into two regions $Y^d_{1}$
and $Y^d_{2}$, 
and impose $\cD_A$ boundary conditions on $Y^d_1$
and $\cD_B$ on $Y^d_2$. As above,
the two regions
meet at the interface
$Z^{d-1} = \partial Y^d_{1} = - \partial
Y^d_{2}$.
This time we construct a single topological interface, described by the action
\be \label{eq_interface_DADB_action}
S_{\rm int}^{\cD_A \, \cD_B} = \frac{1}{2\pi} \int_{Z^{d-1}} 
\bigg[ 
(-1)^{d-p} \Phi_{d-p-2} \wedge d\varphi_p 
+ (-1)^{d-p} \Phi_{d-p-2} \wedge \cA_{p+1}
- \cB_{d-p-1} \wedge \varphi_p 
\bigg] 
\ . 
\ee
We notice the BF coupling between the mode $\varphi_p$ from the $\cD_A$ side,
and the mode $\Phi_{d-p-2}$ from the $\cD_B$ side.

An important property of the $\cD_A$-$\cD_B$ interface is the following.
On the $\cD_A$ side of the interface, we can insert an operator $D^{(\alpha)}_{d-p-1}$. 
If we project it  parallel to the interface, it becomes a c-number.
Alternatively, we can have an open 
$D^{(\alpha)}_{d-p-1}$ operator inside $Y^d_1$
ending perpendicularly on the interface, 
on an operator of the form
$e^{i \alpha \int \Phi_{d-p-2}}$.

Similarly, on $\cD_B$ side of the interface, we can insert an operator $\widetilde D^{(n)}_{p+1}$ (see \eqref{eq_define_tildeDn} for the definition). If we project it parallel to  the interface, it becomes a c-number. Alternatively, we can have an open
$\widetilde D^{(n)}_{p+1}$ operator inside $Y^d_2$ ending perpendicularly on the interface, on an operator of the form
$e^{i n \int \varphi_p}$.

\subsubsection{Scenarios for unbroken and broken symmetry}

In this subsection
we adopt the general picture described in Section 
\ref{sec_corners} and we describe 
the choices of $B^d_{\rm sym}$, $B^d_{\rm lat}$,
$C^{d-1}_{\rm sym}$
that correspond to different scenarios for a $U(1)$ $p$-form symmetry.
For simplicity, 
 we focus on two the extreme scenarios in which the symmetry is either completely unbroken, or completely broken.
Our proposal to describe these scenarios is encoded in the following table 
\be 
\begin{array}{|l|>{\centering}p{1.2cm}| 
>{\centering}p{1.2cm}|
>{\centering\arraybackslash}p{3.2cm}|}
\hline 
&   $B^d_{\rm sym}$ \rule[-.25cm]{0mm}{0.7cm} & 
$B^d_{\rm lat}$
&
$C^{d-1}_{\rm sym}$
\\ 
\hline 
\text{unbroken symmetry \hspace{5mm}}\rule[-.25cm]{0mm}{0.7cm}&  $\cD_A$  & $\cD_B$ & $\cD_A$-$\cD_B$ interface \\ 
\hline
\text{broken symmetry} \rule[-.25cm]{0mm}{0.7cm}& 
$\cD_A$ &  $\cD_A$ & $\cD_A$-$\cD_A$ interface \\ 
\hline
\end{array}
\ee 
In both cases,
on the boundary
$B^d_{\rm sym}$
we have topological operators $D^{(\alpha)}_{d-p-1}$, obtained from parallel projection of the bulk operators
$\mathbf Q^{(\alpha)}_{d-p-1}$. Upon closing the 
interval $z$ direction,
the $D^{(\alpha)}_{d-p-1}$ operators on $B^d_{\rm sym}$
yield the topological 
operators that implement the
global $U(1)$ $p$-form symmetry of the physical theory on $M^d$.
When we close the interval direction, 
$C^{d-1}_{\rm sym}$
$B^d_{\rm lat}$,
and $C^{d-1}_{\rm phys}$
all collapse onto $\partial M^d$, the boundary of the physical theory.
We discuss the two scenarios in turn.
\begin{itemize}
    \item \emph{Unbroken symmetry}.
In this case we have a single $\cD_A$-$\cD_B$ interface
on $C^{d-1}_{\rm sym}$.
Moreover, we know that
this interface can absorb the topological operators
$D^{(\alpha)}_{d-p-1}$
coming from the $\cD_A$ boundary on $B^d_{\rm sym}$,
and projected parallel onto
$C^{d-1}_{\rm sym}$.
When we close the interval direction, 
we interpret the result as a unique boundary condition,
which is invariant under the action of the topological operators
implementing the $U(1)$ $p$-form symmetry. This is corresponds to   the case of unbroken symmetry.

    \item \emph{Broken symmetry}.
In this case we have a family of $\cD_A$-$\cD_A$ interfaces
on $C^{d-1}_{\rm sym}$.
Moreover, we know that
we can change the parameter $\xi_p$ of the 
$\cD_A$-$\cD_A$ interface by 
taking a topological 
 operator
$D^{(\alpha)}_{d-p-1}$
on $B^d_{\rm sym}$
and projecting it 
parallel onto $C^{d-1}_{\rm sym}$.
When we close the interval direction, 
we interpret the result as a family of boundary conditions,
which 
are acted upon transitively
by  the $U(1)$ $p$-form symmetry. This corresponds to the case of completely broken symmetry.

\end{itemize}

\subsubsection{Ward identities
from the SymTFT picture}
\label{sec_ward_from_symtft}

Using the SymTFT picture with boundaries and corners proposed above, we can derive Ward identities for (un)broken $p$-form symmetries.
We discuss four cases:
unbroken 0-form symmetry;
broken 0-form symmetry;
unbroken $p$-form symmetry, $p\ge 1$;
broken $p$-form symmetry, $p\ge 1$.
For ease of exposition,
we assume that the spacetime $M^d$ has the topology of a closed ball, with boundary $\partial M^d \cong S^{d-1}$.

\subsubsection*{Unbroken 0-form symmetry}

Let $\cO$ be a local operator on the physical theory, with charge $n$ under the global $U(1)$ 0-form symmetry.
In the SymTFT picture,
$\cO$ is realized by a bulk Wilson line $\mathbf W^{(n)}_{1}$ that stretches along the interval $z$ direction,
from the boundary $B^d_{\rm sym}$ to the boundary $B^d_{\rm phys}$.
Let us denote the endpoint of
$\mathbf W^{(n)}_{1}$
on $B^d_{\rm sym}$
as $\mathcal E^0$.

\begin{figure}
    \centering
\includegraphics[width=15cm]{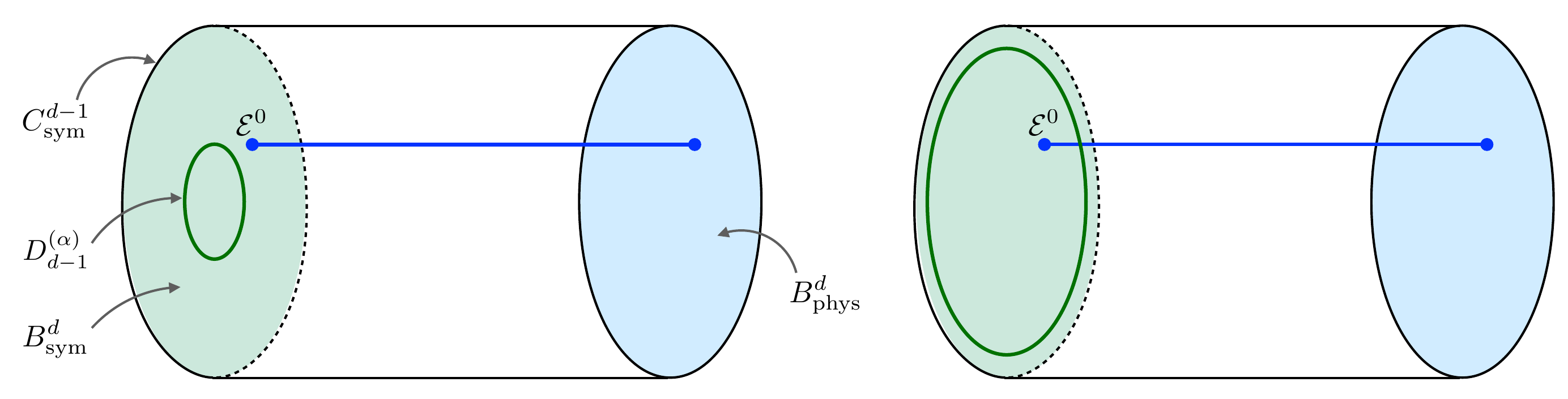}
    \caption{A local operator charged under the 0-form symmetry is realized in the SymTFT construction by a Wilson line $\mathbf W^{(n)}_1$ stretched from
    $B^d_{\rm sym}$ to $B^{d}_{\rm phys}$. On the boundary $B^d_{\rm sym}$,
    which supports $\cD_A$ boundary conditions,
    a defect $D^{(\alpha)}_{d-1}$ is   nucleated.
    Then, it is expanded, until it first crosses the endpoint $\cE^0$ of $\mathbf W^{(n)}_1$, and finally reaches the interface $C^{d-1}_{\rm sym}$.
    }
    \label{fig_0_form_ward}
\end{figure}

Starting from this initial configuration, we perform the following topological moves, see Figure \ref{fig_0_form_ward}.
\begin{enumerate}
    \item We nucleate a topological operator
    $D^{(\alpha)}_{d-1}$
    supported on a small sphere
    $S^{d-1}$
    inside $B^d_{\rm sym}$,
    away from the endpoint
    $\mathcal E^0$ of $\mathbf W^{(1)}_{1}$ on $B^d_{\rm sym}$. 
    \item We expand $D^{(\alpha)}_{d-1}$, remaining inside $B^d_{\rm sym}$. Eventually, we sweep past the endpoint $\mathcal E^0$.
    We pick up the non-trivial linking phase $e^{2\pi i n \alpha}$.
    \item We keep expanding 
    $D^{(\alpha)}_{d-1}$ until it reaches 
    $C^{d-1}_{\rm sym}$.
    \item Since $C^{d-1}_{\rm sym}$ supports the 
    $\cD_A$-$\cD_B$ interface, the operator $D^{(\alpha)}_{d-1}$ gets absorbed.
\end{enumerate}
Retracing the above steps,
we infer the following identity,
\be 
\langle \cO \rangle = e^{2\pi i n \alpha} \langle \cO \rangle  \ . 
\ee 
Since this must hold for generic $\alpha$, we conclude
\be 
\langle \cO \rangle = 0 \ . 
\ee 
Thus, using the SymTFT picture we recover the fact that the 
1-point function 
of a charged operator    must be zero
if we are in a setup with 
a single boundary conditions   that preserves the $U(1)$ symmetry.

\subsubsection*{Broken 0-form symmetry}

The setup is the same as the case just discussed.
We perform the same 
topological moves 1.~to~3.,
but the final point 4.~is now replaced by 
\begin{itemize}
    \item[4.${}^\prime$] Since $C^{d-1}_{\rm sym}$ supports a $\cD_A$-$\cD_A$ interface, the operator $D^{(\alpha)}_{d-1}$ dresses the pre-existing $\cD_A$-$\cD_A$ interface, turning into  a different $\cD_A$-$\cD_A$ interface. 
\end{itemize}
Retracing the above steps,
we infer an  identity of the form
\be 
\langle \cO \rangle_{v} = e^{2 \pi in \alpha} \langle \cO \rangle_{v'}  \ . 
\ee 
We have added a label $v$, $v'$ to stress that we have a collection of distinct boundary conditions. The label $v'$
stands for the boundary condition that is obtained from $v$ by acting with the $U(1)$ group element~$e^{2\pi i\alpha}$.
In this case the SymTFT picture gives a relation among 1-point functions with different boundary conditions. It is compatible with a non-zero value for the 1-point functions. 

\subsubsection*{Unbroken $p$-form symmetry, $p\ge1$}

To make contact with spontaneous breaking
for a 
$U(1)$ $p$-form symmetry ($p  \ge 1$),
we consider 
a charged $p$-dimensional 
operator $\cO_p$ supported on an $S^{p}$ in spacetime $M^d$.
Moreover, we have to consider the limit in which 
the radius of the $S^p$ goes to infinity. 
In our SymTFT construction,
infinity is modeled by the boundary $\partial M^d$ of physical spacetime $M^d$.
Thus, we are led to consider
a $p$-dimensional operator
supported on a disk $B^{p}$
with  boundary $\partial B^p \cong S^{p-1}$ inside
$\partial M^d$.
(For $p=1$, $B^1$ is a closed interval, and $\partial B^1 \cong S^0$ 
is the disjoint union of its two endpoints.)\footnote{\, More explicitly, suppose we describe spacetime $M^d \cong B^d$ as a closed $d$-dimensional ball in $\mathbb R^{d}$, with coordinates $x^1$, $\dots$,
$x^d$ satisfying
\be 
(x^1)^2 + \dots + (x^d)^2 \le 1 \ .  \nn
\ee 
Then, we consider a $p$-dimensional operator whose support $( \cong B^p)$ is the intersection of the above closed ball in $\mathbb R^{d}$ with the locus
\be 
x^{p+1} = x^{p+2} = \dots = x^{d} = 0 \ .  \nn
\ee 
}

The SymTFT realization of the above $p$-dimensional charged operator
involves a bulk topological operator $\mathbf W_{p+1}^{(n)}$, where $n$ is the charge. The support $\Sigma^{p+1}$
of the operator
$\mathbf W_{p+1}^{(n)}$
consists of two pieces,
\be 
\Sigma^{p+1} = \Sigma^{p+1}_1 \cup \Sigma^{p+1}_2 \ .
\ee 
The first piece $\Sigma^{p+1}_1$ extends 
along $B^p$ combined with the interval direction,
\be \label{eq_Sigma_first_piece}
\Sigma^{p+1}_1 \; : \qquad 
B^p \times [z_0,z_1] \subset M^d \times [z_0, z_1] \ . 
\ee 
Notice that $\Sigma_1^{p+1}$ has three boundary components.
The first two are of the form
$B^p \times \{z_0\}$
and $B^p \times \{ z_1\}$.
They are the loci where 
$\Sigma_1^{p+1}$
ends perpendicularly on
$B^d_{\rm sym}$ and $B^d_{\rm phys}$.
The third boundary component
of $\Sigma_1^{p+1}$
is of the form
\be \label{eq_boundary_Sigma1}
\text{a component of }\partial \Sigma_1^{p+1} \; : S^{p-1} \times [z_0,z_1] \subset \partial M^d \times [z_0, z_1] \ .
\ee 
In fact, we can say that 
$\Sigma_1^{p+1}$ also ends perpendicular to the
lateral boundary $B^d_{\rm lat}$.
Recall, however,
that we impose $\cD_B$ boundary conditions on $B^d_{\rm lat}$,
hence a $\mathbf W^{(n)}_{p+1}$ operator cannot end perpendicularly there.
Rather, we must have an L-shaped configuration,
with $\mathbf W^{(n)}_{p+1}$
bending and continuing inside
$B^d_{\rm lat}$.
This shows why we have to consider a second piece 
$\Sigma^{p+1}_2$
for the support of 
$\mathbf W^{(n)}_{p+1}$.
This second piece lies entirely inside $B^d_{\rm lat}$ and is of the form 
\be \label{eq_Sigma_second_piece}
\Sigma^{p+1}_2 \; : \qquad 
B'^p \times [z_0,z_1] \subset \partial M^d \times [z_0, z_1] \ . 
\ee 
Here $B'^p$ denotes
a closed $p$-dimensional ball inside $\partial M^d \cong S^{d-1}$,
such that its boundary
$\partial B'^p$
coincides with the $S^{p-1}$
appearing in $\partial \Sigma_1^{p+1}$, see 
\eqref{eq_boundary_Sigma1}.\footnote{\, Continuing the explicit parametrization of the previous footnote,
$\Sigma^{p+1}_1$ and 
$\Sigma^{p+1}_2$ can be taken to be
\be 
\Sigma^{p+1}_1 \; : \qquad 
\left\{
\begin{array}{l}
(x^1)^2 + \dots +(x^d)^2 \le 1 \\
x^{p+1} = x^{p+2} = \dots = x^d = 0 \\ 
z\in [z_0, z_1]
\end{array} 
\right.  \ ,
\qquad
\Sigma^{p+1}_2 \; : \qquad 
\left\{
\begin{array}{l}
(x^1)^2 + \dots +(x^d)^2 = 1 \\
x^{p+1} \ge  0 \\
x^{p+2} = \dots = x^d = 0 \\ 
z\in [z_0, z_1]
\end{array} 
\right.  \ .
\nn 
\ee 
}
Let us denote as $\mathcal E^p$ the end-locus 
of the first piece
$\Sigma^{p+1}_1$ of $\mathbf W^{(n)}_{p+1}$
on $B^d_{\rm sym}$.
Thus,
\be \label{eq_Ep_endpoint}
\mathcal E^p \; : \qquad 
B^p \times \{z_0\} \subset M^d \times [z_0, z_1] \ . 
\ee

\begin{figure}
    \centering
\includegraphics[width=10cm]{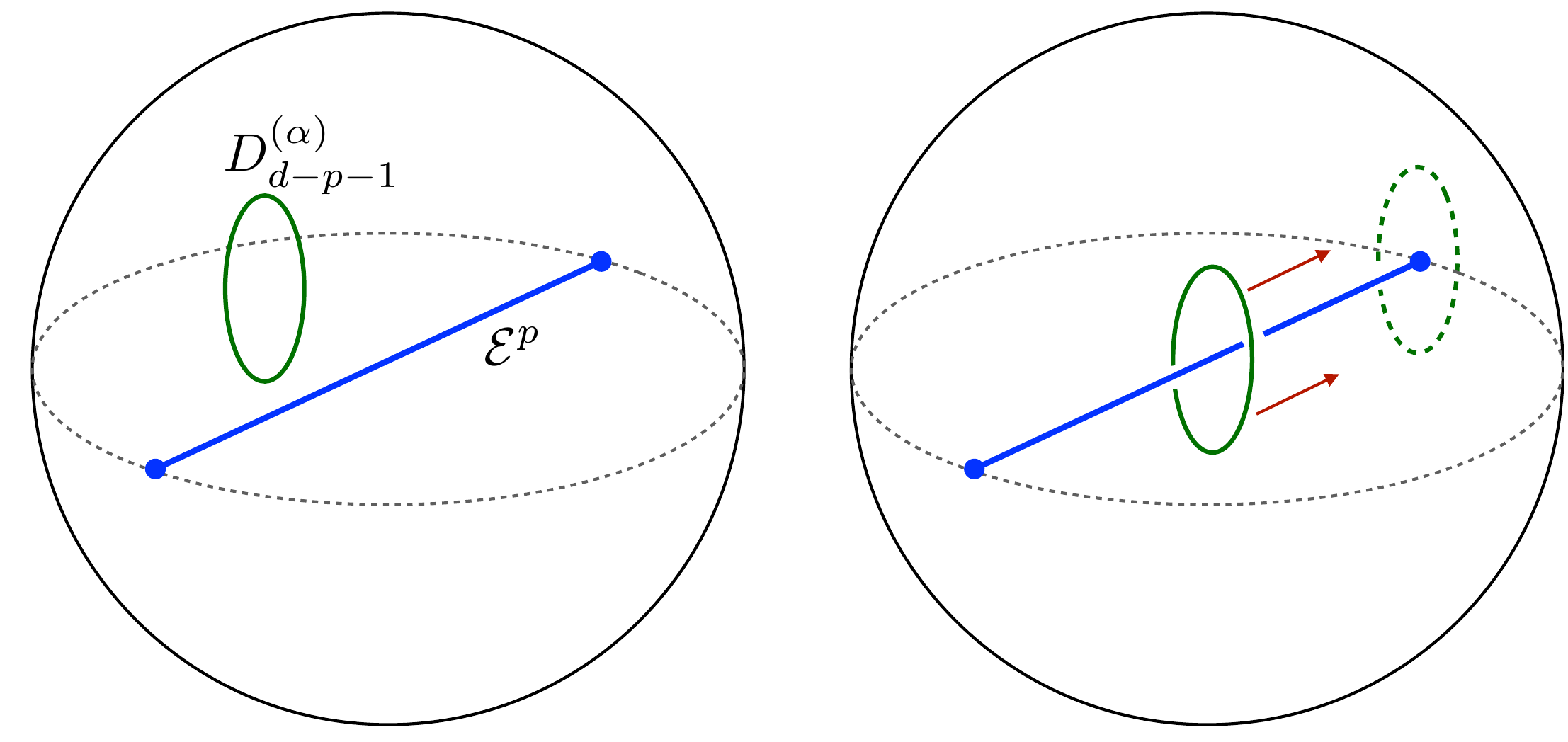}
    \caption{The figure depicts the boundary
    $B^d_{\rm sym}$, with the topology of a $d$-dimensional ball,
    visualized for $d=3$.
    The interval direction $z$ is suppressed.
    The boundary $\partial B^d_{\rm sym} \cong S^{d-1}$ (a 2-sphere in the figure) supports the interface $C^{d-1}_{\rm sym}$.
    The locus $\mathcal E^p$
    is the locus where the bulk operator
    $\mathbf W^{(n)}_{p+1}$
    ends on $B^d_{\rm sym}$. It is visualized here for $p=1$. After nucleating 
    a $D^{(\alpha)}_{d-p-1}$ defects inside $B^{d}_{\rm sym}$, we link it with
    $\cE^{p}$. Finally,
    we slide $D^{(\alpha)}_{d-p-1}$ along $\cE^p$ until it reaches $C^{d-1}_{\rm sym}$.
    }
    \label{fig_p_form_slide}
\end{figure}

Having described the initial configuration,
we now perform the following topological moves. See also Figure \ref{fig_p_form_slide}.
\begin{enumerate}
    \item We nucleate a topological operator $D^{(\alpha)}_{d-p-1}$ 
    inside $B^d_{\rm sym}$.
    Its support is a small sphere $S^{d-p-1}$,
    located away from $\cE^p$.
    \item Remaining inside $B^d_{\rm sym}$, we move the support of 
    $D^{(\alpha)}_{d-p-1}$ until it crosses $\cE^p$.
    As a result, the support
    of $D^{(\alpha)}_{d-p-1}$ is now a small sphere linking $\cE^p$
    in $B^d_{\rm sym}$.
    In transitioning into a linked configuration,
    we pick up the non-trivial linking factor $e^{2\pi in\alpha}$.
    \item We slide  $D^{(\alpha)}_{d-p-1}$
    parallel to $\cE^p$,
    until $D^{(\alpha)}_{d-p-1}$ touches the interface $C^{d-1}_{\rm sym}$.
    \item Since $C^{d-1}_{\rm sym}$ supports the
    $\cD_A$-$\cD_B$ interface,
    the operator $D^{(\alpha)}_{d-p-1}$
    gets absorbed.
\end{enumerate}
Retracing the steps above, we infer the following identity,
\be 
\langle \cO_p \rangle = e^{2\pi in\alpha} \langle \cO_p \rangle \ . 
\ee 
Since this must hold for generic $\alpha$, we conclude
\be 
\langle \cO_p \rangle = 0 \ . 
\ee 
We thus recover the fact that, if we have a single boundary condition that preserves the $U(1)$ $p$-form symmetry,
the 1-point function of a large order parameter scales to zero.

\subsubsection*{Broken $p$-form symmetry, $p\ge1$} 

In this case, the setup is very similar to the one of the previous subsection, up to the following modifications.
To realize the charged $p$-dimensional operator we still use a stretched $\mathbf W^{(n)}_{p+1}$ operator. This time, however,
its support consists only of the first piece $\Sigma^{p+1}_1$, see \eqref{eq_Sigma_first_piece}. As noted above, this piece ends perpendicularly on the lateral boundary
$B^d_{\rm lat}$.
The latter is now equipped with the $\cD_A$ boundary condition, so $\mathbf W^{(n)}_{p+1}$ can end perpendicularly on 
$B^d_{\rm lat}$
and there is no need for the second piece
$\Sigma^{p+1}_2$ in \eqref{eq_Sigma_second_piece}.
We still use the notation
$\cE^p$ for the locus where
$\Sigma^{p+1}_1$ ends inside $B^d_{\rm sym}$, cfr.~\eqref{eq_Ep_endpoint}.

After these preliminaries,
we perform topological moves as in the steps 1.~through 3.~above.
Step 4.~is replaced by 
\begin{itemize}
    \item[4.${}^\prime$]
    Since $C^{d-1}_{\rm sym}$ supports a 
    $\cD_A$-$\cD_A$ interface, the operator $D^{(\alpha)}_{d-1}$ dresses the 
    pre-existing $\cD_A$-$\cD_A$ interface, turning into  a different $\cD_A$-$\cD_A$ interface. 
\end{itemize}
Retracing the above steps,
we infer an  identity of the form
\be \label{eq_ward_p_broken}
\langle \cO_p \rangle_{v} = e^{2\pi in \alpha} \langle \cO_p \rangle_{v'}  \ . 
\ee 
As in the 0-form case,
we have added a label $v$, $v'$ to stress that we have a collection of boundary conditions. The label $v'$
stands for the boundary condition that is obtained from $v$ by dressing with the $U(1)$ 
topological operator
$D^{(\alpha)}_{d-p-1}$.
The support of this operator is a small sphere
$S^{d-p-1}$ that links, inside $\partial M^d \cong S^{d-1}$, with the end-locus
$\partial B^p \cong S^{p-1}$
of $\cO_p$ on the boundary
$\partial M^d$ of physical spacetime.
The identity 
\eqref{eq_ward_p_broken} relates 1-point function with different boundaries and is compatible with non-zero 1-point functions for large $p$-dimensional order parameters.

\subsection{Comments on the non-Abelian case}
\label{sec_corners_nonAb}

Let us now comment on the analog of the analysis of Section \ref{sec_ward_from_symtft} when the 0-form symmetry group $G$ is non-Abelian.
The bulk TQFT is the BF theory with group $G$.
We  have the direct analog of the topological boundary conditions $\cD_A$ and $\cD_B$,   as well $\cD_A$-$\cD_A$ and $\cD_A$-$\cD_B$ interfaces. They are constructed explicitly in Appendix \ref{app_nonAb_boundary_conditions}.

In the non-Abelian case we can still consider
the configuration  in Figure \ref{fig_0_form_ward}.
The Wilson line stretched between $B^d_{\rm sym}$ and $B^d_{\rm phys}$ is labeled by an irreducible representation $\mathbf R$ of $G$.
The operator $D^{(\alpha)}_{d-1}$ of the Abelian setting is replaced now by the operator
$D_{X_0}$, $X_0 \in \mathfrak g$, described
in Section \ref{sec_non_genuine_B}. We emphasize that, contrary to the Abelian case, 
$D_{X_0}$ is not realized by taking a genuine topological operator in the bulk and  projecting it onto $B^d_{\rm sym}$.
(As we have seen in Section \ref{sec_non_genuine_B}, it is possible to think of $D_{X_0}$ as limiting case of a non-genuine bulk operator.)
This point, however, does not modify the logic of the arguments of Section \ref{sec_ward_from_symtft} in the derivation of Ward identities.
Indeed, the essential point in the process depicted in Figure \ref{fig_0_form_ward} is that
we have a non-trivial linking between
the topological operator inside $B^d_{\rm sym}$
and the endpoint $\cE^0$ of the stretched Wilson line.
This is indeed the case, with the expression for the linking factor being
\be 
\frac{\chi_{\mathbf R}(e^{ 2\pi X_0})}{
\chi_{\mathbf R}(1)}  \ , 
\ee 
see \eqref{eq_linking_factor}.
Thus we can repeat the same steps as above in the case of unbroken symmetry.
We get a Ward identity line of the form
\be 
\langle \cO \rangle = \frac{\chi_{\mathbf R}(e^{ 2\pi X_0})}{
\chi_{\mathbf R}(1)}  
\langle \cO \rangle \ . 
\ee 
This should hold for arbitrary $X_0 \in \mathfrak g$, showing that 
the 1-point function of $\langle \cO \rangle$ vanishes.
In a similar fashion, if we repeat the arguments in the case of broken symmetry,
we get a relation of the form
\be 
\langle \cO \rangle_v = \frac{\chi_{\mathbf R}(e^{ 2\pi X_0})}{
\chi_{\mathbf R}(1)}  
\langle \cO \rangle_{v'} \ . 
\ee
The labels $v$, $v'$ corresponds to different boundary conditions, with $v'$ begin obtained from $v$ by the action of the $G$ group element $e^{2\pi X_0}$.
This time, the identity is compatible with a non-zero 1-point function.

\subsection{Remarks on vacuum degeneracies and Goldstone modes} 
\label{sec_Goldstone}

An important feature of the 
 SymTFT construction with boundaries and corners
outlined in Section \ref{sec_corners} is 
the existence of families of topological
interfaces at $C^{d-1}_{\rm sym}$. 
Crucially, in the context of continuous symmetries, we find continuous families
of interfaces.
Upon closing the sandwich, $C^{d-1}_{\rm sym}$ fuses with the physical corner $C^{d-1}_{\rm phys}$,
yielding the boundary condition for the physical theory $\cT_d$.
Let us assume that distinct topological
interfaces on $C^{d-1}_{\rm sym}$ give rise in this way to distinct boundary conditions for $\cT_d$. 
If this is the case, the physical theory possesses a continuous family of boundary conditions at infinite distance.\footnote{\, Here we stress our assumption that the space $M^d$ is non-compact and the boundary $\partial M^d$ is located at infinite distance is key to access the low-energy physics of the system. If one instead is interested in a space $M^d$ with boundaries at finite distance, our construction works \textit{verbatim}, and one still obtains an action of the symmetry on the boundary conditions, but its physical interpretation changes, because $C^{d-1}_{\rm phys}$ is no longer in correspondence with the vacua of $\mathcal T_d$ anymore.} 
Moreover, from the construction outlined above, we have an action of the topological defects corresponding to the continuous symmetry transformations on this collection of boundary conditions. This implies that the resulting Euclidean boundary conditions transform non-trivially with respect to the action of the symmetry, a hallmark of spontaneous symmetry breaking. Since boundary conditions at infinity in the Euclidean formulation are mapped to vacua in the Minkowskian formulation, this is the Euclidean counterpart
of having a continuous family of degenerate vacua in Minkowski spacetime, acted upon by the broken symmetry. Let us consider for definiteness the case of a 
0-form symmetry. Then, it is known that, if the spacetime dimension $d$ is greater than 2,
a continuous family of degenerate vacua
implies the existence of a gapless Goldstone mode.
Heuristically, the latter can be thought 
of  as an excitation
describing an infinitesimal motion 
in the family of vacua.

\medskip

We can use our results in Section \ref{sec_corner_abelian_case} to make the picture outlined above more concrete for Abelian $p$-form symmetries. As discussed there, we have a unique interface corresponding to the unbroken symmetry, $\cD_A$-$\cD_B$, while we find a continuous family of $\cD_A$-$\cD_A$ interfaces \eqref{eq_interface_DA_DA} that correspond to the case of broken symmetry. This is confirmed by the fact that the former can absorb the topological defects corresponding to symmetry transformations while the latter transform non-trivially with respect to such an action. In particular, we propose that the $\cD_A$-$\cD_A$ interfaces are described with an action 
\be
S_{\rm int}^{\cD_A \,\cD_A} = \frac{1}{2\pi} \int_{Z^{d-1}}  B_{d-p-1} \wedge (\varphi_p^{(2)} - \varphi_p^{(1)}
- d\nu_{p-1}
- \xi_p  )   \ . 
\ee 
which is in particular labeled by a $p$-form $\xi_p$. Let's comment here about the physical interpretation of $\xi_p$. We have seen that, if we bring a symmetry operator to the interface, 
the value of the parameter $\xi_p$
is shifted as in \eqref{eq_app_DD_xi_shiftBIS}, which we copy here for convenience
\be
\xi_p \rightarrow \xi'_p = \xi_p + 2\pi \alpha \delta_p(\Sigma^{d-p-1} \subset Z^{d-1}) \ .
\ee 
For simplicity, let us 
switch off the classical backgrounds $\cA_{p+1}^{(1)}$, $\cA_{p+1}^{(2)}$
on the $\cD_A$ boundaries.
Then, the form $\xi_p$ is closed.
Moreover, by a field redefinition of $\nu_{p-1}$
in the path integral for the fields supported on the $\cD_A$-$\cD_A$ interface, we can shift $\xi_{p}$ by any closed form with periods in $2\pi \mathbb Z$. In summary, the actually 
inequivalent interfaces are really parametrized by a closed $p$-form modulo
closed $p$-forms with periods in $2\pi \mathbb Z$. 

\medskip

This is an important property that these interfaces have, which matches our physical intuition in the case of 0-forms. Indeed, let us specialize to $p=0$,
the parameter $\xi_{p=0}$ is a real number and the inequivalent choices of $\cD_A$-$\cD_A$ interfaces in this case are parametrized by $\mathbb R/(2\pi \mathbb Z)$. Upon closing the sandwich, we get a family of boundary conditions
for the theory $\cT_d$ parametrized by
$\mathbb R/(2\pi \mathbb Z)$. This 
is consistent with the fact that, translating to a Minkowskian setting,
the spontaneous breaking of a global $U(1)$ 0-form symmetry leads to a degenerate family of vacua which is copy of $\mathbb R/(2\pi \mathbb Z)$. The resulting Goldstone mode (for $d>2$) is a compact scalar field as expected.
 
\medskip

In the case of higher $p$-form symmetries, the fact that inequivalent boundary conditions in the Euclidean are labeled by a $p$-form modulo closed $p$-forms with periods in $2\pi \mathbb Z$ is consistent with the expectation that the corresponding Nambu-Goldstone mode is a higher $p$-form Abelian gauge field.

\medskip

Non-Abelian 0-form symmetries   proceed in a similar manner. In this work we have constructed the interfaces corresponding to an unbroken symmetry and to a symmetry which is fully broken. Again in the former case there is a single interface, while in the latter case we find interfaces labeled by a $G$-valued parameter. This is consistent with the expected structure for the Nambu-Goldstone sigma model, which in this case is precisely $G$. We present the details in Appendix \ref{app_nonAb_boundary_conditions}. Generalizing this to partial breaking $G\to H$ is an interesting challenge we will address in a 
follow-up work.

\section{Outlook}

The results of this note suggest various possible directions for future investigations.
In Section \ref{sec_higher_linking} we have analyzed a case study 
of higher-linking which is detectable in Horowitz
non-Abelian BF theory. It would be interesting to explore these matters more systematically.
Another compelling open problem is to    
characterize the fusion structure of 
the topological operators of continuous non-Abelian
BF theory.
It could also be interesting to explore potential applications of our proposal for a continuous SymTFT that allows for non-flat background gauge fields.

In the last part of this note we have 
initiated an exploration of
the topic of spontaneous symmetry breaking (SSB)
of continuous symmetries within the framework of
SymTFTs formulated with boundaries and corners.
Several point deserve further study.
For example, 
it would be interesting to revisit the SymTFT description of continuous higher-groups, with the aim to
explore   symmetries with a more complex higher structure. In particular, it can happen that 0-form symmetries are not independent from their higher counterparts, either because of a fusion rule to a condensate of 1-form symmetries, such as in the case of duality defects in four dimensions -- see e.g. \cite{Kaidi:2021xfk,Choi:2021kmx,Choi:2022zal} for abelian cases and \cite{Bashmakov:2022uek,Antinucci:2022cdi} for non-Abelian examples -- or because of a higher group structure mixing 0-form and 1-form symmetries in such a way that only the 1-form symmetry forms a proper subgroup, see e.g. \cite{Kapustin:2013uxa,Benini:2018reh,Cordova:2018cvg,Cordova:2020tij,Lee:2021crt}. In this context we know that there must be a hierarchy in the SSB, meaning that the 1-form symmetry cannot be spontaneously broken if the 0-form symmetry is not.
We plan to explore how the SymTFT with symmetries and corners encodes such hierarchies.

\acknowledgments

We thank Andrea Antinucci, Alberto Cattaneo, and Diego Rodriguez-Gomez for useful discussions. The work of MDZ is supported by the European Research Council (ERC) under the European Union’s Horizon Europe research and innovation program (grant agreement No. 101171852) and by the VR project grant No. 2023-05590. MDZ also acknowledges the VR Centre for Geometry and Physics (VR grant No. 2022-06593) and support from the Simons Foundation Grant \#888984  (Simons Collaboration on Global Categorical Symmetries). 
FB is supported by 
the Program ``Saavedra Fajardo'' 22400/SF/23.
FB also acknowledges support from
Fundaci\'on S\'eneca de la Regi\'on de Murcia FSRM/10.13039/100007801 (22581/PI/24), Espa\~na. Part of this work was completed during the Kavli Institute for
Theoretical Physics (KITP) program “Generalized Symmetries in Quantum Field Theory:
High Energy Physics, Condensed Matter, and Quantum Gravity”, which is supported in
part by grant NSF PHY-2309135 to the KITP.

\appendix
\addtocontents{toc}{\protect\setcounter{tocdepth}{1}}

\section{The example $G=SU(2)$, $H=U(1)$}
\label{app_GH_example}

In this appendix we 
give explicit expressions for the quantities introduced in Section \ref{sec_club_sandwich}
in the illustrative case where
$G=SU(2)$ and $H$ is its standard Cartan torus, consisting of diagonal matrices.

 We choose the anti-hermitian generators 
\be 
T_1 = \tfrac i2 \sigma_x \ , \qquad 
T_2 = \tfrac i2 \sigma_y \ , \qquad 
T_3 = \tfrac i2 \sigma_3 \ , 
\ee
where $\sigma$'s are the Pauli matrices. The decomposition
$\mathfrak g = \mathfrak h \oplus \mathfrak h^\perp$ is 
\be 
\mathfrak h = {\rm span}(T_3) \ , \qquad 
\mathfrak h^\perp = {\rm span}(T_1,T_2) \ . 
\ee 
The $G$ gauge field $A$ is written as
\be 
A = a + \mathscr A \ , \qquad 
a = a^3 T_3 \ , \qquad 
\mathscr A = \mathscr A^1 T_1 + \mathscr A^2 T_2 \ . 
\ee 
We consider a gauge transformation with parameter $g$ restricted in $H \subset G$,
\be  \label{eq_chi_gauge_transf}
g = e^{\chi T_3} \ , \qquad 
A'= g(A+d)g^{-1}  \ , 
\qquad 
\begin{array}{l}
(a^3)' = a^3 - d\chi \  , \\
(\mathscr A^1 + i \mathscr A^2)' = e^{-i \chi} (\mathscr A^1 + i \mathscr A^2) \ . 
\end{array}
\ee 
The flatness condition $F_A=0$ written in terms of the components $a^3$, $\mathscr A^{1,2}$ reads
\be 
da^3 = \mathscr A^1 \wedge \mathscr A^2 \ , \qquad 
d(\mathscr A^1 + i \mathscr A^2) - i a^3 \wedge (\mathscr A^1 + i \mathscr A^2)  = 0 \ .
\ee 
The RHS of the first equation
corresponds to the term
$\tfrac 12 {\rm proj}_{\mathfrak h^\perp} [\mathscr A, \mathscr A]$ in the general discussion around \eqref{eq_total_flatness_pieces}.
In contrast, in this example
the piece $\tfrac 12 {\rm proj}_{\mathfrak h} [\mathscr A, \mathscr A]$ is zero, because $G/H$ is a symmetric space.

Let us also record an explicit parametrization of the $G$-valued 0-form $\cV$,
\be 
\cV = e^{\phi T_3} e^{\theta T_1} e^{\psi T_3} \ , \qquad \phi \in [0,2\pi), \quad 
\theta \in [0,\pi] \ , \quad 
\psi \in [0,4\pi) \ .
\ee 
This leads to $\cV^{-1} d\cV = P+Q$ with
\be 
\ba 
P &= \tfrac 12 e^{-i \psi} (d\theta + i \sin \theta d\phi) (T_1 + i T_2)
+\tfrac 12 e^{i \psi} (d\theta - i \sin \theta d\phi) (T_1 - i T_2) \ , \\
Q & = (d\psi + \cos \theta d\phi) T_3 \ .
\ea
\ee
On $B^d_{\rm sym}$ we have $A = \cV^{-1} d\cV$, which more explicitly gives the identifications
\be 
\text{on $B^d_{\rm sym}$:} \qquad 
a^3 = d\psi + \cos \theta d\phi \ , \qquad 
\mathscr A + i \mathscr A^2
=  e^{i \psi} (d\theta - i \sin \theta d\phi) \ .
\ee 
The leading term in the interface action is proportional to 
\be 
{\rm Tr} (\mathscr A \wedge *_d \mathscr A)
= - \tfrac 12 (\mathscr A^1  \wedge *_d \mathscr A^1 + \mathscr A^2 \wedge *_d \mathscr A^2) \ . 
\ee 
When we close the first slab on the club sandwich,
the above gives rise to the leading term 
in the $G/H$ effective action, namely
the term that captures the sigma model into $G/H \cong S^2$,
\be 
{\rm Tr} (\mathscr A \wedge *_d \mathscr A)={\rm Tr} (P \wedge *_d P) = - \tfrac 12 (d\theta \wedge *_d d\theta + \sin^2 \theta d\phi \wedge *_d d\phi) \ . 
\ee 
We want to be able to explore the entire coset $G/H$ without imposing any constraint on the dynamics of $\theta$, $\phi$. As a result,
$da^3 = - \sin \theta d\theta \wedge d\phi$
is non-zero, and it would be unnatural to 
impose $da^3 =0$ by hand.

On the second slab in this case we have Maxwell
theory with a single $U(1)$ gauge field
$\widehat A$. On the interface, it is matched with $a^3$,  
\be 
\text{on $I^d_{\rm int}$:} \qquad a^3 = \widehat A \  .
\ee 
On the physical boundary we generically have fields $\Psi$ transforming in linear representations of $H$. 
 For example, $\Psi$
could be a complex scalar of charge $q$,
\be 
\widehat A' = \widehat A' - d\chi 
 \ , \qquad \Psi' = e^{i q \chi} \Psi  \ . 
\ee 
Then, $\Psi$ couples to the $H$ gauge field $\widehat A$ in the second slab via the covariant derivative  
\be
d_{\widehat A} \Psi = d\Psi + i q \widehat A \Psi \ . 
\ee 
Ultimately, when both slabs are collapsed,
we have
$\widehat A = a^3 = Q$,
so that $\Psi$ couples to the scalars
$\theta$, $\phi$ via
\be 
d_Q \Psi = d\Psi + i q (d\psi + \cos \theta d\phi)
\Psi  
\ . 
\ee

\section{Boundary conditions and interfaces}
\label{app_Dir_and_Neu}

This appendix discusses in greater detail
the topological boundary conditions and interfaces among them introduced in the main text.

\subsection{Abelian setting}

We consider the bulk TQFT
\be 
S_{\rm bulk} = \frac{1}{2\pi} \int_{X^{d+1}}
B_{d-p-1} \wedge dA_{p+1}  \ , 
\ee 
where $B_{d-p-1}$ is an $\mathbb R$ gauge field,
$A_{p+1}$ is a $U(1)$ gauge field. Their gauge transformations read
\be \label{eq_app_Dir_gauge}
A_{p+1} \rightarrow A_{p+1} + d\Lambda_p \ , \qquad 
B_{d-p-1} \rightarrow B_{d-p-1} + d\Omega_{d-p-2} \ , 
\ee 
with $\Lambda_p$ a $U(1)$ gauge field and
$\Omega_{d-p-2}$ a globally-defined form.
We report here the gauge transformation of the bulk action,
and its variation under arbitrary variations of the fields,
\be \label{eq_app_bulk_reference}
\ba 
\Delta S_{\rm bulk} & = \frac{1}{2\pi} \int_{\partial X^{d+1}} (-1)^{d-p-1} d\Omega A
\ , \\ 
\delta S_{\rm bulk} & = \frac{1}{2\pi} \int_{X^{d+1}} \bigg[ 
\delta B dA + (-1)^{d-p} dB \delta A
\bigg] 
+ \frac{1}{2\pi} \int_{\partial X^{d+1}}
(-1)^{d-p-1} B \delta A \ . 
\ea 
\ee 
Here, and in several expressions below, we suppress form degrees and wedge products for brevity.
As usual, the bulk terms in the variation impose the equations of motion
\be 
\text{bulk:} \qquad 
dA_{p+1} = 0  \ , \qquad 
dB_{d-p-1} = 0  \ .
\ee

\paragraph{$\cD_A$ boundary condition.}
The boundary action reads
\be \label{eq_app_Dir_action}
S_{\rm bdy}^{\cD_A} = \frac{1}{2\pi} \int_{Y^d}
(-1)^{d-p} B_{d-p-1} \wedge (A_{p+1} - 
\cA_{p+1} -
d\varphi_p) \ . 
\ee 
The manifold $Y^d$ can either be the entire boundary $\partial X^{d+1}$, or one of its components. 
The field $\varphi_p$ is localized on
$Y^d$. It is a $U(1)$ gauge field,
with gauge transformation
\be 
\varphi_p \rightarrow \varphi_p + d \sigma_{p-1} + \Lambda_{p} \ , 
\ee 
where $\sigma_{p-1}$ is a $U(1)$ gauge field localized on $Y^d$, while $\Lambda_p$ is the same parameter as in the bulk gauge transformation \eqref{eq_app_Dir_gauge}.
The quantity $\cA_{p+1}$ is 
a non-dynamical form on
$Y^d$, which we take to be closed,
\be
d\cA_{p+1} = 0 \ . 
\ee  
In our approach, $\cA_{p+1}$ does not vary under gauge transformations of the dynamical fields in the bulk and boundary.

Let us report the gauge transformation of the boundary action,
\be 
\Delta S_{\rm bdy}^{\cD_A} =  \frac{1}{2 \pi} \int_{Y^d} (-1)^{d-p} d\Omega A
+ \frac{1}{2 \pi}\int_{\partial Y^d} (-1)^{d-p-1} \Omega(\cA + d\varphi) \ .
\ee 
The piece on $\partial Y^d$ is relevant if 
the boundary $\partial X^{d+1}$ consists of several components,
joined at some interface.
We observe that 
the $d\Omega A$ term cancels against the contribution from the bulk, see \eqref{eq_app_bulk_reference}.
Next, we write the variation of the boundary action under arbitrary variations of the fields,
\be 
\delta S_{\rm bdy}^{\cD_A} = \frac{1}{2 \pi} \int_{Y^d} 
\bigg[ 
(-1)^{d-p} \delta B (A - \cA - d \varphi)
+ (-1)^{d-p} B \delta A
- dB \delta \varphi
\bigg] 
+ \frac{1}{2 \pi}\int_{\partial Y^d} B 
\delta \varphi  \ . 
\ee 
The $B \delta A$ cancels against the bulk, see \eqref{eq_app_bulk_reference}. The term with $\delta \varphi$ confirms the bulk equation of motion $dB=0$. The term with $\delta B$ imposes
\be 
\text{on $Y^d$:} \qquad 
A_{p+1} = \cA_{p+1} + d\varphi_p \ . 
\ee

\paragraph{$\cD_B$ boundary condition.}
The boundary action reads\be \label{eq_app_Neu_action}
S_{\rm bdy}^{\cD_B} = \frac{1}{2\pi} \int_{Y^d}
(-1)^{d-p} (d\Phi_{d-p-2}  + \cB_{d-p-1})\wedge  A_{p+1} \ . 
\ee 
The field $\Phi_{d-p-2}$ is localized on
$Y^d$. It is an $\mathbb R$ gauge field,
with gauge transformation
\be 
\Phi_{d-p-2} \rightarrow \Phi_{d-p-2} + d \rho_{d-p-3} + \Omega_{d-p-2} \ , 
\ee 
where $\rho_{d-p-3}$ is a globally-defined form on $Y^d$, while $\Omega_{d-p-2}$ is the same parameter as in the bulk gauge transformation \eqref{eq_app_Dir_gauge}.
The quantity $\cB_{d-p-1}$ is a non-dynamical form on $Y^d$, which we take to be closed,
\be 
d \cB_{d-p-1} = 0 \ . 
\ee
In our approach, $\cB_{d-p-1}$ does not vary under gauge transformations of the dynamical fields in the bulk and boundary.

The gauge transformation of the boundary action reads
\be 
\Delta S_{\rm bdy}^{\cD_B} = 
\frac{1}{2\pi} \int_{Y^d}  
(-1)^{d-p} d\Omega A
+ \frac{1}{2\pi} \int_{\partial Y^d} \bigg[ 
(-1)^{d-p} \Phi d\Lambda
+(-1)^{d-p} \Omega d\Lambda
- \cB \Lambda
\bigg] \ . 
\ee 
The $d\Omega A$ term cancels against the bulk, see \eqref{eq_app_bulk_reference}.

The variation of the action under arbitrary variations of the fields is 
\be 
\delta S_{\rm bdy}^{\cD_B} = \frac{1}{2\pi} \int_Y \bigg[ 
(-1)^{d-p}(d\Phi + \cB) \delta A
- \delta \Phi dA
\bigg]
+ \frac{1}{2\pi} \int_{\partial Y}
(-1)^{d-p} \delta \Phi A
\ee 
The term with $\delta \Phi$ confirms the bulk equation of motion $dA = 0$. The terms with $\delta A$ combine with the bulk \eqref{eq_app_bulk_reference} to give
\be 
\text{on $Y^d$:} \qquad 
B_{d-p-1} = \cB_{d-p-1} + d\Phi_{d-p-2} \ . 
\ee

\paragraph{$\cD_A$-$\cD_B$ interface.}

Let us now partition the boundary $Y^d$ into two regions, meeting along an interface $Z^{d-1}$,
\be 
Y^d = Y^d_{1} \cup 
Y^d_{2} \ , \qquad 
\partial Y^d_{1}  = Z^{d-1}
= - \partial Y^d_{2} \ . 
\ee 
We impose the $\cD_A$ boundary condition
\eqref{eq_app_Dir_action} on $B^d_{\rm bdy,1}$,
and the $\cD_B$ boundary condition
\eqref{eq_app_Neu_action} on $B^d_{\rm bdy,2}$.
For the interface, we propose the action
\be 
S_{\rm int}^{\cD_A \, \cD_B} = \frac{1}{2\pi} \int_{Z^{d-1}} 
\bigg[ 
(-1)^{d-p} \Phi_{d-p-2} \wedge d\varphi_p 
+ (-1)^{d-p} \Phi_{d-p-2} \wedge \cA_{p+1}
- \cB_{d-p-1} \wedge \varphi_p 
\bigg] 
\ . 
\ee 
Its gauge transformation  reads
\be 
\ba 
\Delta S_{\rm int}^{\cD_A \, \cD_B} = \frac{1}{2\pi} \int_{Z^{d-1}}
\bigg[ 
(-1)^{d-p} \Omega (d\varphi + \cA)
+ (-1)^{d-p} \Phi d\Lambda
+ (-1)^{d-p} \Omega d\Lambda
- \cB \Lambda 
+ (-1)^{d-p} d\rho \cA
- \cB d\sigma
\bigg]  \ . 
\ea 
\ee
The term with $\rho$ can be removed by integration by parts ($Z^{d-1}$ has no boundary) using closure of $\cA$ and the fact that $\rho$ is globally defined.
To deal with the term with $\sigma$,
we require that $\sigma$ is topologically trivial
(globally defined form) on the interface.
To get the  gauge transformation of the total action we also have to include the 
 $\partial Y^d$ terms from
the gauge transformations
of $S_{\rm bdy}^{\cD_A}$,
$S_{\rm bdy}^{\cD_B}$
(with a relative minus sign, due to orientation),
\be 
\frac{1}{2\pi} \int_{Z^{d-1}} \bigg[ 
(-1)^{d-p-1} \Omega(\cA + d\varphi)
+(-1)^{d-p-1} \Phi d\Lambda
+(-1)^{d-p-1} \Omega d\Lambda
+ \cB \Lambda
\bigg]  \ . 
\ee 
We verify a total cancellation, showing that the full action is gauge invariant.

Next, we consider the variation of the interface action under variation of the fields,
\be 
\delta S_{\rm int}^{\cD_A \, \cD_B} = \frac{1}{2\pi} \int_{Z^{d-1}} \bigg[ 
(-1)^{d-p} \delta \Phi
d\varphi
- d\Phi \delta \varphi
+ (-1)^{d-p} \delta \Phi \cA
- \cB \delta \varphi
\bigg]  \ . 
\ee 
We also need to add the $\partial Y^d$ terms from the variations of the $\cD_A$ and $\cD_B$ boundary actions,
\be 
\frac{1}{2\pi} \int_{Z^{d-1}} 
\bigg[ 
B \delta \varphi
+ (-1)^{d-p-1} \delta \Phi A
\bigg]  \ . 
\ee 
Combining all terms, we get 
\be 
\frac{1}{2\pi} \int_{Z^{d-1}} 
\bigg[ 
(B - \cB - d\Phi) \delta \varphi
+ (-1)^{d-p-1} \delta \Phi (A - \cA - d\varphi)
\bigg]  \ . 
\ee 
It follows that the interface terms in the variation simply confirm the relations
$A = \cA + d\varphi$,
$B =\cB + d\Phi$ on the two components of the boundary, without introducing any new   constraint.

\paragraph{$\cD_A$-$\cD_A$ interface.}

We proceed as above, but we impose the $\cD_A$ boundary condition
\eqref{eq_app_Dir_action} on both $B^d_{\rm bdy,1}$ and  $B^d_{\rm bdy,2}$.
Now we have two localized fields
$\varphi_p^{(1,2)}$ on $B^d_{\rm bdy,1}$ and  $B^d_{\rm bdy,2}$, with gauge transformations
\be 
\varphi^{(i)}_p \rightarrow 
\varphi^{(i)}_p  + d\sigma^{(i)}_{p-1}
+ \Lambda_p \ , 
\ee 
with the same $\Lambda_p$ from the bulk,
but  distinct $\sigma_{p-1}^{(i)}$.
We also have two classical quantities
$\cA^{(i)}$.
For the interface, we propose the action
\be \label{eq_app_DirDir}
S_{\rm int}^{\cD_A \, \cD_A} = \frac{1}{2\pi} \int_{Z^{d-1}}  B_{d-p-1} \wedge (\varphi_p^{(2)} - \varphi_p^{(1)}
- d\nu_{p-1}
- \xi_p  )   \ . 
\ee 
Here $\nu_{p-1}$ is a $U(1)$ gauge field localized on $Z^{d-1}$, with gauge transformation
\be 
\nu_{p-1} \rightarrow \nu_{p-1} + du_{p-2} + \sigma^{(2)}_{p-1} - \sigma^{(1)}_{p-1} \ ,
\ee 
where the parameter $u_{p-2}$ is a $U(1)$ gauge field on $Z^{d-1}$.
The quantity $\xi_p$ in \eqref{eq_app_DirDir} 
is 
a non-dynamical form on $Z^{d-1}$, which we take such that
\be \label{eq_app_dxi}
\text{on $Z^{d-1}$:}
\qquad 
d\xi_p = \cA^{(1)}_{p+1} - \cA^{(2)}_{p+1} \ . 
\ee 
We regard $\xi_p$ as a parameter that 
labels various possible $\cD_A$-$\cD_A$ interfaces.

The gauge variation of the action $S_{\rm int}^{\cD_A \, \cD_A}$ is 
\be 
\Delta S_{\rm bdy}^{\cD_A \, \cD_A} = 
\frac{1}{2\pi} \int_Z 
d\Omega (\varphi^{(2)} - \varphi^{(1)}
- d\nu - \xi
) \ , 
\ee 
or equivalently, upon integration by parts,
\be 
\Delta S_{\rm bdy}^{\cD_A \, \cD_A} = 
\frac{1}{2\pi} \int_Z 
(-1)^{d-p-1}\Omega (d\varphi^{(2)} - d\varphi^{(1)}
- \cA^{(1)} + \cA^{(2)}
) \ . 
\ee 
These terms combine with the $\partial Y^d$
terms from the two boundary components,
\be 
\frac{1}{2\pi} \int_{Z^{d-1}} \bigg[ 
(-1)^{d-p-1} \Omega (\cA^{(1)} + d\varphi^{(1)}
- \cA^{(2)} - d\varphi^{(2)})
\bigg]  \ . 
\ee 
We observe a complete cancellation, showing that the total action is gauge invariant.

Now we consider the variation of $S_{\rm int}^{\cD_A \, \cD_A}$ under arbitrary variations of the fields,
\be 
\delta S_{\rm int}^{\cD_A \, \cD_A}=
\frac{1}{2\pi} \int_{Z^{d-1}} \bigg[ 
\delta B (\varphi^{(2)}
- \varphi^{(1)}
- d\nu - \xi
)
+ B (\delta \varphi^{(2)}
- \delta \varphi^{(1)}
)
+(-1)^{d-p-1} dB \delta \nu
\bigg]  \ . 
\ee 
This combines with the $\partial Y^d$ terms from the two boundary components,
\be 
\frac{1}{2\pi} \int_{Z^{d-1}} \bigg[ 
B \delta \varphi^{(1)}
- B \delta \varphi^{(2)}
\bigg] \ . 
\ee 
The terms with $\delta \varphi^{(i)}$ cancel.
The term with $\delta \nu$ confirms the bulk equation of motion $dB=0$. The terms with $\delta B$ enforce
\be 
\text{on $Z^{d-1}$:} \qquad 
\varphi^{(2)}_p = \varphi^{(1)}_p + d\nu_{p-1} + \xi_p \ . 
\ee

Suppose we dress a $\cD_A$-$\cD_A$ interface with parameter $\xi_p$ with a closed Wilson operator for $B_{d-p-1}$, of the form
$e^{i \alpha \int_{\Sigma^{d-p-1}} B_{d-p-1}}$,
with $\Sigma^{d-p-1}$ a cycle in $Z^{d-1}$. The new action for the interface is 
\be 
\frac{1}{2\pi} \int_{Z^{d-1}}\bigg[  B_{d-p-1} \wedge (\varphi_p^{(2)} - \varphi_p^{(1)} - \xi_p  )
+ B_{d-p-1} \wedge 2\pi \alpha \delta_p(\Sigma^{d-p-1} \subset Z^{d-1})
\bigg] \ . 
\ee 
This is equivalent to a new $\cD_A$-$\cD_A$ interface with parameter
\be \label{eq_app_DD_xi_shift}
\xi'_p = \xi_p + 2\pi \alpha \delta_p(\Sigma^{d-p-1} \subset Z^{d-1}) \ .
\ee 
If $\alpha$ is an integer, the new term can be
alternatively reabsorbed by a redefinition of $\nu_{p-1}$. (This is only possible for integer $\alpha$ because after the redefinition we still have that $d\nu_{p-1}$ has periods in $2\pi \mathbb Z$.) 
We observe that the shift  from $\xi_p$
to $\xi_p'$ preserves 
the requirement~\eqref{eq_app_dxi}.

\subsection{Non-Abelian setting}
\label{app_nonAb_boundary_conditions}

We now consider the bulk TQFT
\be 
S_{\rm bulk} = \frac{1}{2\pi} \int_{X^{d+1}}
{\rm Tr}(B \wedge F_A) \ . 
\ee 
For convenience, we repeat the bulk gauge transformations,
\be 
A \rightarrow g(d+A)g^{-1} \ , \qquad 
B \rightarrow g(B-d_A\tau) g^{-1} \ . 
\ee
We also record the gauge transformation of the bulk action,
as well as its variation under arbitrary variations of the fields,
\be \label{eq_app_bulk_refernce_nonAb}
\ba
\Delta S_{\rm bulk} &= \frac{1}{2\pi} \int_{\partial X^{d+1}} {\rm Tr} (- \tau  F_A) \ .
\\
\delta S_{\rm bulk} &= \frac{1}{2\pi} \int_{X^{d+1}}{\rm Tr} (
\delta B F_A + \delta A d_A B
) + \frac{1}{2\pi} \int_{\partial X^{d+1}}
\delta A B \ . 
\ea 
\ee

\paragraph{$\cD_A$ boundary conditions.}
We propose the boundary action
\be \label{eq_app_nonAb_Dir}
S_{\rm bdy}^{\cD_A} = \frac{1}{2\pi} \int_{Y^d} {\rm Tr} \bigg( 
(\cV^{-1} d\cV +\cV^{-1} \cA \cV  -A) \wedge (B + d_A \beta)
+ \beta \wedge F_A
\bigg) \ . 
\ee 
The fields $\cV$ and $\beta$ are localized on $Y^d$, with $\cV$ a $G$-valued 0-form
and $\beta$ a $\mathfrak g$-valued $(d-2)$-form.
Their gauge transformations are 
\be 
\cV \rightarrow \cV g^{-1}  \ , \qquad 
\beta \rightarrow g (\beta + \tau) g^{-1} \ . 
\ee 
The quantity $\cA$ is a non-Abelian $\mathfrak g$-valued 1-form
on $Y^d$ which we take to satisfy 
\be 
0 = \cF_\cA := d\cA + \tfrac 12 [\cA,\cA] \ . 
\ee 
In our approach, $\cA$ does not vary under gauge transformations of the dynamical fields in the bulk and boundary.

The field $\cV$ is the analog of $\varphi_p$ in the Abelian case, while $\beta$ has no analog
(for $G$ Abelian, the $\beta$ terms in the action recombine into a total derivative.)

We define the shorthand notation
\be 
\mathfrak L = \cV^{-1} d\cV +\cV^{-1} \cA \cV  -A \ . 
\ee 
One can verify that, under a gauge transformation,
\be 
\mathfrak L  \rightarrow g \mathfrak L g^{-1} \ . 
\ee 
It follows that 
in \eqref{eq_app_nonAb_Dir} the term
$\mathfrak L (B + d_A \beta)$ is gauge invariant. The second term gives the gauge transformation
\be 
\Delta S_{\rm bdy}^{\cD_A} = \frac{1}{2\pi} \int_{Y^d} {\rm Tr}(\tau F_A) \ . 
\ee 
This cancels the bulk contribution in \eqref{eq_app_bulk_refernce_nonAb}. Thus, the total action is gauge invariant.

Next, we study the variation of $S_{\rm bdy}^{\cD_A}$. We get 
\be 
\ba 
\delta S_{\rm bdy}^{\cD_A} & = \frac{1}{2\pi} \int_{Y^d} {\rm Tr} \bigg[ 
\mathfrak L \delta B
+ d_A\mathfrak L   \delta \beta 
+ \delta \beta F_A
- \delta A B
- \cV^{-1} \delta \cV d_{\mathfrak L -A}(B + d_A \beta)
\bigg] 
\\
& + \frac{1}{2\pi} \int_{\partial Y^d} {\rm Tr} \bigg[ 
- \mathfrak L \delta \beta
+ \delta A \beta
+ \cV^{-1} \delta \cV (B + d_A \beta)
\bigg] \ . 
\ea 
\ee 
A useful identity is 
\be 
d_A \mathfrak L = - \mathfrak L^2 + \cV^{-1} \cF_{\cA} \cV
- F_A \ . 
\ee 
Some obervations are in order. The term with $\delta A$ cancels against the bulk \eqref{eq_app_bulk_refernce_nonAb}.
The term with $\delta B$ imposes $\mathfrak L=0$, namely
\be 
\text{on $Y^d$:} \qquad 
A = \cV^{-1} \cA \cV + \cV^{-1} d\cV \ . 
\ee 
A corollary of this condition is 
$F_A = \cV^{-1} \cF_\cA \cV$, which is compatible with the bulk equation of motion
$F_A=0$ because we demand $\cF_\cA=0$.
The terms with $\delta \beta$ impose 
\be 
0 = d_A \mathfrak L + F_A = - \mathfrak L^2
+ \cV^{-1} \cF_\cA \cV\ , 
\ee 
which is already satisfied  using $\mathfrak L = 0$ and 
$\cF_\cA = 0$.
Finally, the terms with $\delta \cV$, using $\mathfrak L = 0$, impose
\be 
d_A(B + d_A \beta) = 0 \ . 
\ee 
This relation, however, is automatically satisfied thanks to the bulk equations of motion $d_AB = 0$, $F_A = 0$, and the identity
$d_A d_A \beta = [F_A, \beta]$.

We notice that an open Wilson line operator in the representation $\mathbf R$ of $G$ can end on $Y^d$, on an operator of the form $D_{\mathbf R}(\cV)$. A closed Wilson loop, projected parallel onto $Y^d$,   becomes a c-number, because $A$ is gauge-equivalent to the c-number $\cA$  on~$Y^d$.

\paragraph{$\cD_B$ boundary conditions.}
We propose the boundary action
\be 
S_{\rm bdy}^{\cD_B} =  \frac{1}{2\pi} \int_{Y^d} {\rm Tr} \Big( 
 \Phi \wedge F_A
\Big) \ . 
\ee 
The field $\Phi$ is localized on $Y^d$.
It is  $\mathfrak g$-valued $(d-2)$-form,
with   gauge transformation 
\be 
\Phi \rightarrow g (\Phi + \tau) g^{-1} \ . 
\ee 
This field $\Phi$ is the analog
of $\Phi_{d-p-2}$ in the Abelian setting.

The gauge variation of $S_{\rm bdy}^{\cD_B}$
cancels against the boundary term of the gauge variation of $S_{\rm bulk}$, exactly as in the previous case.
Under arbitrary variations of the fields,
\be 
\delta S_{\rm bdy}^{\cD_B} = \frac{1}{2\pi} \int_{Y^d} {\rm Tr} \Big( 
\delta \Phi F_A + \delta A d_A \Phi 
\Big) +
\frac{1}{2\pi} \int_{\partial Y^d} {\rm Tr} \Big( 
\delta A \Phi 
\Big)  \ .
\ee 
The term with $\delta A$ combines with the $\delta A$ term from the bulk variation, to give
$B + d_A\Phi = 0$. We have no new information
from $\delta \Phi$. All in all,
on-shell we have
\be 
\text{on $Y^d$:} \quad 
B + d_A \Phi =0  \ . 
\ee 
We observe that the boundary value of $A$ must be flat, but is otherwise unconstrained.
In contrast, the boundary value of $B$
is a gauge transform of $B = 0$
(with parameters $g=1$, $\tau = \Phi$).

In this non-Abelian setting, it is not obvious how to turn on a non-zero value $\cB$ for $B$ on the boundary.
The boundary value should obey $0=d_A 
\cB = d\cB +[A,\cB]$,
but this equation contains $A$, which is fluctuating/summed over on the boundary.

\paragraph{$\cD_A$-$\cD_B$ interface.}
We propose the action
\be 
S_{\rm int}^{\cD_A \, \cD_B} = \frac{1}{2\pi} \int_{Z^{d-1}} {\rm Tr} (\cV^{-1} d\cV
+\cV^{-1} \cA \cV
-A) (\beta -\Phi) \ . 
\ee 
The total action
(bulk, two boundary components, interface)
is gauge invariant.
For the variational problem, we start collecting the terms
that originate from $S_{\rm bdy}^{\cD_A}$
and $S_{\rm bdy}^{\cD_B}$ when $\partial Y^d$ is nonempty. They read
\be 
\frac{1}{2\pi} \int_{Z^{d-1}}
{\rm Tr}\bigg[ 
- \mathfrak L \delta \beta
+ \delta A \beta
+ \cV^{-1} \delta \cV (B + d_A \beta)
- \delta A \Phi
\bigg]  \  . 
\ee 
These term combine with
\be 
\delta S_{\rm int}^{\cD_A \, \cD_B} = \frac{1}{2\pi} \int_{Z^{d-1}} {\rm Tr}
\bigg(
- \delta A(\beta - \Phi)
+\mathfrak L(\delta \beta - \delta \Phi)
- (\cV^{-1} \delta \cV) d_{\mathfrak L -A}(\beta - \Phi)
\bigg) \ . 
\ee 
We observe that the terms with $\delta A$ and $\delta \beta$ cancel. 
The term with $\delta \Phi$ confirms $\mathfrak L = 0$.
The terms with $\delta \cV$ impose
\be 
0= B + d_A \beta - d_{\mathfrak L-A} \beta
+ d_{\mathfrak L-A} \Phi \ , 
\ee 
but this is automatically zero by using the relations $\mathfrak L=0$
and $B + d_A\Phi=0$ that we have already established on the two components of the boundary. In total, the $Z^{d-1}$ terms 
in the variation of the total action
do not impose any new constraint on the fields.

\paragraph{$\cD_A$-$\cD_A$ interfaces.}
In this setting we have the fields
$\cV_{(1)}$, $\beta_{(1)}$ living on the first component of the boundary,
and $\cV_{(2)}$, $\beta_{(2)}$ living on the second component.
We also have the two classical quantities
$\mathcal A_{(1)}$,
$\mathcal A_{(2)}$.
On the interface, we assume that there exists a classical $G$-valued scalar $\mathcal K$ such that
\be \label{eq_app_NonAb_cK}
\text{on $Z^{d-1}$:} \qquad 
\mathcal A_{(2)}  = \cK^{-1} \mathcal A_{(1)} \cK +  \cK^{-1} d \cK \ . 
\ee 
In other words, at the interface, the value of
$\mathcal A_{(2)}$ 
 is a gauge transform of the value of $\mathcal A_{(1)}$.
The quantity $\cK$ is analogous to the parameter $\xi$ in the Abelian $\cD_A$-$\cD_A$ interface \eqref{eq_app_DirDir},
cfr.~\eqref{eq_app_dxi}, and so in this case we have a $G$-valued family of $\cD_A$-$\cD_A$ interfaces.

At the interface, we
impose the following gluing condition for the values of $\cV_{(1)}$,
$\cV_{(2)}$,
\be \label{eq_app_DD_nonAb_constraint}
\cV_{(1)} \cV^{-1}_{(2)} = \mathcal K  \ . 
\ee 
By taking a variation and a derivative of the constraint \eqref{eq_app_DD_nonAb_constraint}, and using \eqref{eq_app_NonAb_cK}, we get
\be \label{eq_from_VV_constraint}
\cV^{-1}_{(1)} \delta \cV_{(1)} = \cV^{-1}_{(2)} \delta \cV_{(2)} \ , \qquad 
\cV^{-1}_{(1)} d \cV_{(1)} + \cV^{-1}_{(1)} \cA_{(1)} \cV_{(1)} = \cV^{-1}_{(2)} d \cV_{(2)}
+ \cV^{-1}_{(2)} \cA_{(2)} \cV_{(2)}\ .
\ee 
The second relation is compatible with the fact that, on each of the two $\cD_A$ components of the boundary, we have
$A = \cV^{-1}_{(i)} d\cV_{(i)}
+ \cV^{-1}_{(i)} \cA_{(i)}\cV_{(i)}
$.
Next, we revisit the 
expression for $\delta S_{\rm bdy}^{\cD_A}$ in \eqref{eq_app_nonAb_Dir} and we collect the terms on $\partial Y^d$,
with a relative minus sign for the two $\cD_A$ components of the boundary.
We get the terms
\be 
\frac{1}{2\pi} \int_{Z^{d-1}} {\rm Tr} \bigg( 
\delta A \beta_{(1)}
- (\cV^{-1}_{(1)} d\cV_{(1)}
+\cV^{-1}_{(1)} \cA_{(1)}\cV_{(1)}
-A) \delta \beta_{(1)}
+ (\cV^{-1}_{(1)} \delta \cV_{(1)}) (B+d_A\beta_{(1)})
\bigg) - (1 \leftrightarrow 2 ) \ . 
\ee 
If we use \eqref{eq_from_VV_constraint}, these can be cast as
\be \label{eq_app_DD_nonAb_terms}
\frac{1}{2\pi} \int_{Z^{d-1}} {\rm Tr} \bigg( 
\delta A (\beta_{(1)} -\beta_{(2)})
- (\cV^{-1}_{(1)} d\cV_{(1)}
+ \cV^{-1}_{(1)} \cA_{(1)}\cV_{(1)}  -A) (\delta \beta_{(1)}
- \delta \beta_{(2)})
+ (\cV^{-1}_{(1)} \delta \cV_{(1)})d_A(\beta_{(1)}-\beta_{(2)})
\bigg)  \ . 
\ee 
To deal with these terms, we propose to add the following action on the interface,
\be 
S_{\rm int}^{\cD_A \, \cD_A} = \frac{1}{2\pi} \int_{Z^{d-1}} {\rm Tr}\Big[ (\cV^{-1}_{(1)} d\cV_{(1)}
+\cV^{-1}_{(1)} \cA_{(1)}\cV_{(1)}
-A) \wedge (\beta_{(1)} - \beta_{(2)})
\Big] \ . 
\ee 
We could have equivalently written this action using $\cV_{(2)}$ and $\cA_{(2)}$.
The action is invariant under $g$ and $\tau$ transformations.
Its variation under arbitrary variations of 
$A$, $\cV_{(1)}$, $\beta_{(i)}$ reads
\be 
\ba 
\delta S_{\rm int}^{\cD_A \, \cD_A} &= \frac{1}{2\pi} \int_{Z^{d-1}} {\rm Tr} \bigg( 
-\delta A (\beta_{(1)} -\beta_{(2)})
+ (\cV^{-1}_{(1)} d\cV_{(1)}
+ \cV^{-1}_{(1)} \cA_{(1)}\cV_{(1)}
-A) (\delta \beta_{(1)}
- \delta \beta_{(2)})
\\
& - (\cV^{-1}_{(1)} \delta \cV_{(1)})d_{\cV^{-1}_{(1)} d\cV_{(1)}
+ \cV^{-1}_{(1)} \cA_{(1)}\cV_{(1)}
}(\beta_{(1)}-\beta_{(2)})
\bigg) \ . 
\ea 
\ee 
Recalling that
$A = \cV_{(1)}^{-1} d\cV_{(1)}
+ \cV^{-1}_{(1)} \cA_{(1)}\cV_{(1)}$, we have a total cancellation
against the terms in~\eqref{eq_app_DD_nonAb_terms}.

We close with a comparison with the Abelian setting.
A crucial role in the Abelian $\cD_A$-$\cD_A$ interface \eqref{eq_app_DirDir} is played by terms of the form ${\rm Tr}(B_{d-p-1} \varphi^{(i)}_p)$.
There is no obvious analog of such terms in the non-Abelian case. This is why we 
impose the constraint
\eqref{eq_app_DD_nonAb_constraint}. 
Moreover,
we expect a similar phenomenon to \eqref{eq_app_DD_xi_shift},
in which a $\cD_A$-$\cD_A$ interface ``absorbs'' a $B$-defect
and turns into a different 
$\cD_A$-$\cD_A$ interface.

\subsection{$\mathbb Q/\mathbb Z$ symmetry}
\label{app_QmodZ}

The bulk SymTFT is 
\be \label{eq_QmodZ_bulk2}
S_{\rm bulk} = \frac{1}{2\pi} \int_{X^5} \bigg[
B_3 \wedge dA_1 + B_2 \wedge dA_2 + \tfrac 12 \kappa A_1 \wedge B_2 \wedge B_2
\bigg] \ ,
\ee
with the gauge transformations reported in \eqref{eq_QmodZ_gauge}.
The gauge transformation of the action gives a boundary term,
\be 
\Delta S_{\rm bulk} = \frac{1}{2\pi} \int_{\partial X^5} \bigg[ 
\Omega_1 dA_2
+ \Omega_2 dA_1
+ \frac 12 \kappa A_1 \Omega_1 d\Omega_1
- \frac 12 \kappa \Omega_1 d\Lambda_0 d\Omega_1
- \frac 12 \kappa \Lambda_0 B_2 B_2
- \kappa \Lambda_0 B_2 d\Omega_1
\bigg]  \ .  
\ee 
The variation of the action under an arbitrary variations of the fields reads
\be 
\ba 
\delta S_{\rm bulk} &=
\frac{1}{2\pi} \int_{X^5} \bigg[ 
\delta A_1 \Big( 
dB_3 + \tfrac 12 \kappa B_2 B_2
\Big)
- \delta A_2 dB_2 
+ \delta B_2 (dA_2 + \kappa A_1 B_2)
+ \delta B_3 dA_1
\bigg] 
\\
& + \frac{1}{2\pi} \int_{\partial X^5}
\bigg[ \delta A_2 B_2 + \delta A_1 B_3
\bigg]  \ . 
\ea 
\ee 
Next, we discuss the analog of the $\cD_A$ boundary condition for $(A_1, A_2)$.

\paragraph{$\cD_{A_1} \oplus \cD_{A_2}$ boundary.}
The boundary action reads 
\be 
S_{\rm bdy}^{\cD_A} = \frac{1}{2\pi} \int_{Y^4} \bigg[ 
B_3 (A_1  - d\varphi_0)
- B_2 (A_2- \cA_2 - d \varphi_1)
- \frac 12 \kappa \varphi_0 B_2 B_2
\bigg] \ . 
\ee 
The gauge transformations of $\varphi_0$, $\varphi_1$ are reported in \eqref{eq_QmodZ_varphi_gauge}.
The gauge variation of the boundary action, combined with the residual
gauge variation from the bulk action,
gives a total derivative,
\be 
\ba 
\Delta S_{\rm bdy}^{\cD_A}
+ \Delta S_{\rm bulk} = \frac{1}{2\pi} \int_{\partial Y^4} \bigg[ 
A_1 \Omega_2
- A_2 \Omega_1
+ \frac 12 \kappa \varphi_0 \Omega_1 d\Omega_1
+ \frac 12 \kappa \Lambda_0 \Omega_1 d\Omega_1
- \Omega_2 d\varphi_0
+ \Omega_1 d\varphi_1
+ \cA_2 \Omega_1
\bigg]  \ . 
\ea 
\ee 
As above, the terms on $\partial Y^4$ are relevant to study interfaces, and can be dropped if $\partial X^5$ consists of a single component.

The variation of the boundary action combines with the term from the bulk to give
\be 
\ba 
\delta S_{\rm bulk} + \delta S_{\rm bdy}^{\cD_A} & = \frac{1}{2\pi} \int_{Y^4} \bigg[ 
\delta B_3(A_1 - d\varphi_0)
- \delta B_2(A_2  
- \cA_2 - d\varphi_1 + \kappa B_2 \varphi_0)
\\
&
- \delta \varphi_0 (dB_3 + \tfrac 12 \kappa B_2 B_2)
+ \delta \varphi_1 dB_2
\bigg]
 + \frac{1}{2\pi} \int_{\partial Y^4} \bigg[
B_2 \delta \varphi_1 + B_3 \delta \varphi_0
\bigg]
\ea 
\ee 
The $\varphi_0$, $\varphi_1$ variations confirm the bulk equations of motion.
The $B_2$, $B_3$ variations impose
\be 
\text{on $Y^4$:} \qquad
A_1 = d\varphi_0 \ , \qquad 
A_2 = \cA_2 + d\varphi_1 - \kappa B_2 \varphi_0 \ . 
\ee 
These relations are compatible with the bulk EOMs
\be 
dA_1 = 0 \  , \qquad 
dA_2 + \kappa A_1 B_2 = 0 \ .
\ee 
Here we recall that $\cA_2$
is closed.

\bibliographystyle{JHEP}
\bibliography{refs}

\end{document}